\documentclass[aps, 10pt, notitlepage, twocolumn, superscriptaddress,
nofootinbib,longbibliography]{revtex4-2}

\usepackage{amsmath}
\usepackage{amssymb}
\usepackage{amsfonts}
\usepackage[utf8]{inputenc}
\usepackage[T1]{fontenc}
\usepackage{mathrsfs}
\usepackage{anyfontsize}

\usepackage[linktocpage,breaklinks]{hyperref}
\usepackage[usenames,dvipsnames]{xcolor}

\usepackage{txfonts}
\usepackage{tensor}
\usepackage{bm}

\usepackage{graphicx}
\usepackage{epsfig}
\usepackage{epstopdf}

\usepackage{natbib}
\usepackage{hyperref}

\hypersetup{colorlinks=true, citecolor=Purple, linkcolor=Purple,
urlcolor=Purple}

\usepackage{tikz}

\begin{document}

\title{Imaging compact boson stars with hot-spots and thin accretion disks}

\author{Jo\~{a}o Lu\'{i}s Rosa} \email{joaoluis92@gmail.com}
\affiliation{University of Gda\'{n}sk, Jana Ba\.{z}y\'{n}skiego 8, 80-309 Gda\'{n}sk, Poland}
\affiliation{Institute of Physics, University of Tartu, W. Ostwaldi 1, 50411 Tartu, Estonia}
\author{Caio F. B. Macedo} \email{caiomacedo@ufpa.br}
\affiliation{Faculdade de F\'isica, Campus Salin\'opolis, Universidade Federal do Par\'a, 68721-000, Salin\'opolis, Par\'a, Brazil}
\author{Diego Rubiera-Garcia} \email{drubiera@ucm.es}
\affiliation{Departamento de F\'isica Te\'orica and IPARCOS,
	Universidad Complutense de Madrid, E-28040 Madrid, Spain}

\begin{abstract}
In this work we consider the observational properties of compact boson stars with self-interactions orbited by isotropically emitting (hot-spot) sources and optically thin accretion disks. We consider two families of boson stars supported by quartic and sixth-order self-interaction potentials, and choose three samples of each of them in growing compactness; only those with large enough compactness are capable to hold light-rings, namely, null bound orbits. For the hot-spots, using inclination angles $\theta=\{20^\circ, 50^\circ, 80^\circ \}$ we find a secondary track plunge-through image of photons crossing the interior of the boson star, which can be further decomposed into additional images if the star is compact enough. For accretion disks we find that the latter class of stars actually shows a sequence of additional secondary images in agreement with the hot-spot analysis, a feature absent in typical black hole space-times. Furthermore, we also find a shadow-like central brightness depression for some of these stars in both axial observations and at the inclination angles above. We discuss our findings in relation to the capability of boson stars to effectively act as black hole mimickers in their optical appearances as well as potential observational discriminators.
\end{abstract}

\maketitle


\section{Introduction}

The Kerr hypothesis establishes the universality of the celebrated Kerr solution \cite{Kerr:1963ud} to describe the end-state of full gravitational collapse in terms of the formation of a black hole entirely described for external observers by its mass and angular momentum \cite{joshi_2007}. Analytical and computational investigations upon its background successfully reproduces current observations of gravitational wave profiles out of binary mergers \cite{LIGOScientific:2016aoc,LIGOScientific:2017vwq} as well as the features of the strong light deflection around the supermassive objects at the heart of the M87 \cite{EventHorizonTelescope:2019dse} and Milky Way \cite{EventHorizonTelescope:2022wkp} galaxies. This way, the existence of black holes are taken as yet another success of Einstein's General Relativity to accurately describe our Universe and the objects living in it \cite{Will:2014kxa,Yagi:2016jml}.

Nonetheless, given the (arguable) impossibility of directly observing the most salient feature of black holes -- its event horizon -- as well as the theoretical and observational uncertainties and inherent bias in the interpretation of gravitational waves and shadow observations, in the last few years an entire field of {\it black hole mimickers} has blossomed  (see \cite{Cardoso:2019rvt} for a review of their observational status). These black holes mimickers usually are ultra-compact objects potentially capable of disguise themselves as black holes despite not having an event horizon. Among them, boson stars -- hypothetical macroscopic Bose-Einstein condensates -- bear a special place. This is so because they are supported by complex scalar fields with canonical kinetic and potential terms but, more importantly, because mechanisms for their dynamical generation are known \cite{Liebling:2012fv,Brito:2015yga,Brito:2015yfh}, thus overlooking the main criticisms applied to other popular black hole mimickers such as wormholes. Furthermore, they allow for a large flexibility in their implementation with scalar and  vector fields \cite{Brito:2015pxa} sustained by different classes of self-interactions, and with important phenomenological repercussions in X-ray spectroscopy \cite{Cao:2016zbh}, the dark matter problem \cite{Sharma:2008sc,Hui:2016ltb}, or gravitational wave signatures \cite{Palenzuela:2017kcg} and echoes \cite{Cardoso:2017cqb}. 

While gravitational wave observations currently focus on stellar-mass black holes only (while we await for the arrival of LISA and Einstein Telescope devices), shadow observations explore an entire different mass range, namely that of millions of solar masses upwards. Recently many studies have recognized the great opportunity to look for observational hints of black hole mimickers (including boson stars) hidden in shadow images \cite{Atamurotov:2013sca, Vincent:2015xta,Olivares:2018abq,Abdikamalov:2019ztb,Vincent:2020dij,Herdeiro:2021lwl,Lima:2021las}. Such images are created by highly-bent trajectories of light rays issued by the accretion disk partnering the compact object, and which consists on a wide ring of radiation (the light-ring) enclosing a central brightness depression (the shadow). Such features are strongly linked to yet another salient feature of the Kerr black hole, namely the existence of unstable null bound orbits, identified as light-rings, which are a generic feature of asymptotically flat black holes \cite{Cunha:2020azh} but can also be sustained by other ultra-compact objects. Whether i) a black hole mimicker without such bound orbits can still mimic the observed light-ring/shadow features in the images of a black hole and ii) new features may arise that effectively act as observational discriminators between a black hole and its mimicker, are under heavy scrutiny in the literature. The most succulent feature to carry out these tasks is the sequence of additional images in which the secondary images and the light-ring can be decomposed into, depending on the properties of the orbiting material \cite{Gralla:2019xty,Johnson:2019ljv,Wielgus:2021peu}, and the presence/absence of a shadow including its (calibrated) size \cite{Chael:2021rjo,Vagnozzi:2022moj}. 

The main aim of this work is to analyze the features above by imaging two families of spherically symmetric compact boson stars supported by quartic and sixth-order self-interaction terms using two methods: the observational properties of hot-spots (bright regions associated to temperature anisotropies of the non-homogeneous accretion flow \cite{Vos:2022yij}) orbiting the boson star, and those of optically and geometrically thin accretion disks, emitting isotropically. This analysis brings out the succulent observational features of these objects. Indeed, for the first method we observe a secondary track, plunge-through trajectory, on top of the primary-track of the hot-spot in the integrated fluxes of those boson stars, which can be further decomposed into additional secondary tracks for large enough compactness, a feature that becomes more acute for larger observational angles, in agreement with previous results \cite{Rosa:2022toh}. This neatly distinguishes some of our configurations from canonical black holes and triggers new observational opportunities. Indeed, when using the second method for such very compact boson stars when illuminated by well-motivated intensity profiles, we find two interesting features of some of these objects: i) a sequence of new secondary images that do not appear in their black hole counterparts and ii) a shadow-like feature, i.e., a central brightness depression. While the first such feature is in agreement with the one found in the hot-spots and may act as a clear observational discriminator between boson stars and black holes, the second allows such boson stars to effectively act as black hole mimickers, even at large observation inclination angles. 

This work is organized as follows: in Sec. \ref{sec:II} we set our theoretical framework and specify the three plus three configurations of boson stars supported by quartic and sixth-order self-interactions with different compactnesses, respectively, develop the equations for geodesic motion and discuss the stability of time-like orbits. In Sec. \ref{sec:III}, we consider the observational properties of hot-spots analyzing the integrated fluxes and astrometrical quantities at observation angles $\theta=\{20^\circ, 50^\circ, 80^\circ \}$. In Sec. \ref{sec:IV} we consider the observational properties of these boson stars when illuminated by a geometrically and optically thin accretion disk, placing the focus on the multi-ring structure and the shadow-like mimicking features of some of these stars at both axial inclination and the observational inclinations mentioned above. In Sec. \ref{sec:V} we conclude with a summary and critical discussion of our results.

\section{Theoretical framework} \label{sec:II}

\subsection{Models and configurations}

Let us a consider a (complex) scalar field $\Phi$ minimally coupled to the gravitational field via the action ($a=0,1,2,3$)
\begin{equation} \label{eq:action}
	\mathcal{S}=\int d^4x\sqrt{-g}\left[\frac{R}{16\pi}-\frac{1}{2}\partial_a\Phi^*\partial^a\Phi-\frac{1}{2}V(|\Phi|^2)\right],
\end{equation}
where $g$ is the determinant of the space-time metric $g_{\mu\nu}$ written in terms of a coordinate system $x^a$, $R$ is the Ricci scalar, a star $\Phi^*$ denotes a complex conjugate, and $V$ is the scalar potential. We have adopted a system of geometrized units for which $G=c=1$. The corresponding field equations are obtained by varying Eq. (\ref{eq:action}) with respect to the metric $g_{ab}$ and the scalar field $\Phi$ to yield
\begin{align}
	G_{ab}=8\pi T_{ab},\label{eq:einstein}\\
	\nabla_a\nabla^a\Phi= \frac {dV}{d|\Phi|^2}\Phi,\label{eq:scalar}
\end{align}
where $T_{ab}$ is the stress-energy tensor of the complex scalar field $\Phi$ and it is given by
\begin{equation}
	T_{ab}=\frac{1}{2}(\nabla_a\Phi^*\nabla_b\Phi+\nabla_b\Phi^*\nabla_a\Phi)-\frac{1}{2}g_{ab}(\nabla_c\Phi^*\nabla^c\Phi+V),
\end{equation}
where $\nabla_c$ denotes covariant differentiation. We are interested here in considering static and spherically symmetric boson stars, and thus we consider the ansatz for the metric
\begin{align}
	ds^2&=-A(r) dt^2+B(r)^{-1}dr^2+r^2d\Omega^2,\\
	\Phi&=\phi(r) e^{-i \omega t},
\end{align}
where we have introduced the metric functions $A(r)$, $B(r)$, while $\phi(r)$ characterizes the radial part of the scalar field, with $\omega$ denoting its frequency, and $d\Omega^2$ denotes the line-element on the two-sphere. Replacing this ansatz into the field equations in Eq. \eqref{eq:einstein} and \eqref{eq:scalar} leads to the equations of motion of the system (a prime denotes a radial derivative):
\begin{align}
	&\frac{r B'+B-1}{r^2}=-
	2\pi \left(\frac{w^2 \phi^2}{A}+B \phi'^2+V\right),\\
	&\frac{\frac{B r A'}{A}+B-1}{r^2}=2\pi  \left(\frac{w^2 \phi^2}{ A}+B\phi'^2\right),\\
	&\frac{1}{2} \phi' \left(B \left(\frac{A'}{A}+\frac{4}{r}\right)+B'\right)+\phi \left(\frac{w^2}{A}-\frac{dV}{d|\phi|^2}\right)+B \phi''=0.
\end{align}
This is a highly non-linear system whose resolution demands the employment of suitable numerical methods. To this end, we supply asymptotic boundary conditions by demanding asymptotic flatness of the geometry via a Schwarzschild-like behaviour, and a vanishing radial scalar field at $r \to \infty$, that is 
\begin{align}
	A(r\to\infty)&=1-\frac{2M}{r}\label{eq:bc1},\\
	B(r\to\infty)&=1-\frac{2M}{r} \label{eq:bc2},\\
	\phi(r\to \infty)&=0,
\end{align}
where $M$ is the total mass of the boson star. At the origin we demand the metric functions to be normalized to a finite value, and the radial scalar field to a target value $\phi_c$, i.e., 
\begin{align}
	A(r\approx 0)&=A_c,\label{eq:bco1}\\
	B(r\approx 0)&=1,\label{eq:bco2}\\
	\phi(r\approx 0)&=\phi_c,~{\rm and}~\phi'(r\approx0)=0.\label{eq:bco3}
\end{align}
In practice, we can always set $A_c=1$ through a time reparametrization. This leads to a background that does not asymptotically match Eq. \eqref{eq:bc1}. Upon finishing the integration, we rescale the time coordinate (changing $\omega$ and $A$) to obtain the spacetime solution in the usual Schwarzschild-like coordinates. Therefore, we integrate the differential equations Eqs.~\eqref{eq:einstein} and~\eqref{eq:scalar} for static and spherically symmetric backgrounds from the origin using the boundary conditions in Eqs. \eqref{eq:bco1}-\eqref{eq:bco3}. In order to do so, we must first specify the scalar potential. In what follows, we consider two well-motivated potentials:

\begin{itemize}
	\item \textit{$V=\mu^2|\Phi|^2+\Lambda|\Phi|^4$}~\cite{Colpi:1986ye}, where $\mu$ is the mass term and $\Lambda$ is a coupling constant. Boson stars (BS) presenting quartic self-interactions can be highly massive objects as the maximum mass configuration scales with $\Lambda$ as $M_{\rm max}\sim \Lambda^{1/2}m_{\rm p}^3/\mu^2$ for large values of $\Lambda$. However, the compactness of these solutions saturates for large $\Lambda$, with the boson stars radius $R$ never being smaller than $6M$. This implies that all circular orbits with orbital radii $r_{o}>0$ are stable independently of the value of $\Lambda$~\cite{Amaro-Seoane:2010pks}. We shall denote this class of models as $\Lambda$BS.
	
	\item \textit{$V=\mu^2|\Phi|^2(1+|\Phi|^2/\alpha^2)^2$}~\cite{Lee:1986ts} where $\alpha$ is a constant parameter. Potentials of this type allow for (degenerate) vacuum configurations. This potential is usually labeled as solitonic, and its self-gravitating solutions as solitonic boson stars (SBS) as it is one of simplest potentials that feature non-topological solitons in the absence of gravity. In this case,  ultra-compact solutions can be achieved in the limit $\alpha\to0$, with the minimum radius being $R\approx 2.81M$ \cite{Lee:1991ax,Cardoso:2021ehg}. Because of their compactness, solitonic BS can have light-rings, being an interesting candidate as a spherical black hole mimicker~\cite{Macedo:2013jja}.
\end{itemize}

\begin{figure*}
	\includegraphics[width=\linewidth]{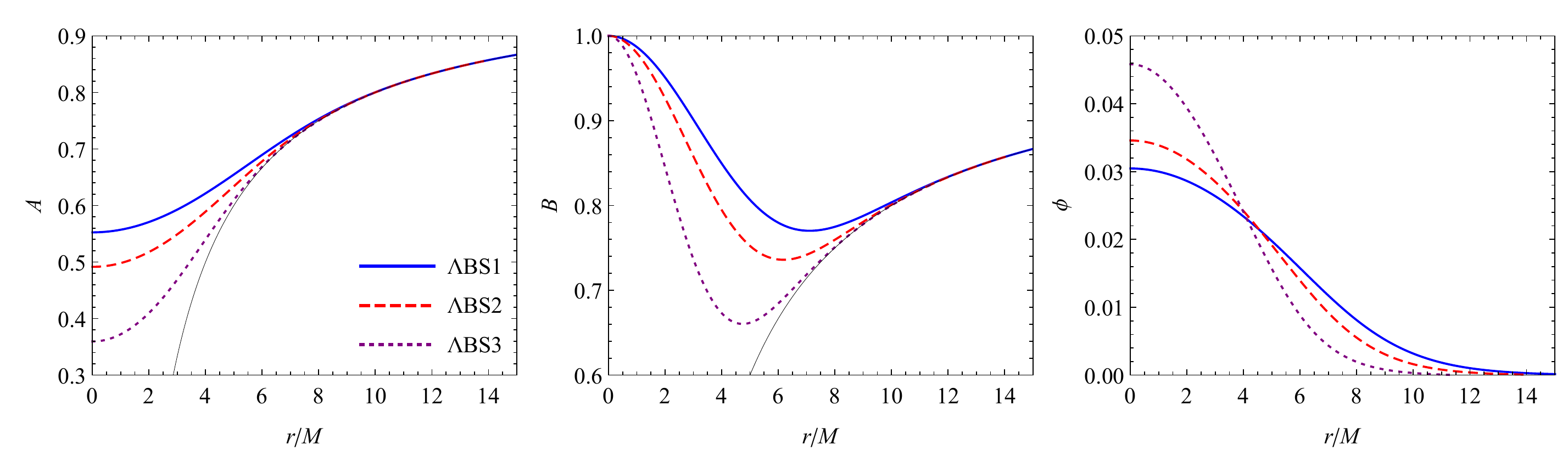}
	\includegraphics[width=\linewidth]{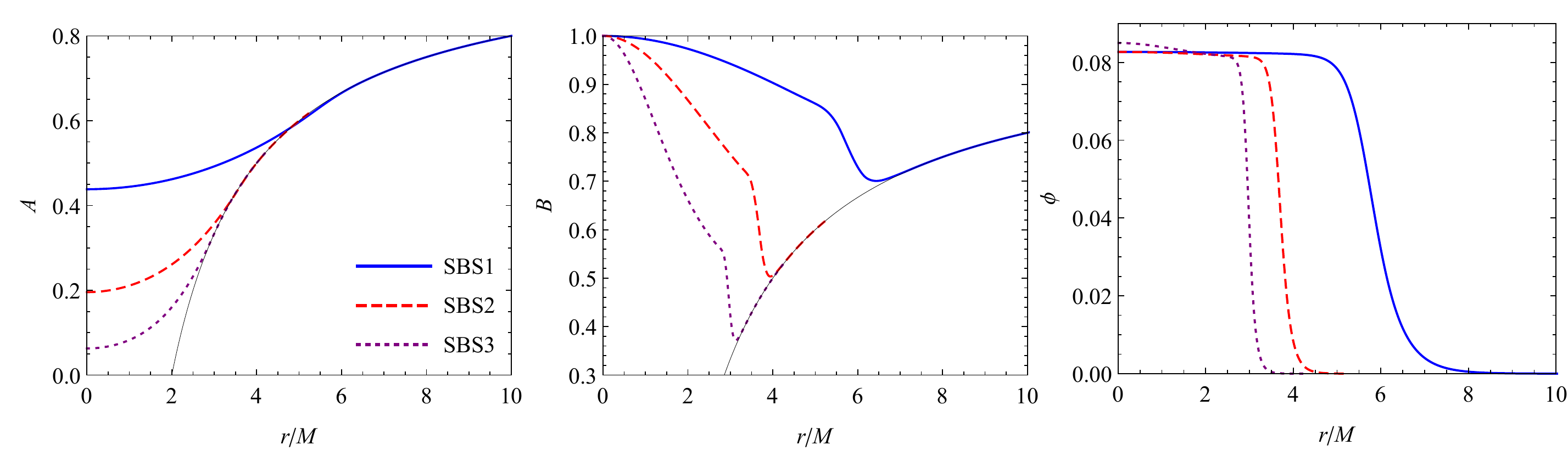}
	\caption{Background quantities for the metric coefficients and the scalar field for the three $\Lambda$BS (top) and the three SBS (bottom) compared to the Schwarzschild black hole (thin black line).} \label{fig:functions}
\end{figure*}

\begin{figure*}
	\includegraphics[width=\columnwidth]{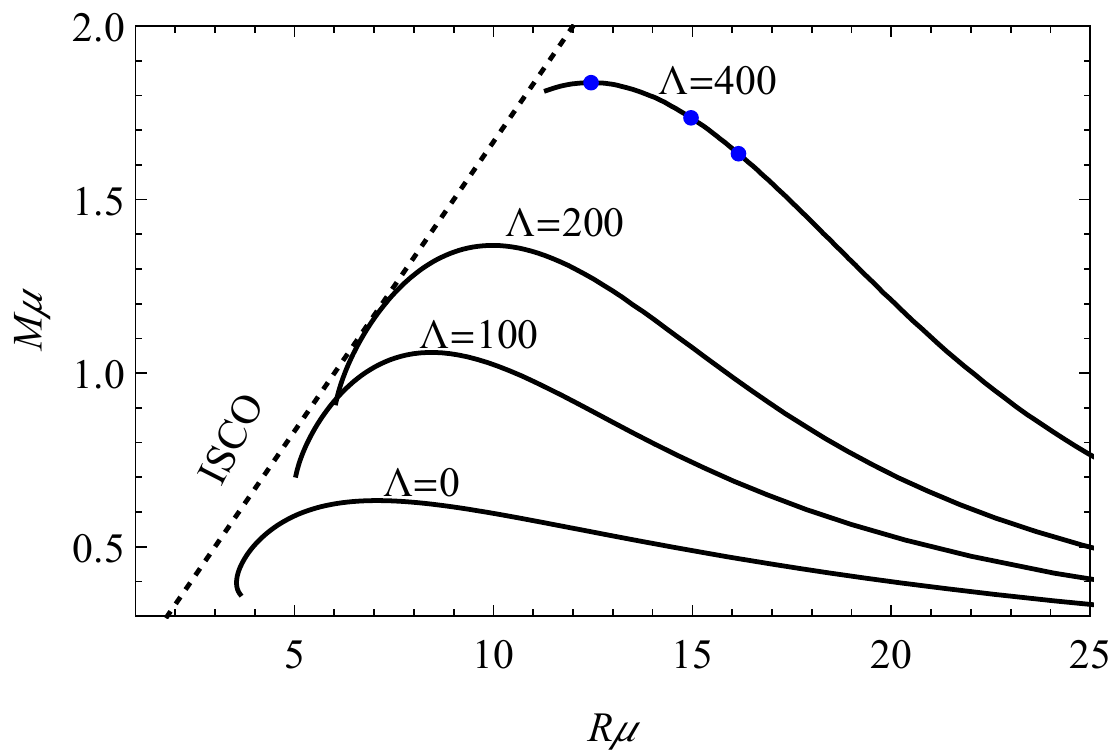}\includegraphics[width=\columnwidth]{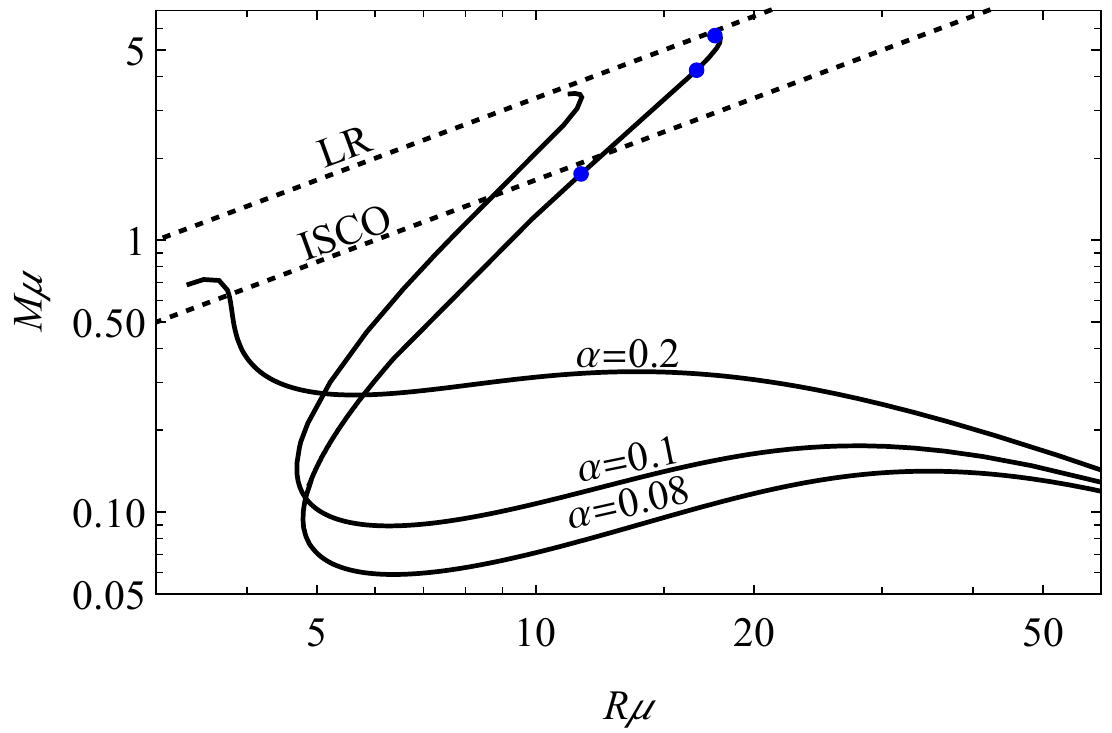}
	\caption{Mass-radius curves for $\Lambda$BS (left) and SBS (right) configurations considering different values of the parameters. While for the $\Lambda$BS we see that the radius never crosses the ISCO region (given by $R=6M$), the solitonic potential can generate solutions very close to the corresponding light-ring in the Schwarzschild black hole ($R=3M$). The blue markers indicate the solutions used in this paper.}\label{fig:mass_radius}
\end{figure*}

In this work we consider three candidates belonging to each $\Lambda$BS and SBS class, whose chosen parameters and main features are displayed in Table ~\ref{tab:solutions}. We focus on the values $\Lambda=400$ and $\alpha=0.08$, picking three solutions for each potential, all of which are linearly stable against radial perturbations. We depict in Fig. \ref{fig:functions} the behaviour of the metric and scalar field functions, and in Fig.~\ref{fig:mass_radius} the corresponding mass-radius relations, highlighting with markers the solutions explored in this paper.  As the scalar field $\phi(r)$ decays exponentially for $r\gg\mu^{-1}$ but never vanishes, one can only define an effective radius for the BS, which we define in such a way as to encompass $98\%$ of its total mass. The mass function $m(r)$ can be found through
\begin{equation}
	B(r)=1-\frac{2m(r)}{r},
\end{equation}
and, therefore the radius is defined by $m(R)=0.98M$. Numerical solutions for very small values of $\alpha$ and very large values of $\Lambda$ are challenging to find with the usual shooting methods, but can be found via alternative semi-analytical approximations \cite{Lee:1991ax,Amaro-Seoane:2010pks}. Nonetheless, we shall use the full numerical solutions in order to analyze the space-time, with the aid of analytical fits for the image computations.

\begin{table}
\begin{center}
	\caption{Boson star configurations used in this paper. We select three different solutions for each model: $ \{ \Lambda$BS1,  $\Lambda$BS2,  $\Lambda$BS3$\}$ for the $\Lambda$ stars, and $ \{$SBS1, SBS2, SBS3$\}$ for the solitonic stars. The parameters are $\phi_c$ for the central scalar field, $\mu$ for the mass term, $R$ for the radius of the star, $M$ for its mass, $\mathcal{C} \equiv M/R$ for the compactness (for a Schwarzschild black hole, $\mathcal{C}=1/2$), and $\omega$ for the scalar field frequency.}\label{tab:solutions}
	\begin{tabular}{c c c c c c}
		\hline\hline
		Configuration & $\phi_c$ & $\mu M$ & $\mu R$  & $\mathcal{C}$ & $\omega/\mu$ \\
		\hline\hline
		 $\Lambda$BS1 & 0.03045  & 1.6321 & 16.1577 & 0.10101 & 0.88124 \\  
		 $\Lambda$BS2 & 0.03457 & 1.7356 & 14.9648 & 0.11597 & 0.86410\\
		 $\Lambda$BS3 & 0.04582 & 1.8368 & 12.4524 & 0.14750 & 0.82786\\
		 SBS1 & 0.0827 &  1.7531 & 11.5430 & 0.1518 & 0.25827 \\
		 SBS2 & 0.0827 & 4.220 &  16.6520 & 0.25342 & 0.17255 \\
		 SBS3 & 0.0850 &  5.655 & 17.6470 & 0.32045 & 0.13967 \\
		\hline\hline
	\end{tabular}
\end{center}
\end{table}

\subsection{Time-like circular geodesics, marginally stable orbits and light-rings}

Before going into details about the imaging of boson stars, it is instructive to study the geodesic structure of these space-times. The Lagrangian describing orbits at the equatorial plane $\theta=\pi/2$ (something we can fix without loss of generality due to the spherical symmetry of the system) is given by
\begin{equation}
	2\mathcal{L}=-A \dot{t}^2+B^{-1}\dot{r}^2+r^2\dot\varphi^2=-\delta,
\end{equation}
where the overdot indicates derivative with respect to the geodesic affine parameter, $\varphi$ is the azimuthal angle, and $\delta=1$ (0) for time-like (null) geodesics. Introducing the following definitions for the specific energy and angular momentum per unit mass, i.e.:
\begin{equation}
	\varepsilon=-\frac{\partial \mathcal{L}}{\partial \dot{t}}=A\dot{t},~~{\rm and}~~\ell=\frac{\partial \mathcal{L}}{\partial \dot{\varphi}} =r^2 \dot{\varphi}
\end{equation}
the equation of motion for the radial coordinate is given by the effective balance equation
\begin{equation}
	\frac{A}{B}\dot{r}^2=\varepsilon^2+V_{\rm eff},
\end{equation}
where
\begin{equation} \label{pot}
	V_{\rm eff}=A\left(\delta+\frac{\ell^2}{r^2}\right),
\end{equation}
is the effective potential describing the geodesics. Let us consider the case of time-like circular orbits. We can use the above equations, together with the first derivatives of the effective potential, to find the specific energy and angular momentum, obtaining
\begin{equation}
 \varepsilon^2=\frac{2A}{2A-r_o A'},\qquad l_P^2=\frac{r_o^3A'}{2A-r_o A'},
\end{equation}
where $r_o$ is the radius of the circular orbit and all the quantities above are evaluated at $r=r_o$. 

\begin{figure}
	\includegraphics[width=\columnwidth]{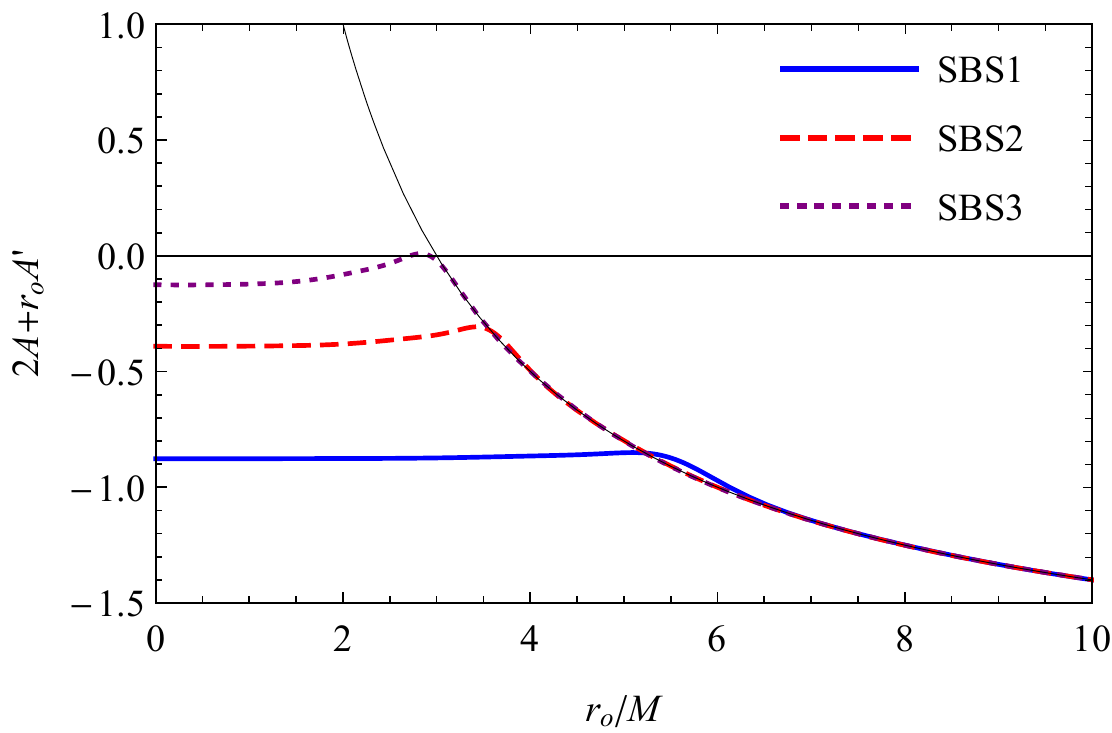}
	\caption{Discriminator for the existence of light-rings. The roots (i.e. the zeroes of $2A(r_o)-r_o A'(r_o)$), when present, correspond to the light-ring positions.}\label{fig:light-ring_pos}
\end{figure}

Note that both the specific energy and angular momentum diverge at a (possible) orbit for which
\begin{equation}
2A(r_o)-r_oA'(r_o)
\label{eq:light-ring_c}
\end{equation}
vanishes. If the above equation is satisfied for real values of $r_o$, this corresponds to the position of null circular orbits (in the Schwarzschild space-time this results in $r_o=3M$). Note that ultra-compact objects without event horizons are known for having light-rings that come in pairs \cite{Cunha:2017qtt,Cunha:2022gde}. Therefore, it is instructive to track  Eq.~\eqref{eq:light-ring_c} to search for possible light-rings. We focus on SBS as these are the most compact objects explored in this paper. We display the evolution of this quantity with the ratio $r_o/M$ for the three SBS configurations presented in this work in Fig. \ref{fig:light-ring_pos}. Among all models (including the $\Lambda$BS), only the SBS3 one presents such light-rings (around $r_o/M\approx 3$).

\begin{figure}
	\includegraphics[width=\columnwidth]{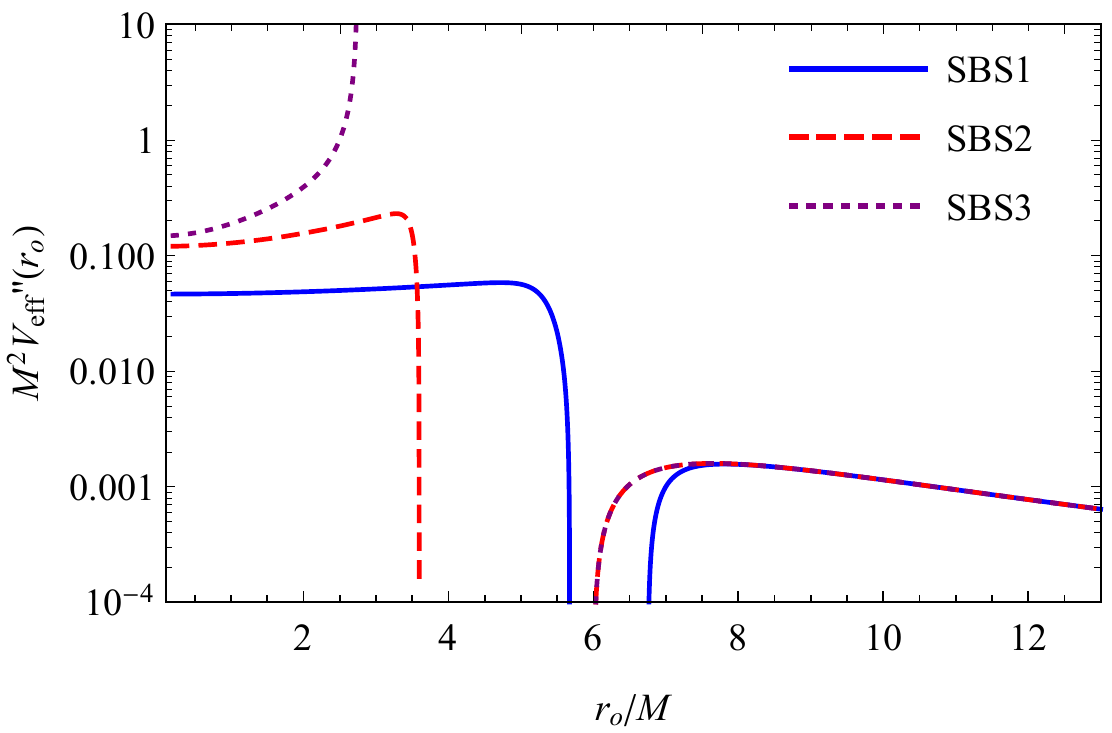}
	\caption{Second derivative of the effective potential in Eq. (\ref{pot}) as function of the orbital radius. Even the least compact SBS solution presents a window of unstable orbits, with radius larger than the corresponding innermost stable circular orbit in the Schwarzschild black hole.}\label{fig:ddpotential}
\end{figure}

Although it is physically possible to place a particle in a circular orbital motion close to light-rings, this analysis does not tell us anything about the stability of such orbits. The stability of time-like circular orbits can be analyzed through the sign of the effective potential in Eq. (\ref{pot}), i.e.,
\begin{equation}
	V_{\rm eff}''
	\left\{
	\begin{array}{l l}
		>0, &~\text{Stable},\\
		=0, &~\text{Marginally stable (ISCO for BH)},\\
		<0, &~\text{Unstable},
	\end{array}
	\right.
\end{equation}
provided that $(\varepsilon,\ell)$ are real for the orbit to exist. By analyzing the models investigated in this paper, we find that for $\Lambda$BS all circular orbits are stable, meaning that accretion disks may extend all the way down to the center of the star. For the SBS models, however, we find that there is a window in which either the orbits do not exist or are unstable. This is illustrated in Fig.~\ref{fig:ddpotential}, where we plot the second derivative of the effective potential for the SBS models. We plot this quantity logscale to illustrate solely the stable orbits. For SBS2 and SBS3, the outer marginally stable circular orbit is located very close to $6M$, similarly to the Schwarzschil black hole case. This is not surprising, as these boson stars have radius such that $R/M<6$. Surprisingly, for SBS1, even though the configuration have radius bigger than $6M$, unstable orbits still exist. This illustrates that naively looking into the compactness only to search for marginally stable circular orbits or light-rings in BSs might lead to wrong results.

\begin{figure*}
	\includegraphics[width=\columnwidth]{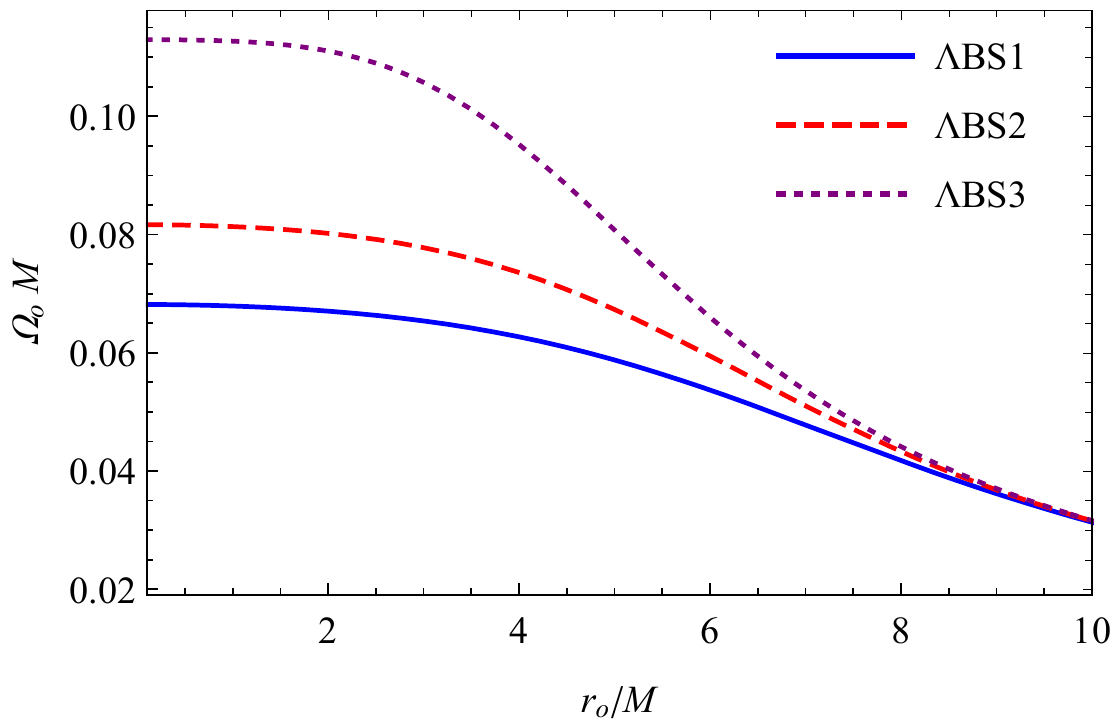}\includegraphics[width=\columnwidth]{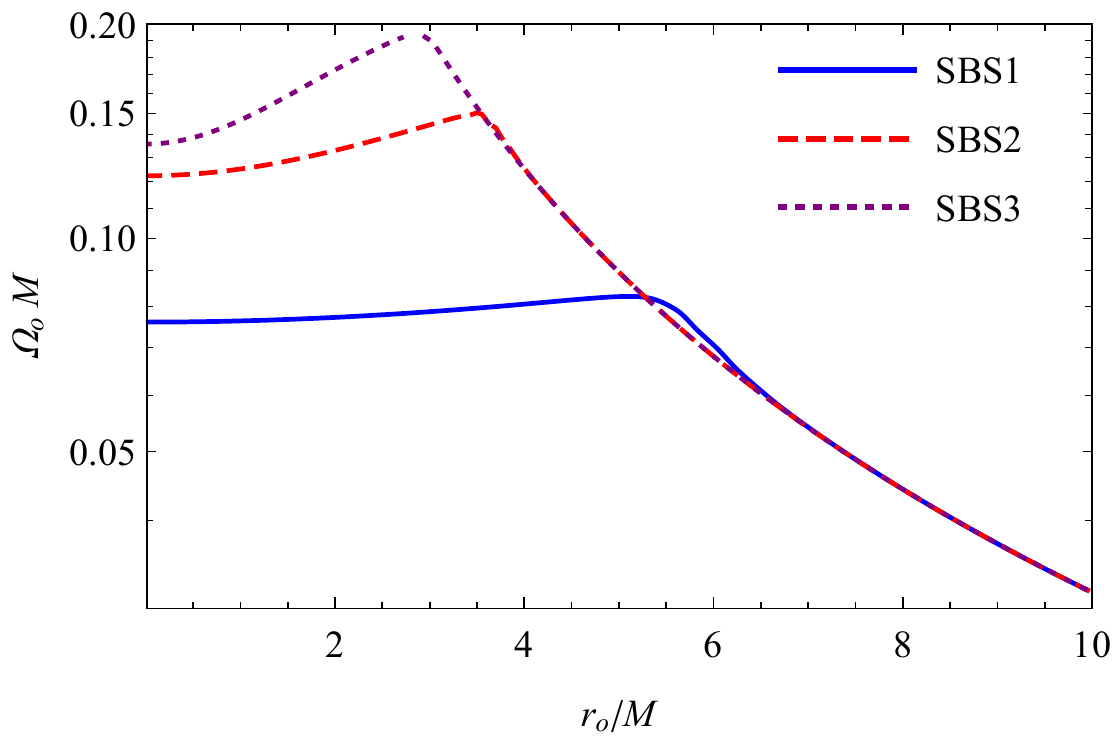}
	\caption{Angular frequency for time-like circular geodesic motion for the different boson star configurations explored in this paper. \textit{Left panel:} for $\Lambda$BS all time-like circular geodesics are stable and present a maximum close to the center of the star. \textit{Right panel:} for SBS however, the maximum frequency is always at some finite radius. We stress here that not all of these orbits are stable (as indicated in Fig.~\ref{fig:ddpotential}). Moreover, for the SBS3 orbits there is a forbidden region in between the light-rings.}\label{fig:frequency}
\end{figure*}

In Ref.~\cite{Olivares:2018abq} it was pointed out that having stable time-like circular orbits inside the BS is not enough to determine whether a physical light source may exist in those orbits (see also Ref.~\cite{Herdeiro:2021lwl}). A central point is the existence of a maximum in the angular frequency of the time-like geodesics $\Omega_o$ at some radius, which introduces a scale for the inner edge of the accretion disks. The angular frequency for time-like geodesics can be computed through
\begin{equation}
	\Omega_o=\left.\frac{\dot\varphi}{\dot t}\right|_{r=r_o}=\sqrt{\frac{A'(r_o)}{2r_o}}.
\end{equation} 
In Fig.~\ref{fig:frequency} we show the angular frequency for the BSs explored in this paper. We see that for the $\Lambda$BS cases (left panel of Fig.~\ref{fig:frequency}) the maximum of the frequency is located near the center of the star, indicating that it would be difficult for accretion disks to have a Schwarzschild-like structure. However, for all SBS explored in this paper, a maximum in the frequency is observed (left panel of Fig.~\ref{fig:frequency}). This indicates that SBS are more likely to produce accretion disk structures similar to those of black holes.

\begin{figure*}
	\includegraphics[width=\columnwidth]{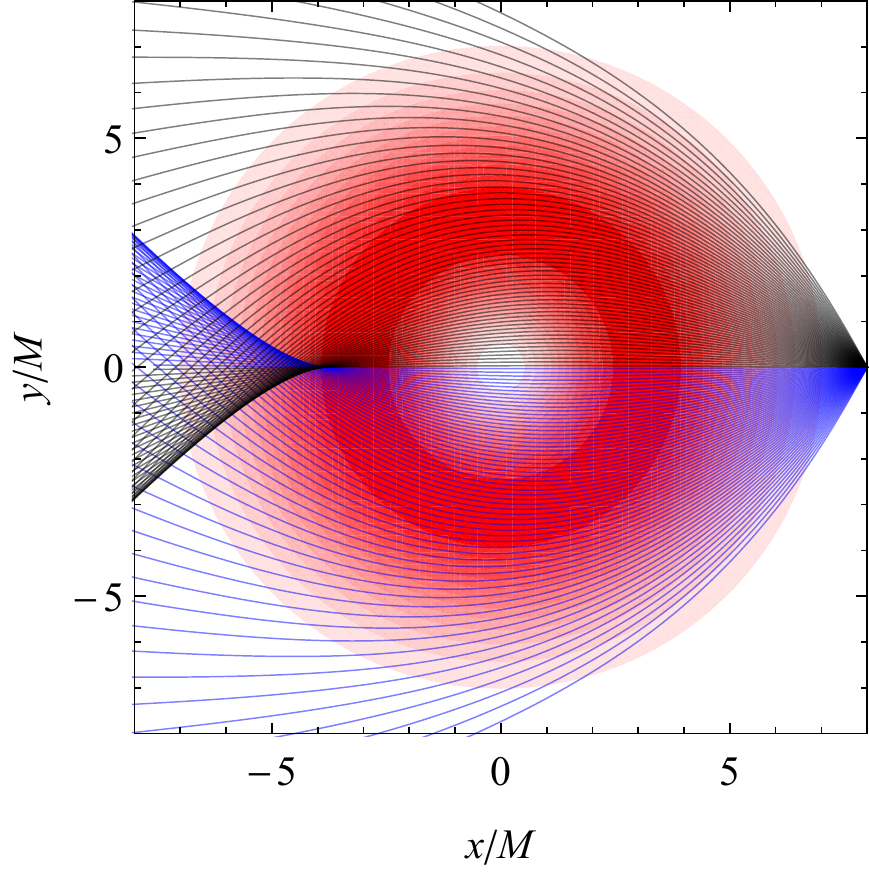}\includegraphics[width=\columnwidth]{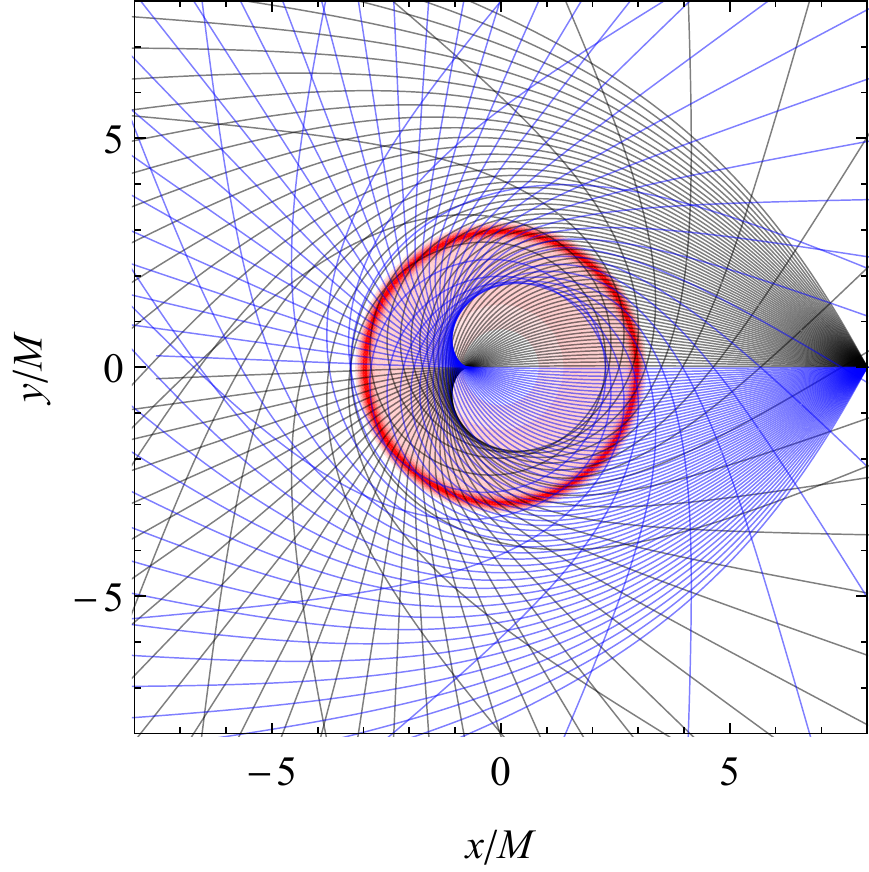}\\
	\includegraphics[width=\columnwidth]{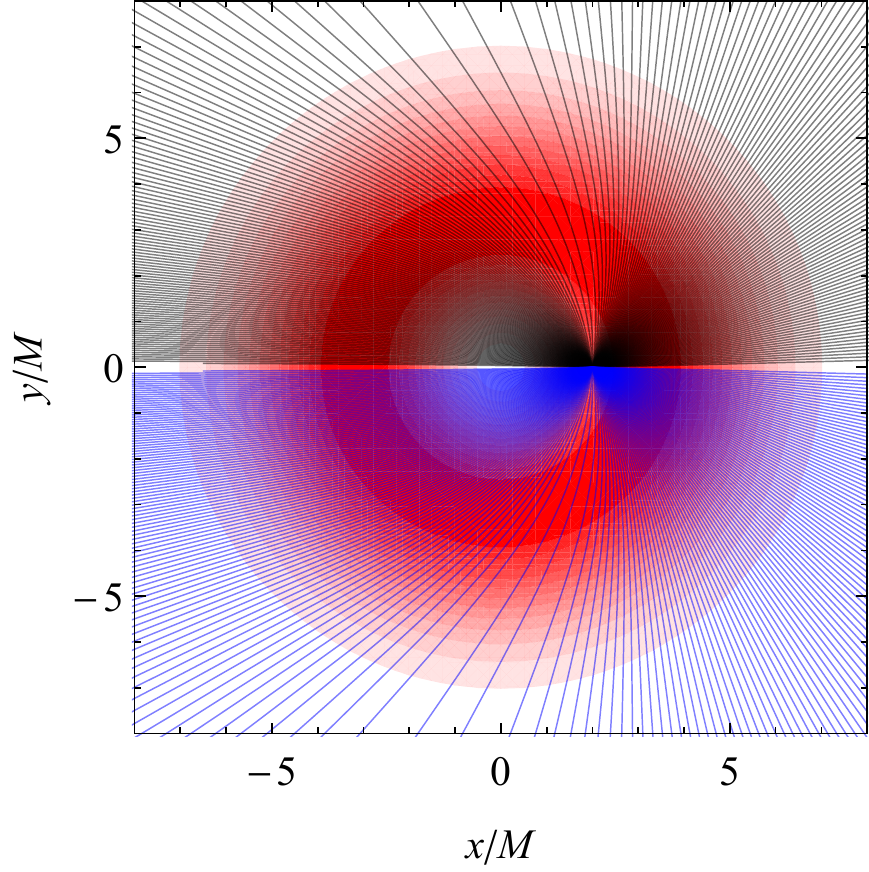}\includegraphics[width=\columnwidth]{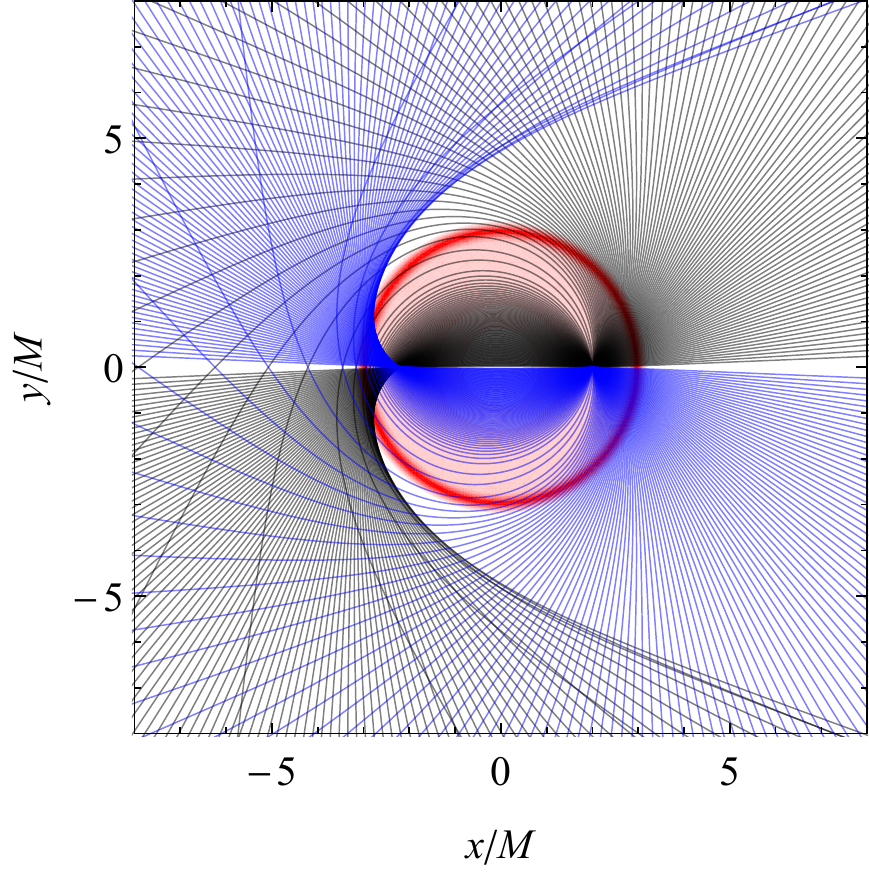}
	\caption{Light-ray deflections by the $\Lambda$BS 3 (left panel) and the SBS3 (right panel) assuming an emitter located at $(x,y)=(8M,0)$ (top panels) and $(x,y)=(2M,0)$ (bottom panels). The contour plot represents the derivative of the mass function, which decreases exponentially near the stellar radius.} \label{fig:ray-tracing}
\end{figure*}

Finally, from directly integrating the geodesic equation we can illustrate how strong the lensing can be in the BSs explored in this paper. Let us focus on the most compact models for each self-interaction, i.e., $\Lambda$BS3 and SBS3. We consider a single emitter located at either $(x,y)=(2M,0)$ or $(x,y)=(8M,0)$, to illustrate the behavior of a source located inside and outside the star, respectively. The result is shown in Fig.~\ref{fig:ray-tracing}. For a source outside the star, the $\Lambda$BS3 produces a caustic effect inside the star (something that was observed in other compact stellar models~\cite{Stratton:2019deq}) and the SBS3 model have strong deflections which is consistent with the fact that such configuration have light-rings. For the source inside the star, we see that the effect in the $\Lambda$BS3 is weaker, indicating that possible observational effects from that region should be mostly due to the gravitational redshift. For the SBS3 model, however, we can see strong deflections even for a source located inside the star, which combined with the strong redshift due to its compactness should provide a strong candidate to mimic black hole images. We shall further investigate more physical sources in the next sections.

\section{Orbits and hot-spots} \label{sec:III}

Let us now analyze the observational properties of hot-spots orbiting a central bosonic star, the latter described by the $\Lambda$BS and SBS configurations described in the previous section. For this purpose, we recur to the ray-tracing open-source code GYOTO \cite{Vincent:2011wz}, on which we model the hot-spot as an isotropically emitting spherical source orbiting the central object at some constant orbital radius $r_0$ and at the equatorial plane, i.e., $\theta=\pi/2$. As a run test, we use this software to ray-trace light trajectories in the two most compact configurations considered in this work, namely, $\Lambda$BS3 and SBS3, which are depicted in Fig. \ref{fig:ray-tracing} (left and right plots, respectively).

For the sake of this work, we have set the orbital radius to $r_o=8M$ and the radius of the hot-spot to $r_H=M$, where $M$ is the ADM mass of the background space-time. Under these assumptions, GYOTO outputs a 2-dimensional matrix of specific intensities $I^\nu_{lm}$ at a given time instant $t_k$. This matrix can be interpreted as an observed image, where each of the pixels $\{m,l\}$ is associated with an observed intensity. The simulation is then repeated through several time instants $t_k\in\left[0,T\right[$, where $T$ is the orbital period of the hot-spot, to obtain cubes of data $I_{klm}=\delta\nu I^\nu_{lm}$, where $\Delta\nu$ is the spectral width. We use these simulated cubes of data to produce three observables, namely:
\begin{enumerate}
\item Time integrated fluxes:
\begin{equation}
\left<I\right>_{lm}=\sum_k I_{klm},
\end{equation}
\item Temporal fluxes:
\begin{equation}
F_k=\sum_l\sum_m\Delta\Omega I_{klm},
\end{equation}
\item Temporal centroids:
\begin{equation}
\vec{c}_k=\frac{1}{F_k}\sum_l\sum_m\Delta\Omega	I_{klm}\vec{r}_{lm},
\end{equation}
\end{enumerate}   
where $\Delta\Omega$ is the solid angle of a single pixel and $\vec{r}_{lm}$ is a vector representing the displacement of the pixel $\{l,m\}$ with respect to the center of the observed image. A more popular astronomical observable, the temporal magnitude $m_k$, can then be obtained from the temporal flux $F_k$, as
\begin{equation}
m_k=-2.5\log\left(\frac{F_k}{\min\left(F_k\right)}\right).
\end{equation}
In Figs. \ref{fig:LBSflux} and \ref{fig:SBSflux} we show the integrated fluxes for the three $\Lambda$BS and the three SBS, respectively, as observed through three different inclination angles with respect to the vertical axis, chosen conveniently as $\theta=\{20^\circ,50^\circ,80^\circ\}$. The magnitudes for the same solutions and observation angles are plotted in Fig. \ref{fig:magnitudes}, and the corresponding centroids are plotted in Fig. \ref{fig:centroids}. In these two latter figures, we also provide a comparison between the three $\Lambda$BS solutions and the three SBS solutions for different inclination angles, in order to allow one to observe how an increase in the compactness of the bosonic star configuration affects its observables. In the following subsections, we analyze separately the integrated fluxes and the astrometrical observables.

\subsection{Integrated fluxes}

\begin{figure*}[t!]
\includegraphics[scale=0.35]{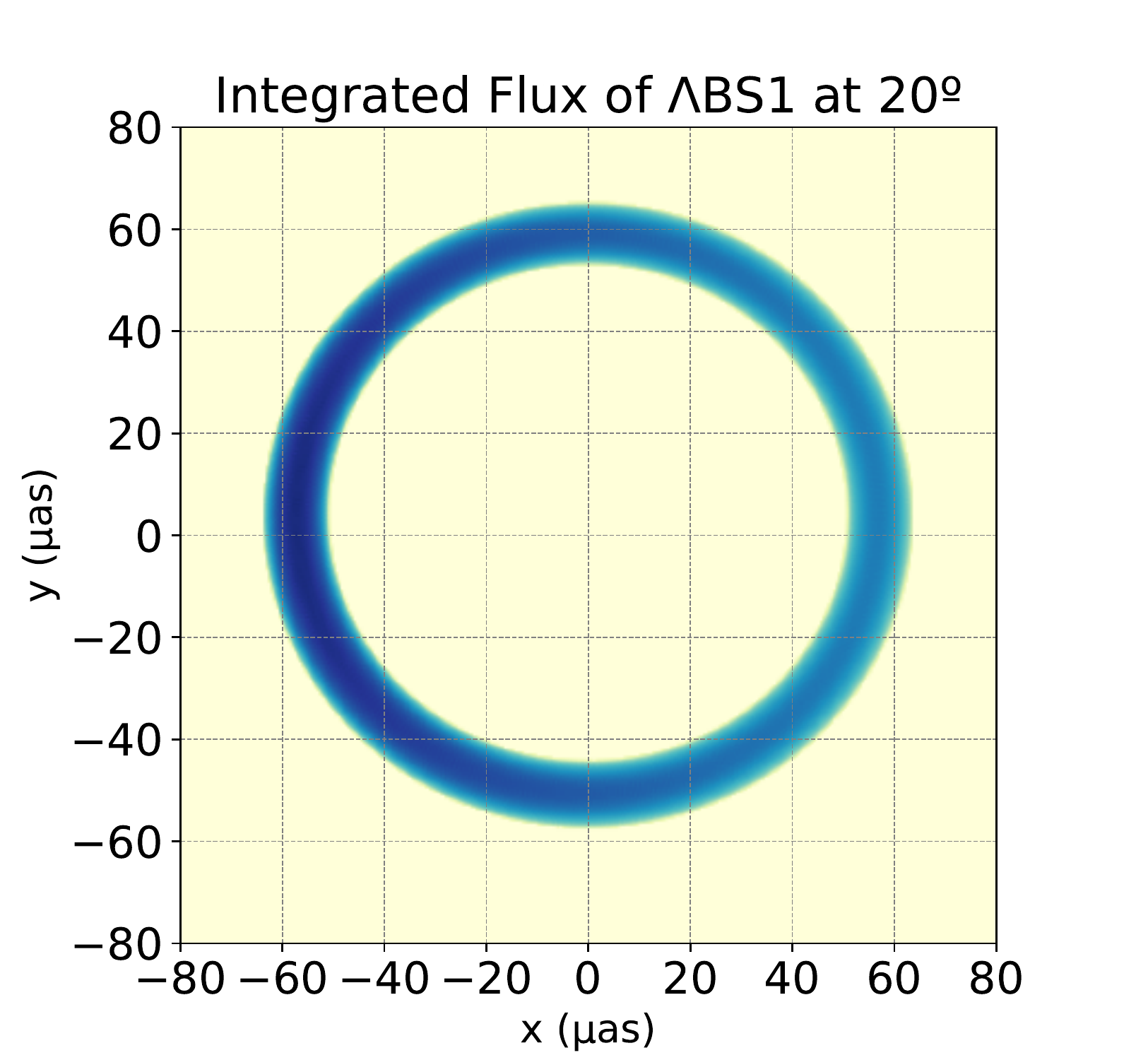}
\includegraphics[scale=0.35]{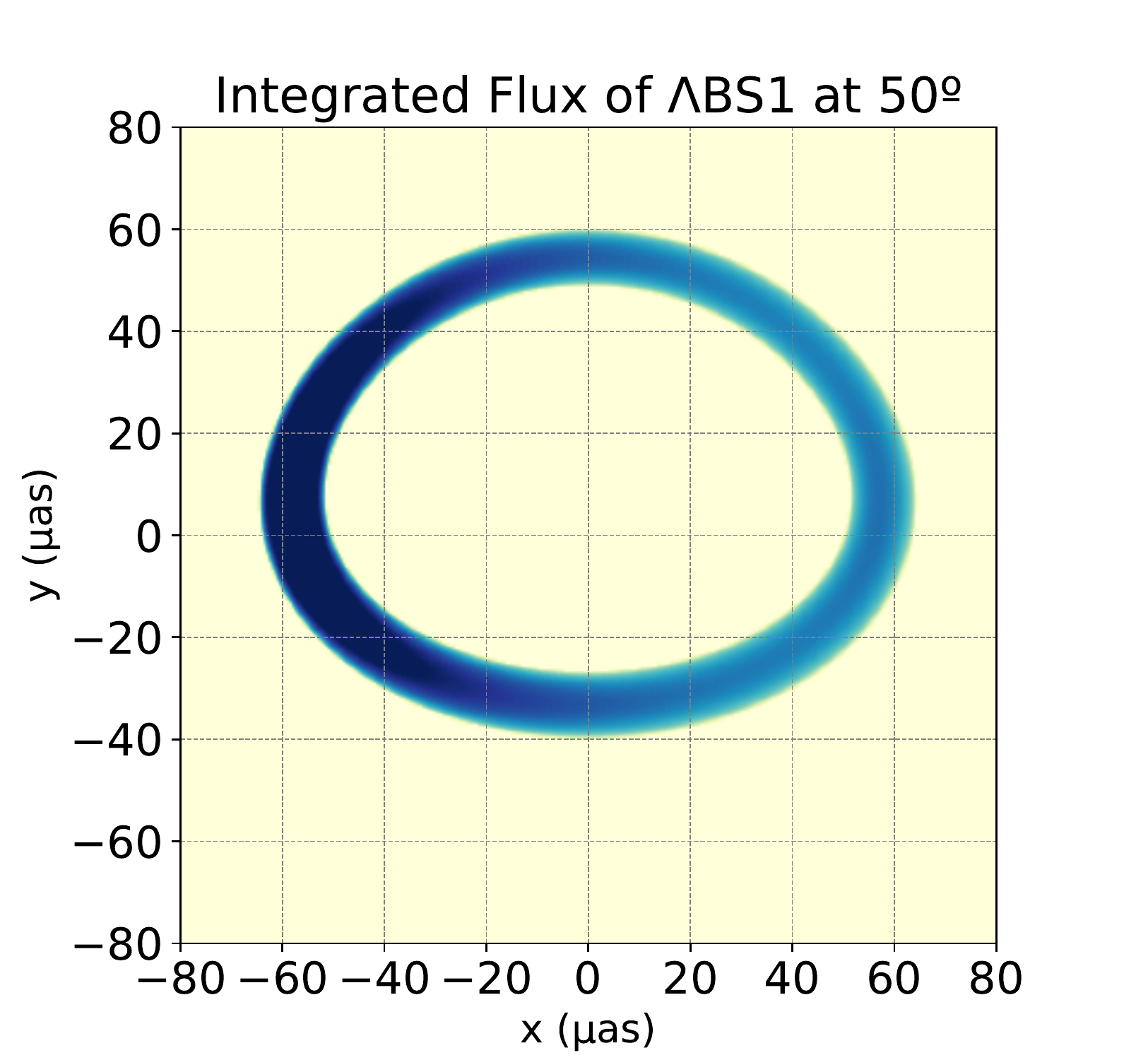}
\includegraphics[scale=0.35]{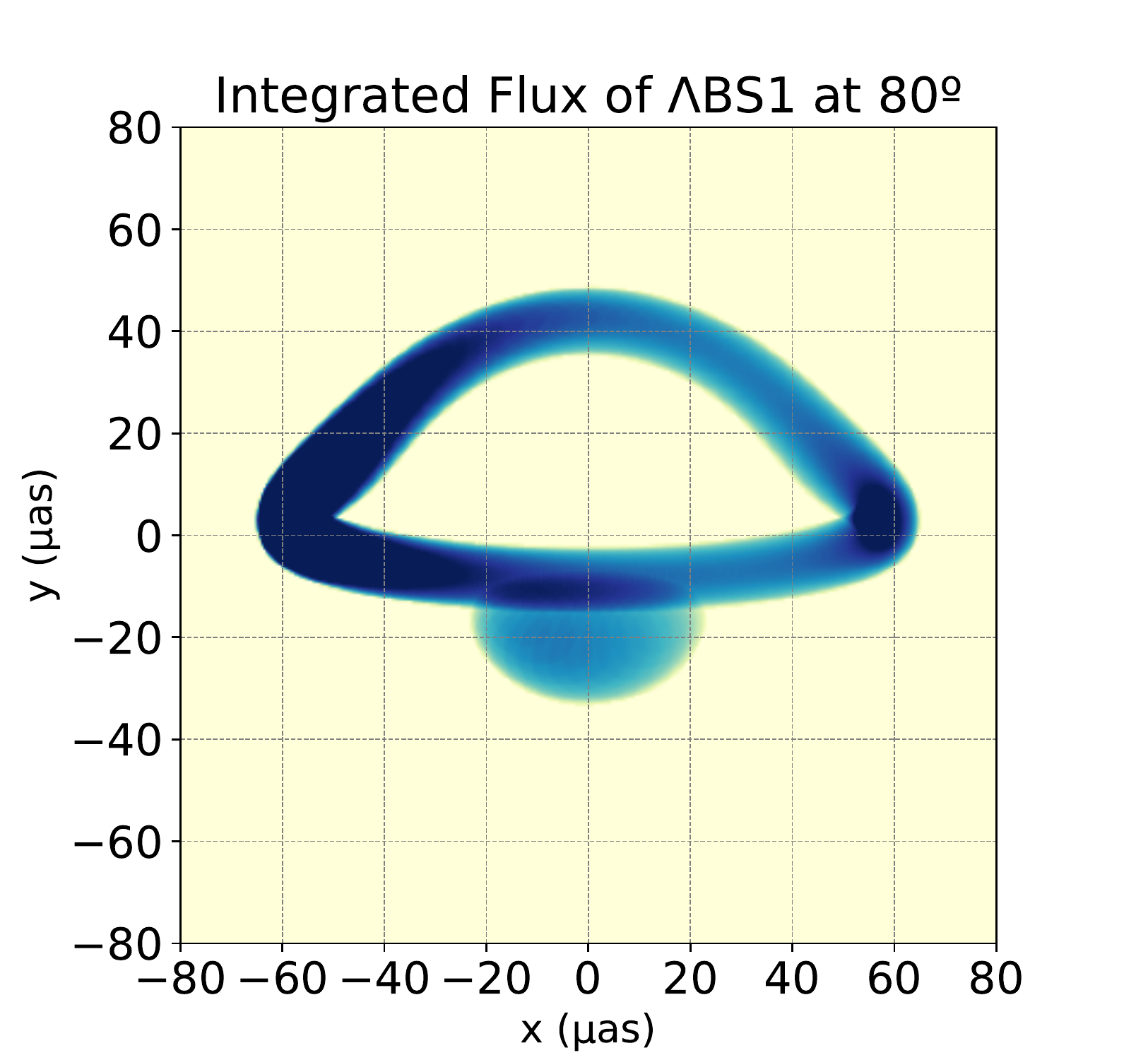}\\
\includegraphics[scale=0.35]{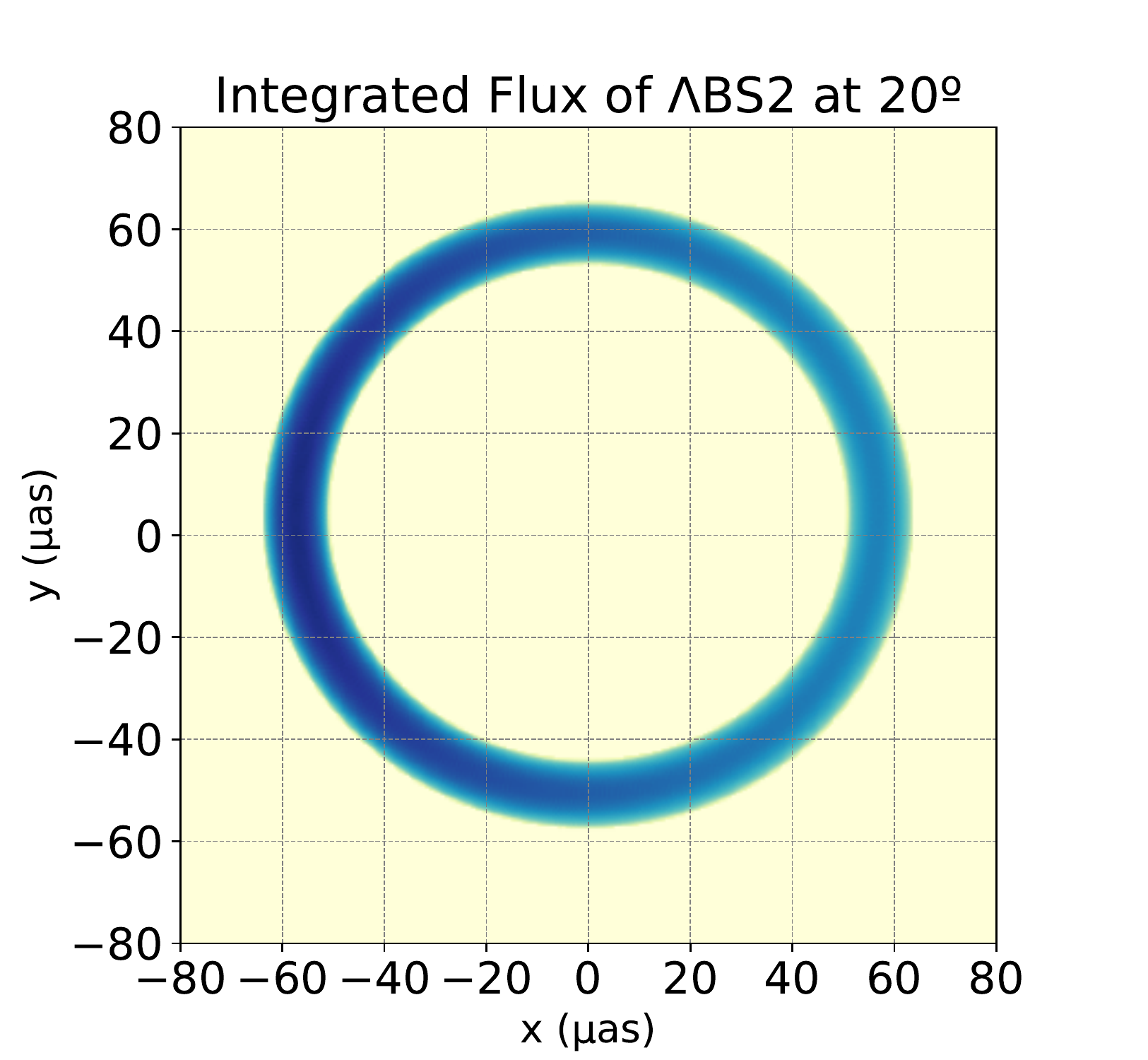}
\includegraphics[scale=0.35]{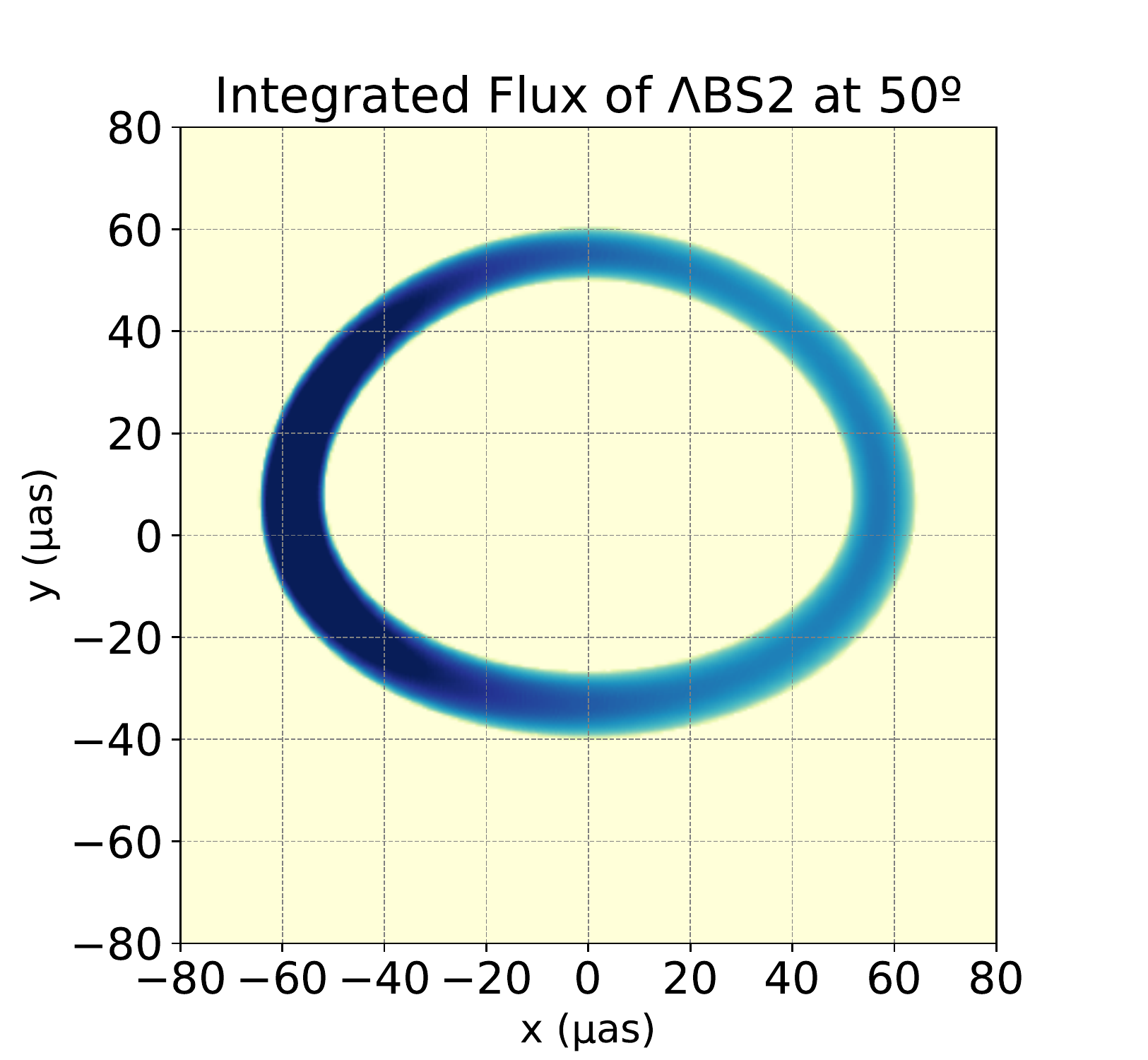}
\includegraphics[scale=0.35]{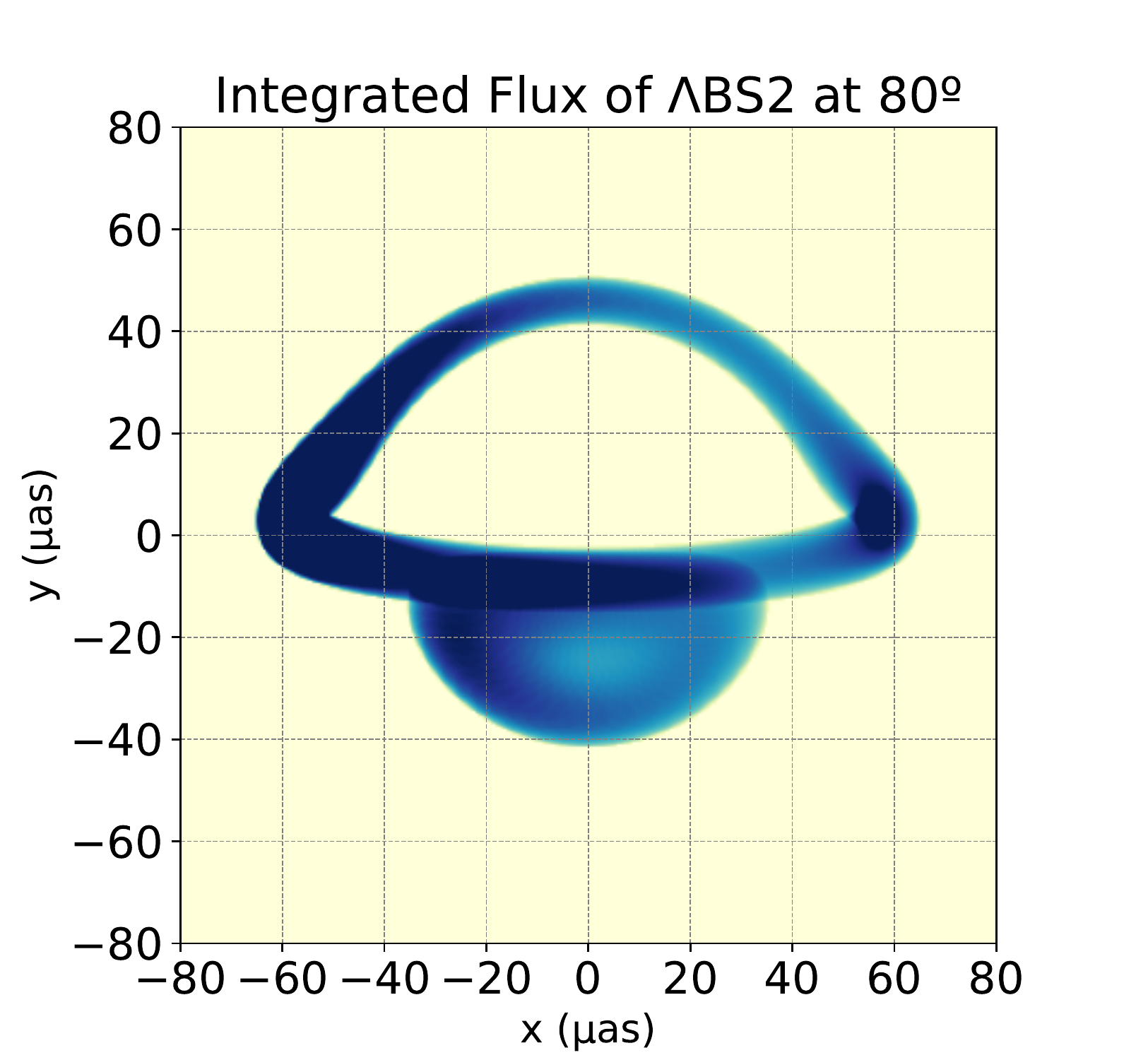}\\
\includegraphics[scale=0.35]{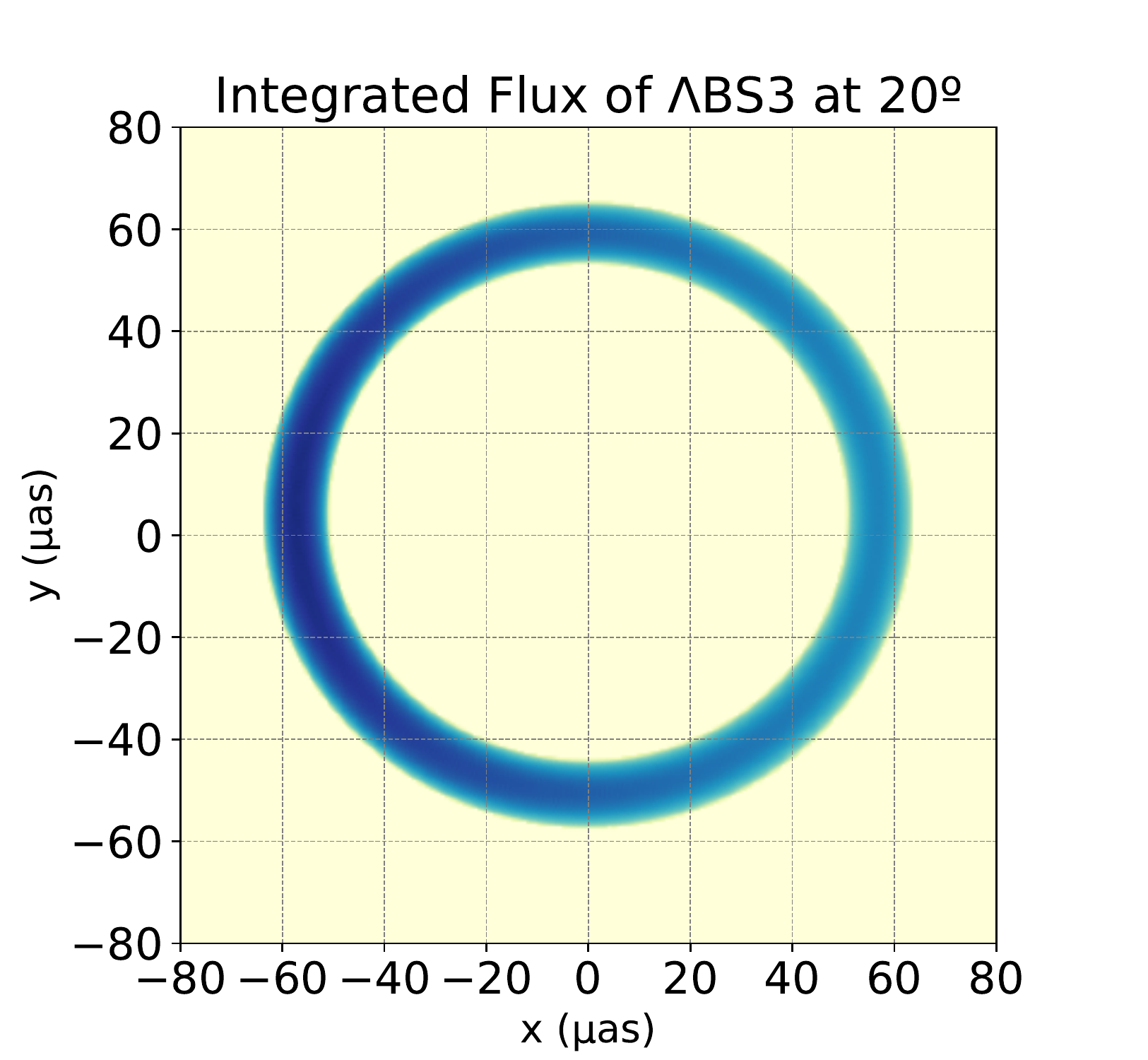}
\includegraphics[scale=0.35]{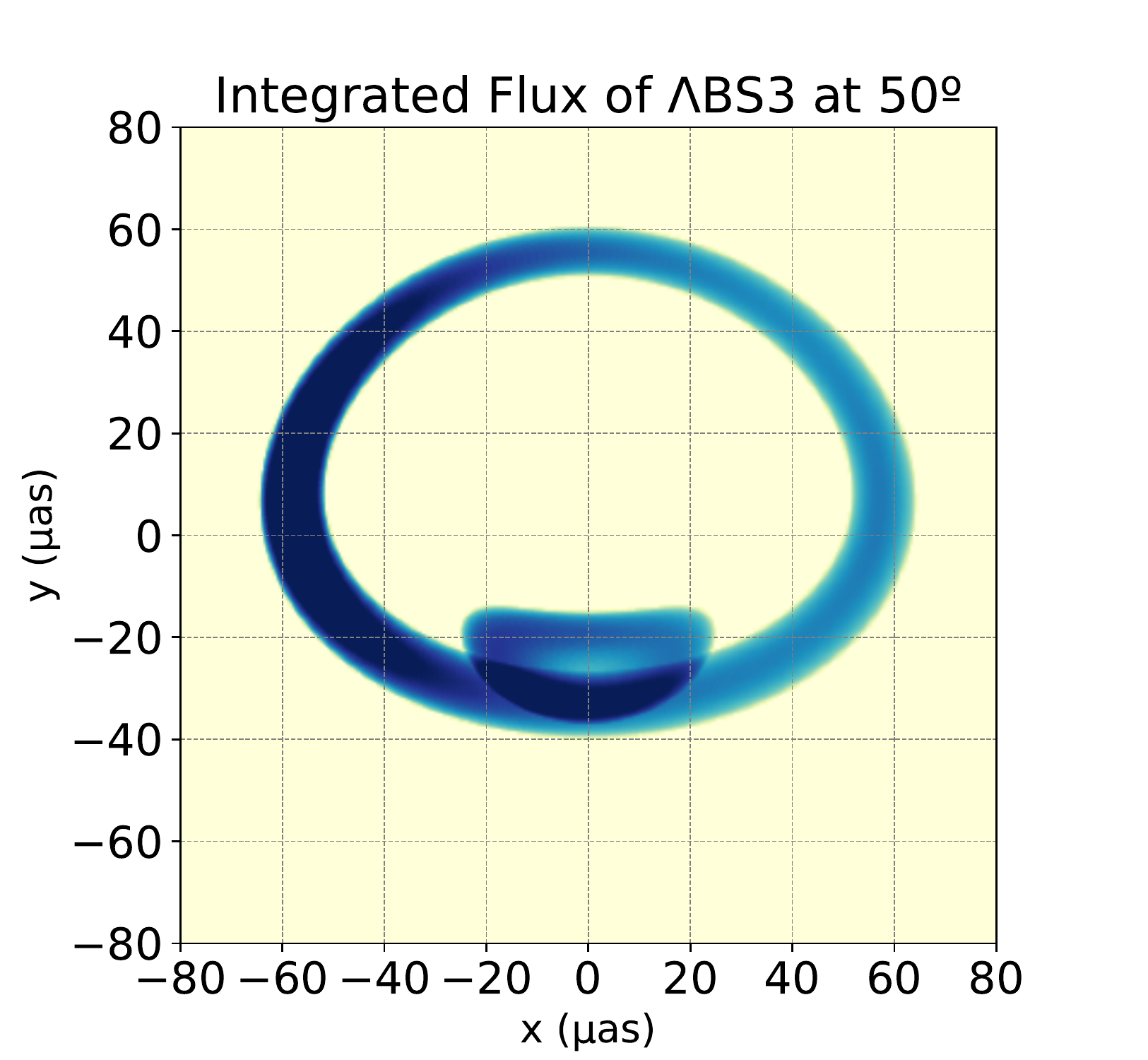}
\includegraphics[scale=0.35]{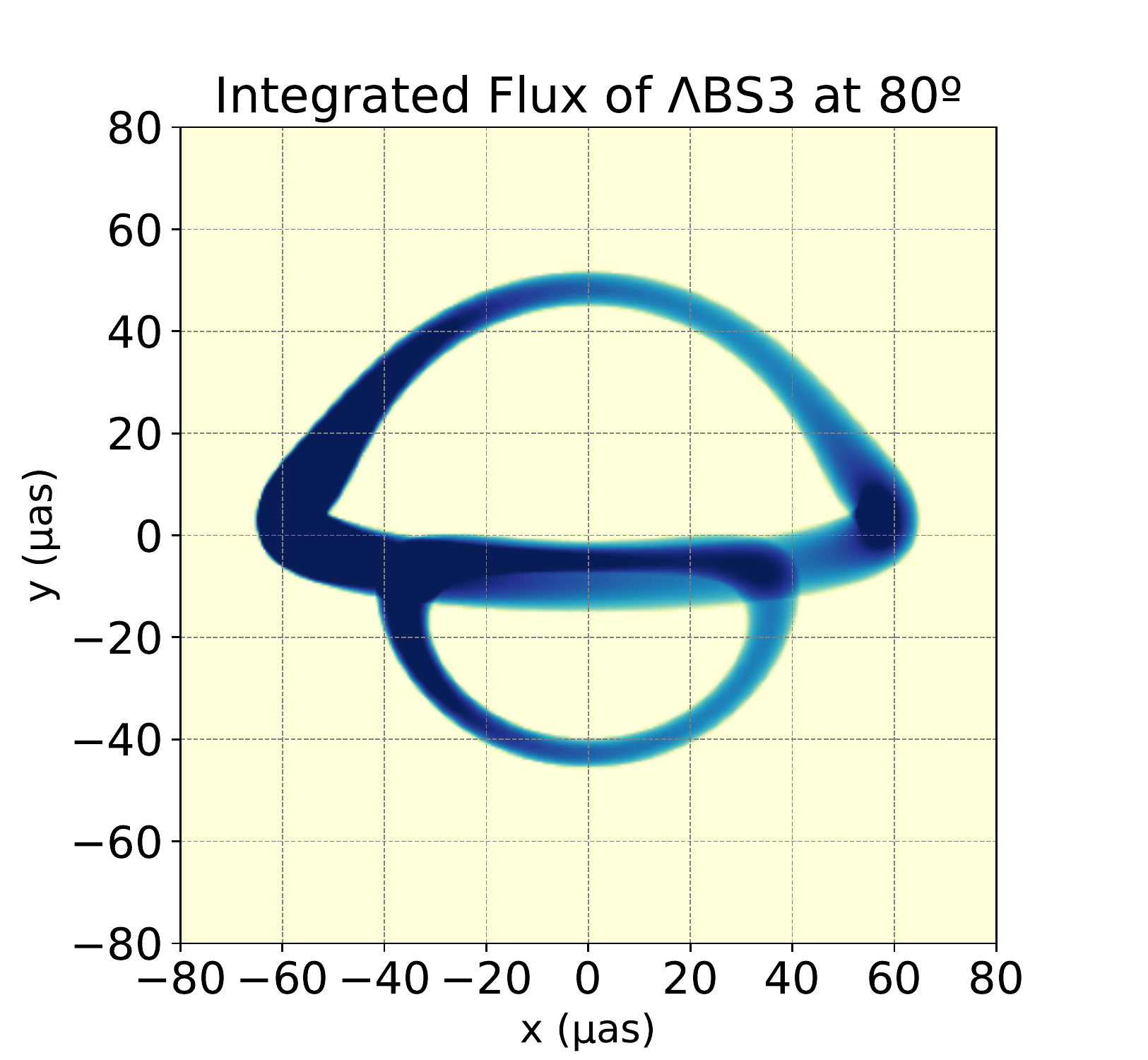}
\caption{Normalized integrated flux of the observations for the $\Lambda$BS models, namely $\Lambda$BS1 (top row), $\Lambda$BS2 (middle row), and $\Lambda$BS3 (bottom row), for an observation inclination of $\theta=20^\circ$ (left column), $\theta=50^\circ$ (middle column), and $\theta=80^\circ$ (right column).}
\label{fig:LBSflux}
\end{figure*}

\begin{figure*}[t!]
\includegraphics[scale=0.35]{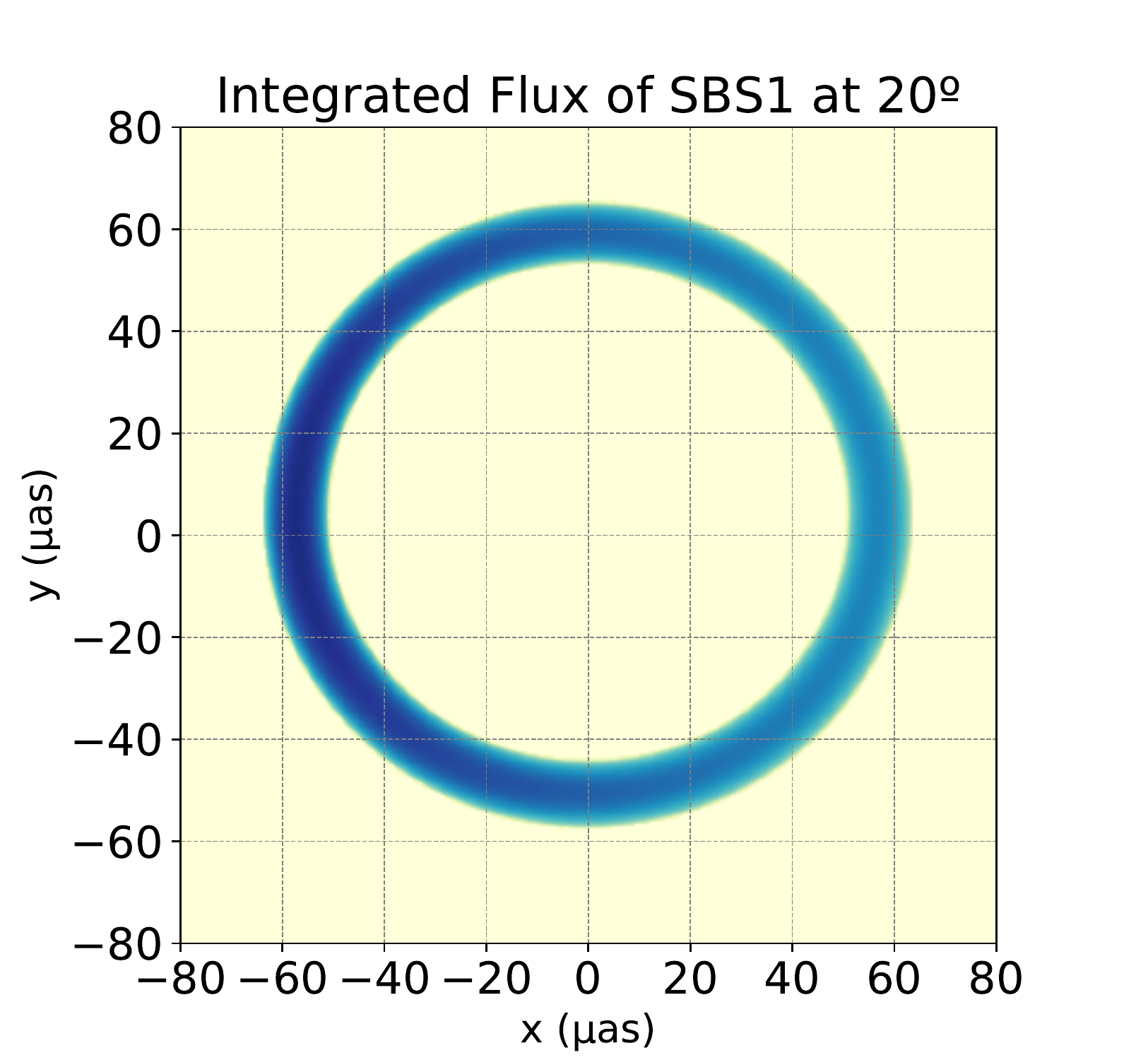}
\includegraphics[scale=0.35]{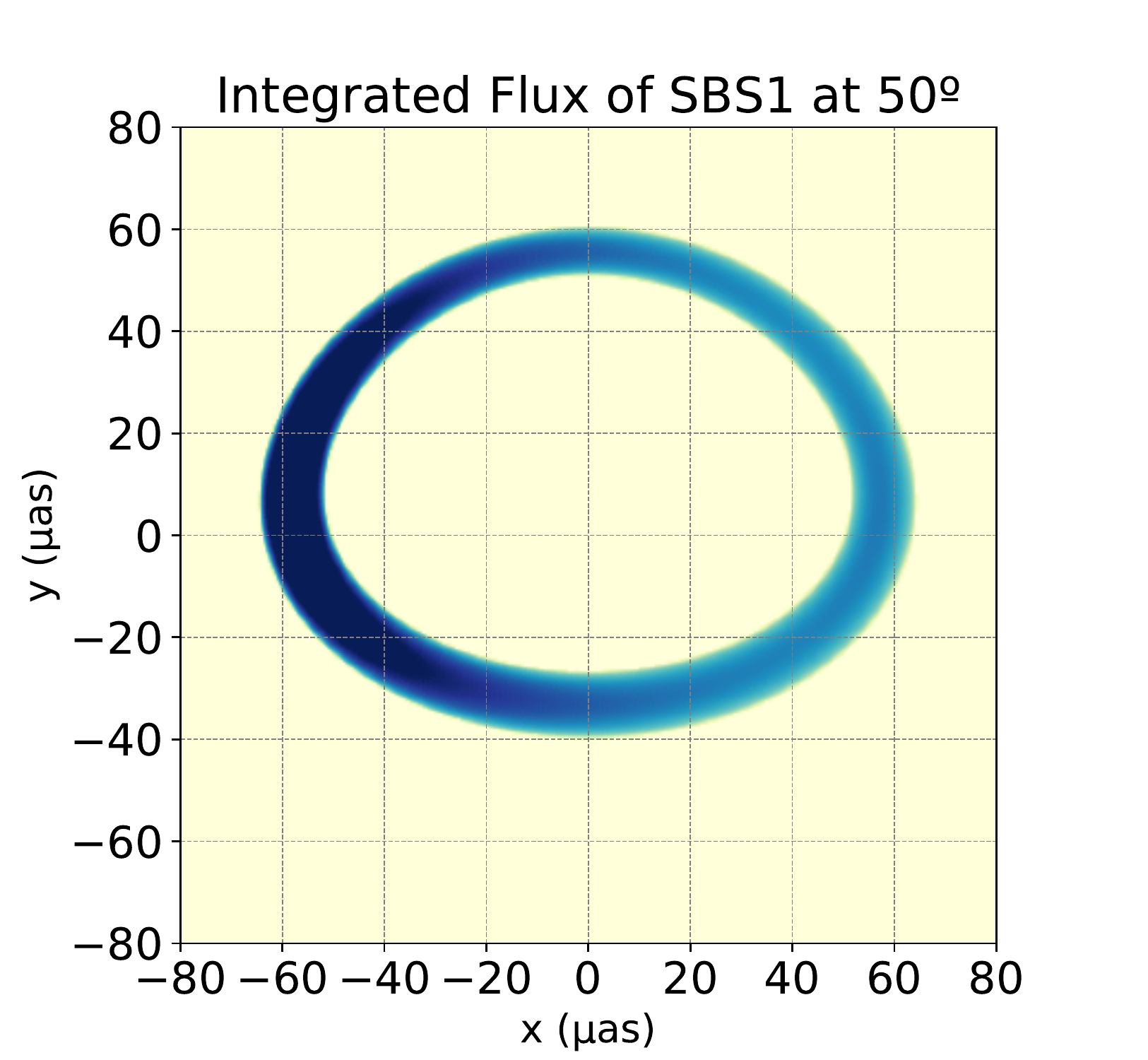}
\includegraphics[scale=0.35]{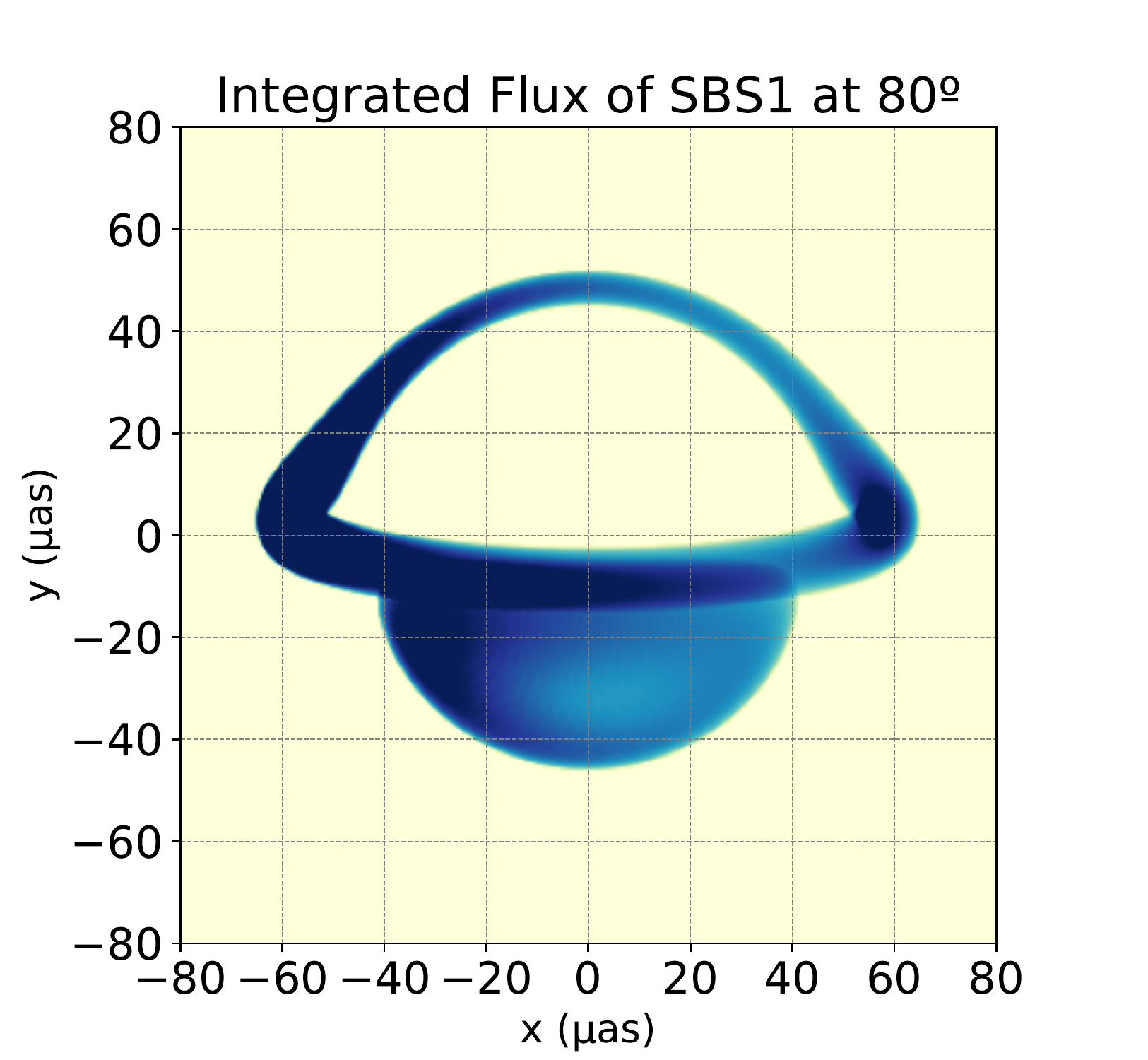}\\
\includegraphics[scale=0.35]{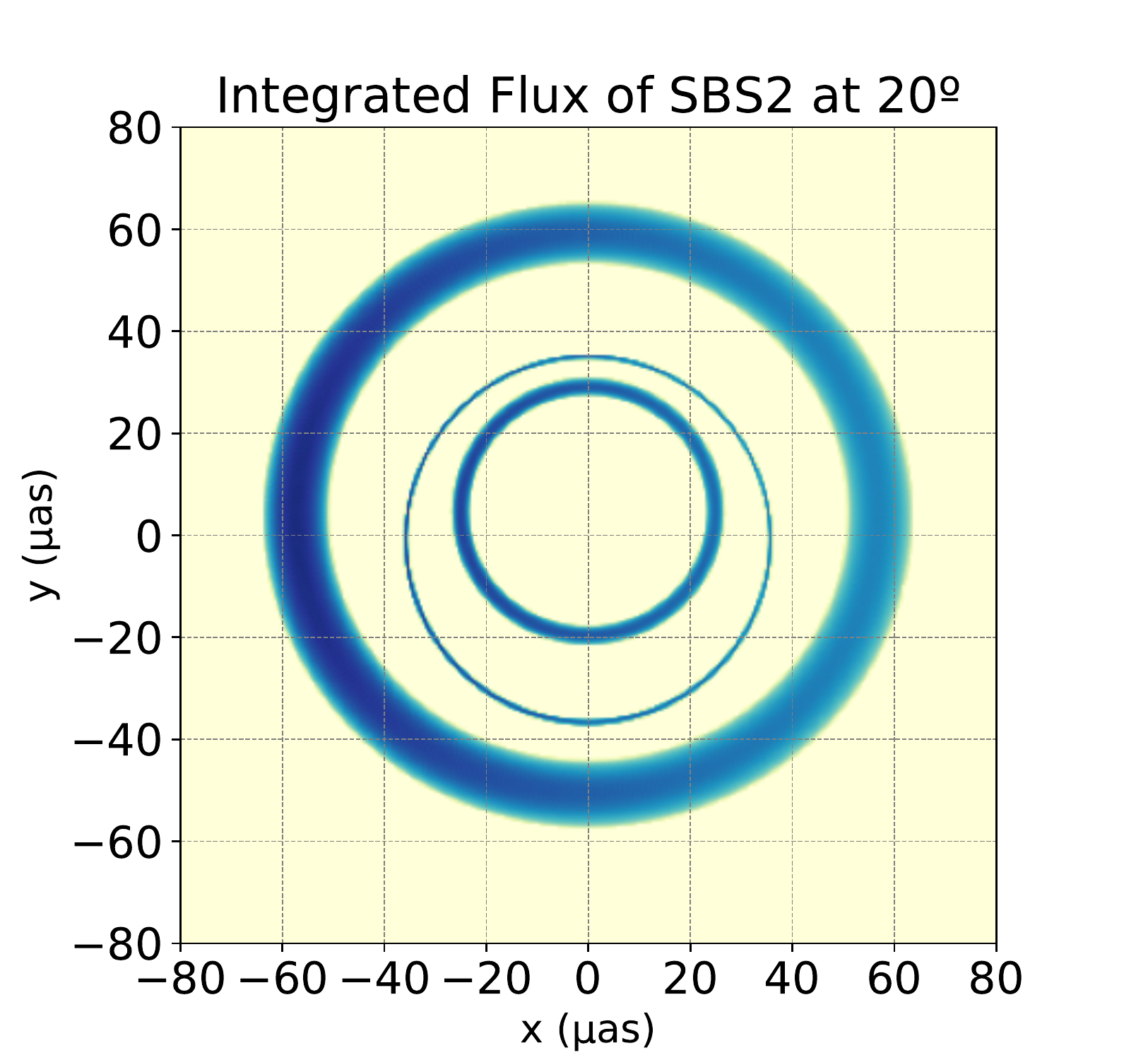}
\includegraphics[scale=0.35]{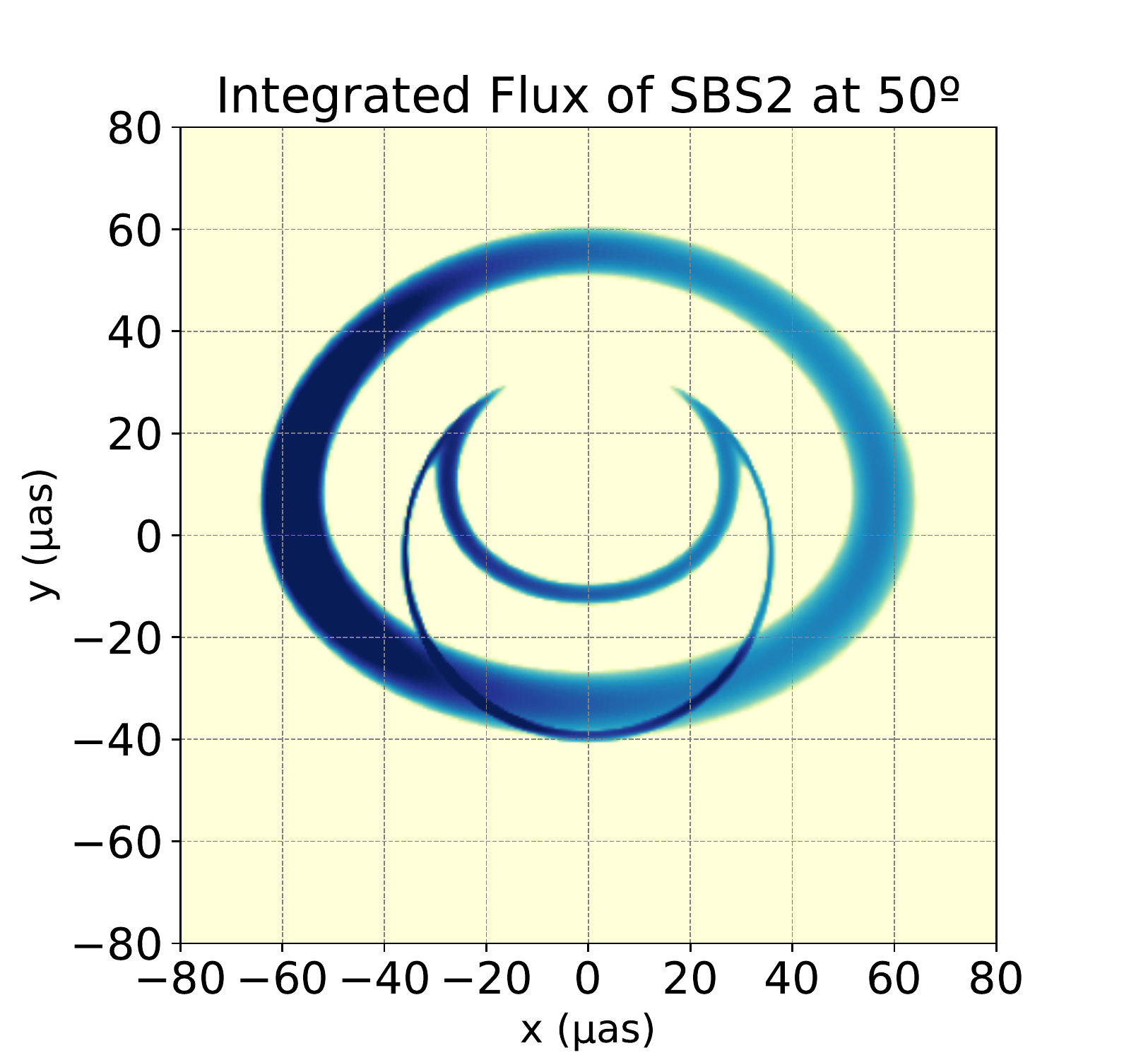}
\includegraphics[scale=0.35]{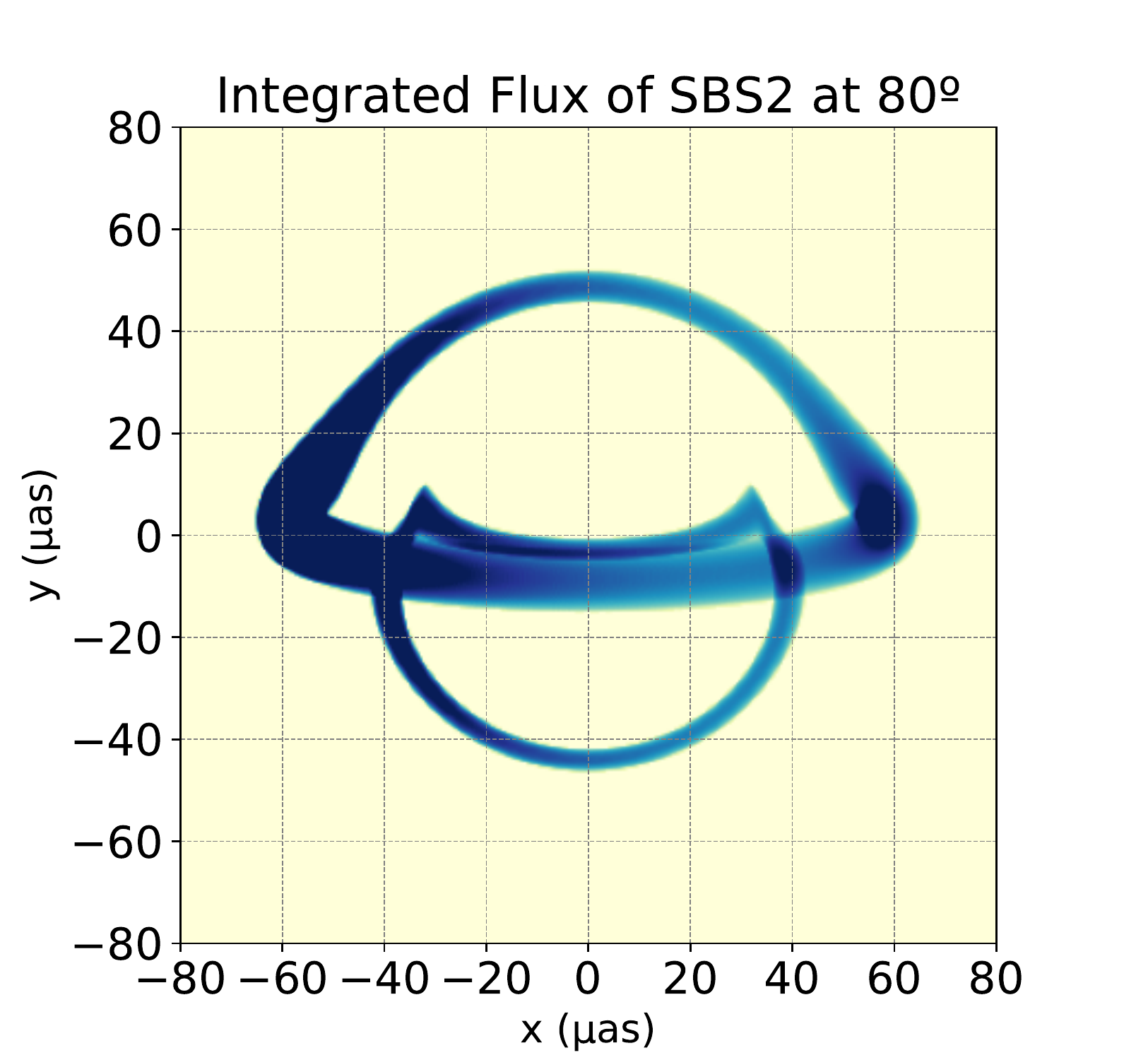}\\
\includegraphics[scale=0.35]{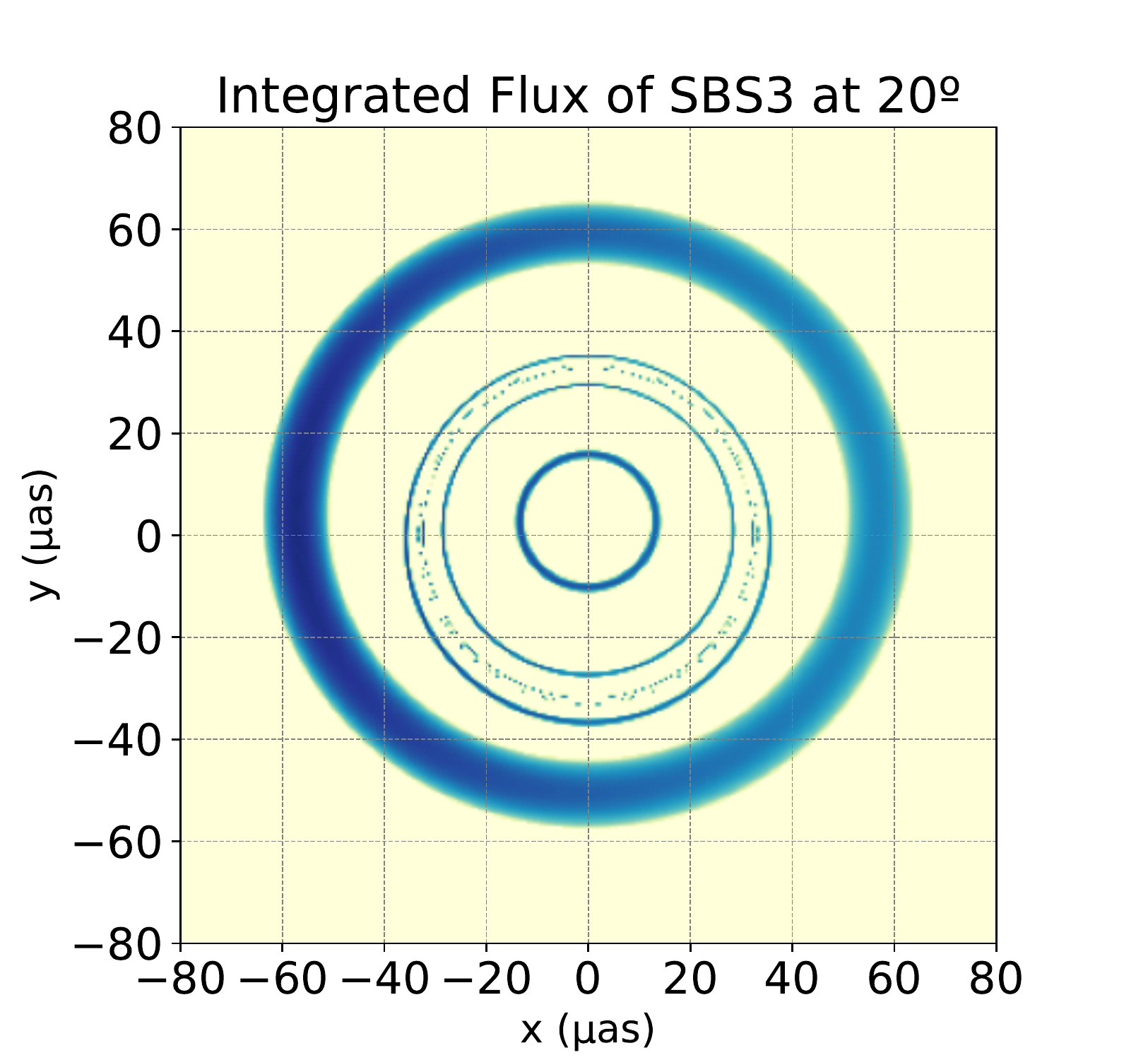}
\includegraphics[scale=0.35]{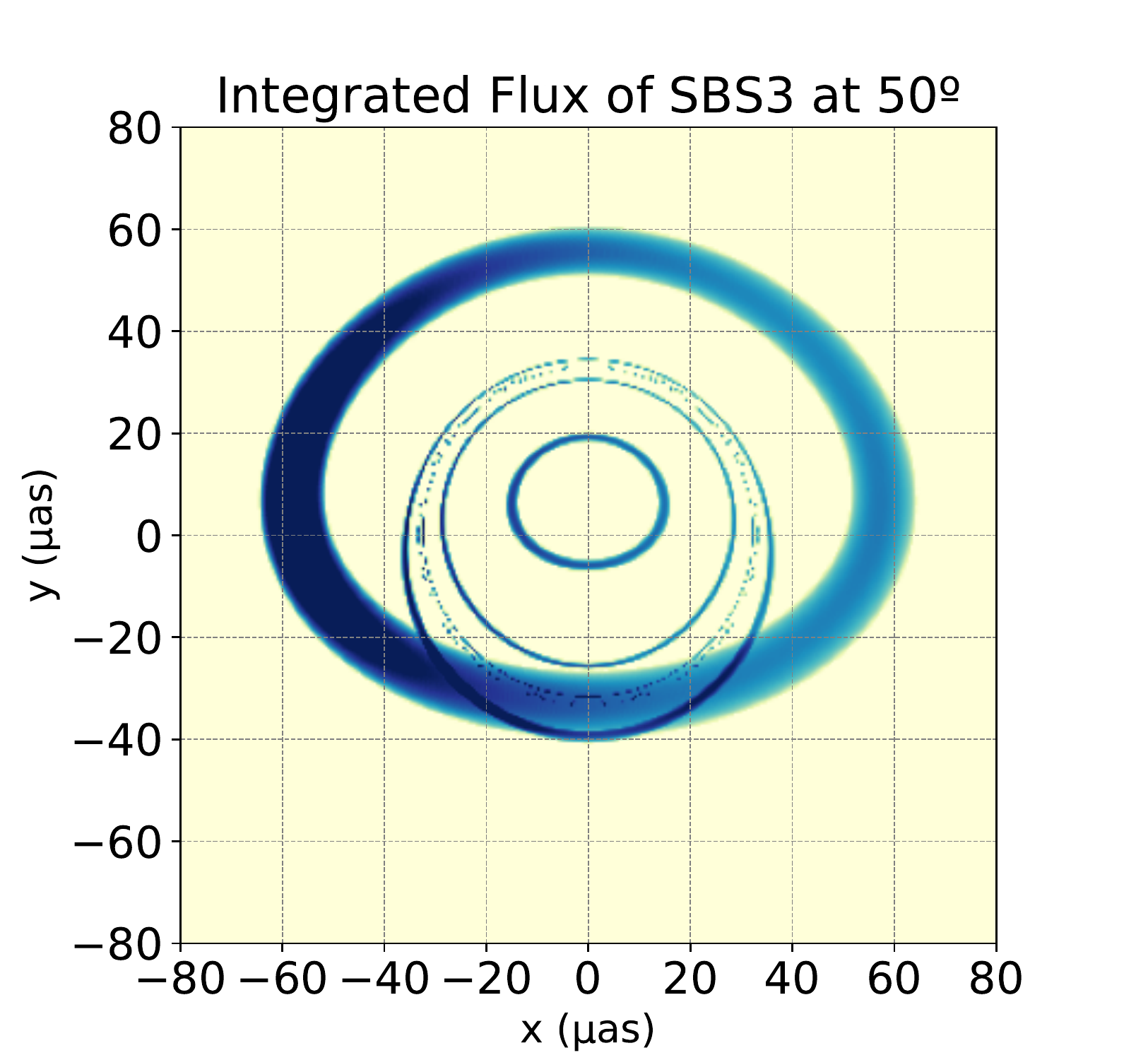}
\includegraphics[scale=0.35]{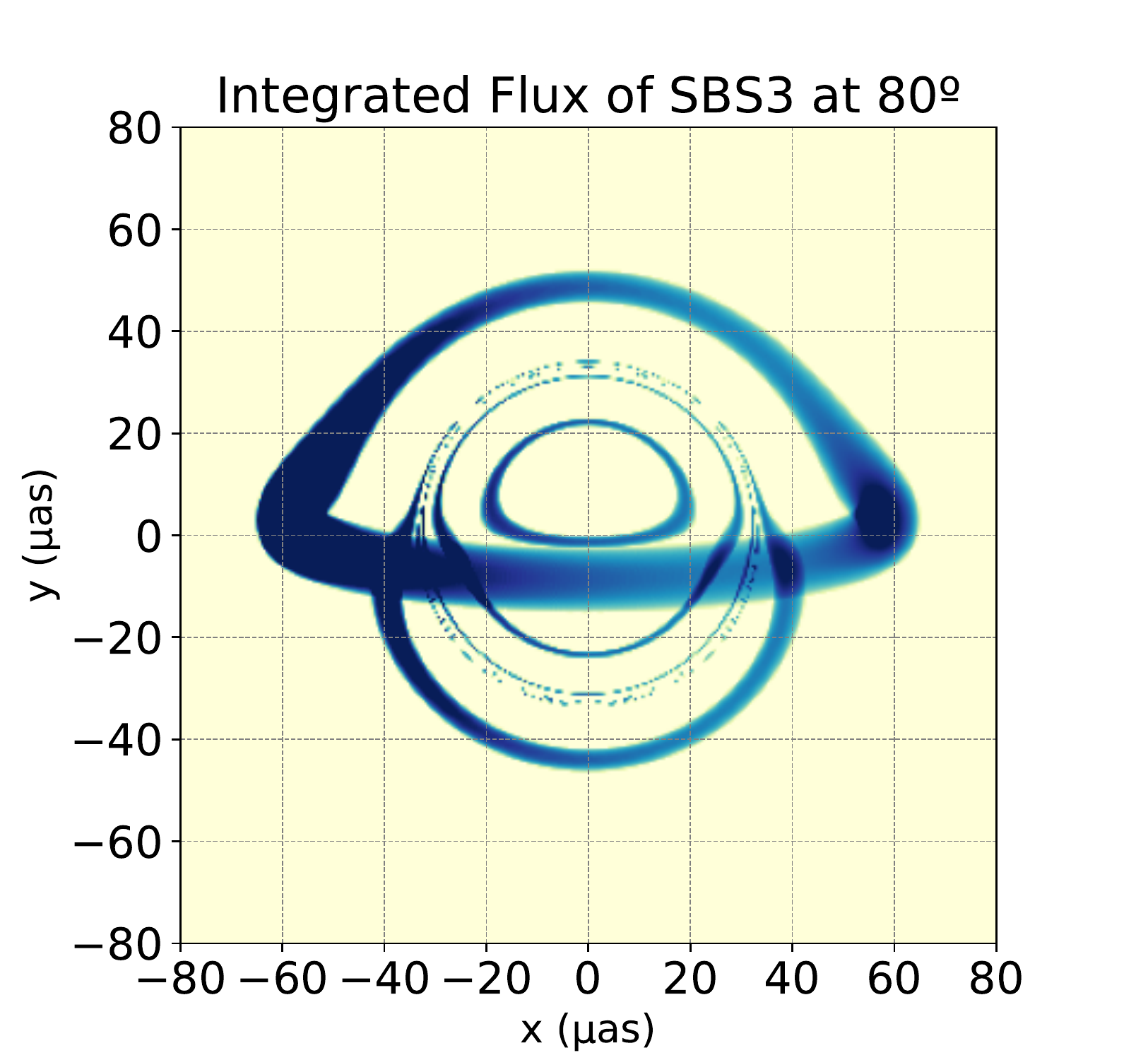}
\caption{Normalized integrated flux of the observations for the SBS models, namely SBS1 (top row), SBS2 (middle row), and SBS3 (bottom row), for an observation inclination of $\theta=20^\circ$ (left column), $\theta=50^\circ$ (middle column), and $\theta=80^\circ$ (right column).}
\label{fig:SBSflux}
\end{figure*}

The integrated fluxes are depicted in Fig. \ref{fig:LBSflux} for the $\Lambda$BS configurations, and in Fig. \ref{fig:SBSflux} for the SBS ones. For the $\Lambda$BS models, one verifies that the results are qualitatively similar to the ones obtained in a previous publication \cite{Rosa:2022toh} for bosonic stars without self-interactions, a result that is somewhat expected since the space-time properties of these configurations are also similar, i.e., these stars are not compact enough to have neither a light-ring or an ISCO, and they do not feature event horizons either. Indeed, for small observation angles one can only observe the primary track of the hot-spot. The secondary track eventually becomes observable as one increases the compactness of the star and/or the observation angle. Such a secondary track features two components, the usual secondary image also observed in black hole space-times, and a plunge-through image corresponding to the photons crossing the interior of the bosonic star before reaching the observer. The latter component is absent in black hole space-times due to the existence of an event horizon and consequent impossibility of photons to escape from the interior of the space-time. 

For the SBS models, more interesting and qualitatively different results arise. Whereas for the SBS1 model the integrated flux images are again qualitatively similar to the ones previously obtained for the $\Lambda$BS models, i.e., only the primary track is observed for a low inclination angle and the secondary track, composed by the usual secondary plus the plunge-through components, eventually arises as one increases the inclination angles. However, the situation drastically changes for the SBS2 and the SBS3 models. Indeed, for such models not only the secondary track is always present, but also several additional tracks can be observed. For the SBS2 model, for an observation angle of $\theta=20^\circ$, two additional closed tracks can be observed besides the primary image. These two extra tracks eventually merge into a single secondary track with two components, the usual secondary and the plunge-through components, for larger observation angles. The SBS3 model features an even more complex structure of sub-images: a third additional track and the light-ring contributions can also be observed for low observation angles, and these contributions do not merge as one increases the observation angle.

These results indicate that the qualitative properties of the observed integrated fluxes depend strongly on the compactness of these horizonless compact objects, a feature that was already hinted by a previous publication on relativistic fluid stars \cite{Rosa:2023hfm}. Three different regimes can thus be identified:
\renewcommand{\theenumi}{\roman{enumi}}
\begin{enumerate}
\item If the light deflection is not strong enough, the secondary track can only be observed for certain inclination angles, i.e., there is a critical observation angle $\theta_c^{(1)}$ such that if $\theta<\theta_c^{(1)}$ the secondary track is absent;
\item For a stronger light deflection, the secondary track is present independently of the observation angle but its two components, the usual secondary and the plunge-through, might be observed as independent tracks for some observation angles, i.e., there is another critical observation angle $\theta_c^{(2)}$, such that if $\theta<\theta_c^{(2)}$ the secondary and the plunge-through components are independent tracks, and if $\theta>\theta_c^{(2)}$ the secondary and the plunge-through components merge into a single secondary track;
\item For a very large light deflection, there is a further split of the secondary track into three independent tracks, independently of the observation angle. 
\end{enumerate}

In this work, the third component of the secondary and the light-ring components are only visible for the SBS3 model. We note, however, that this does not mean that these two contributions always arise simultaneously. Indeed, in the previous work on relativistic fluid spheres referenced above, several examples for which the light-ring contributions are present without the third splitting of the secondary track are provided. 

\subsection{Astrometrical properties}

\begin{figure*}[t!]
\includegraphics[scale=0.28]{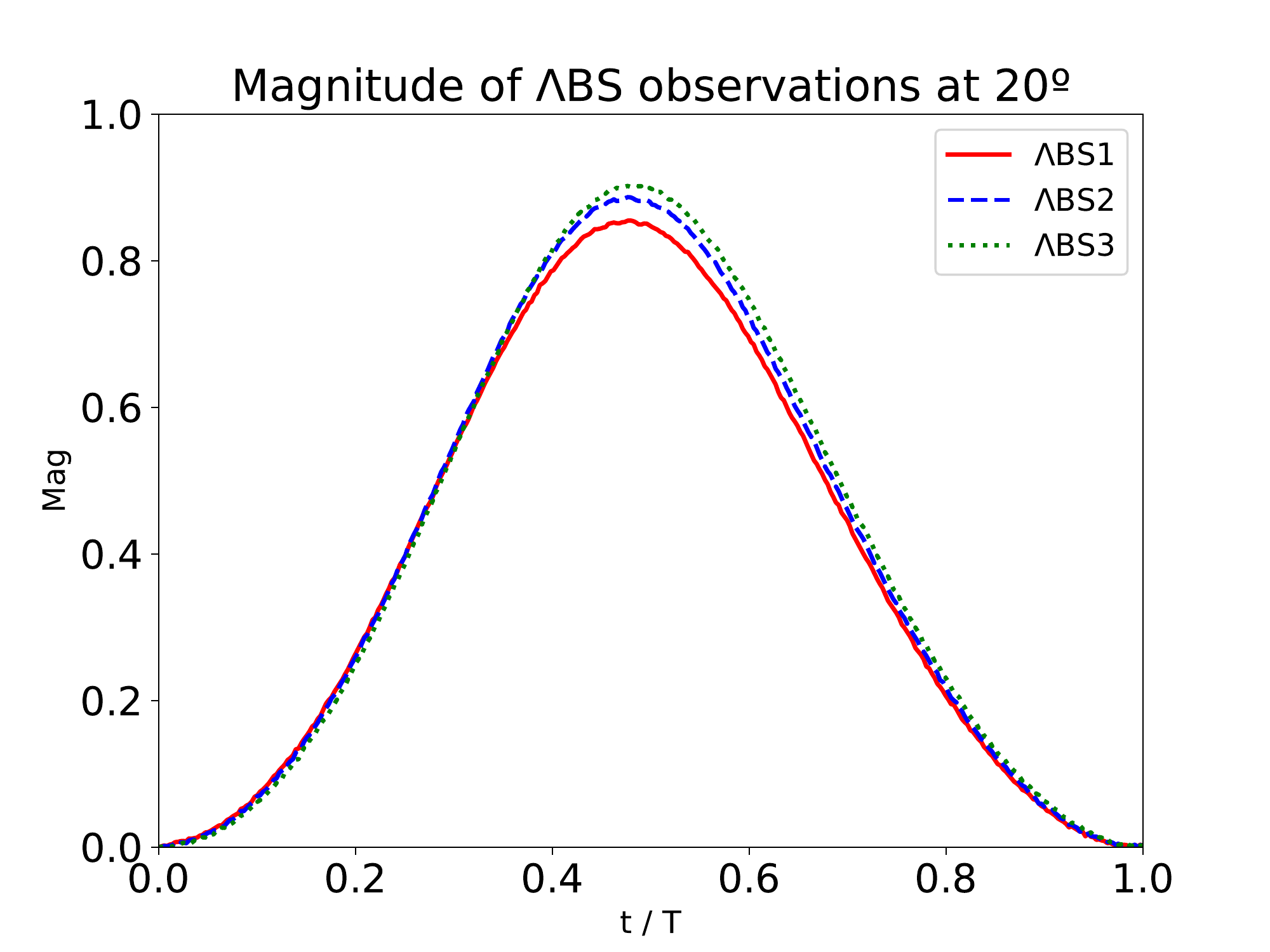}
\includegraphics[scale=0.28]{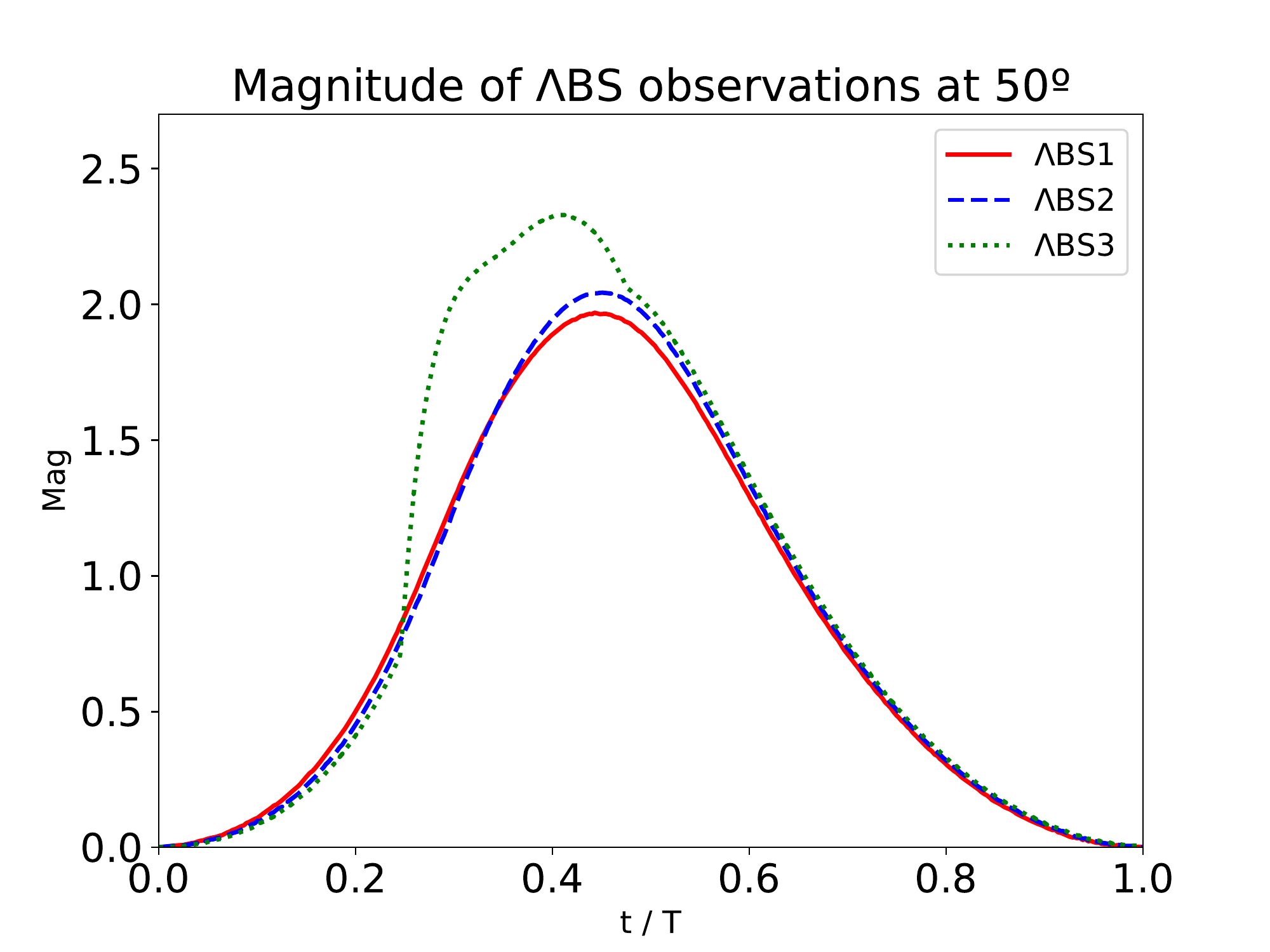}
\includegraphics[scale=0.28]{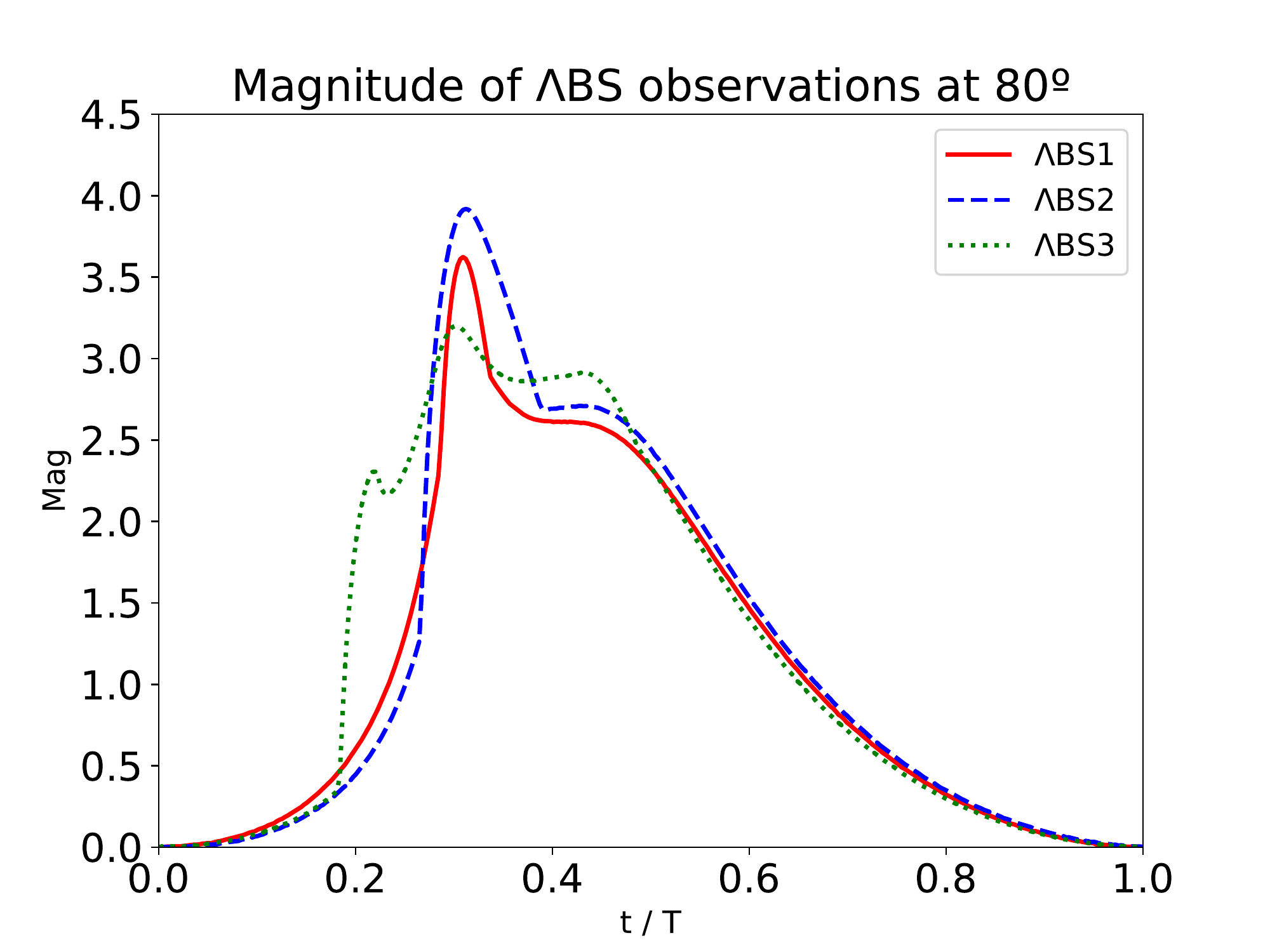}\\
\includegraphics[scale=0.28]{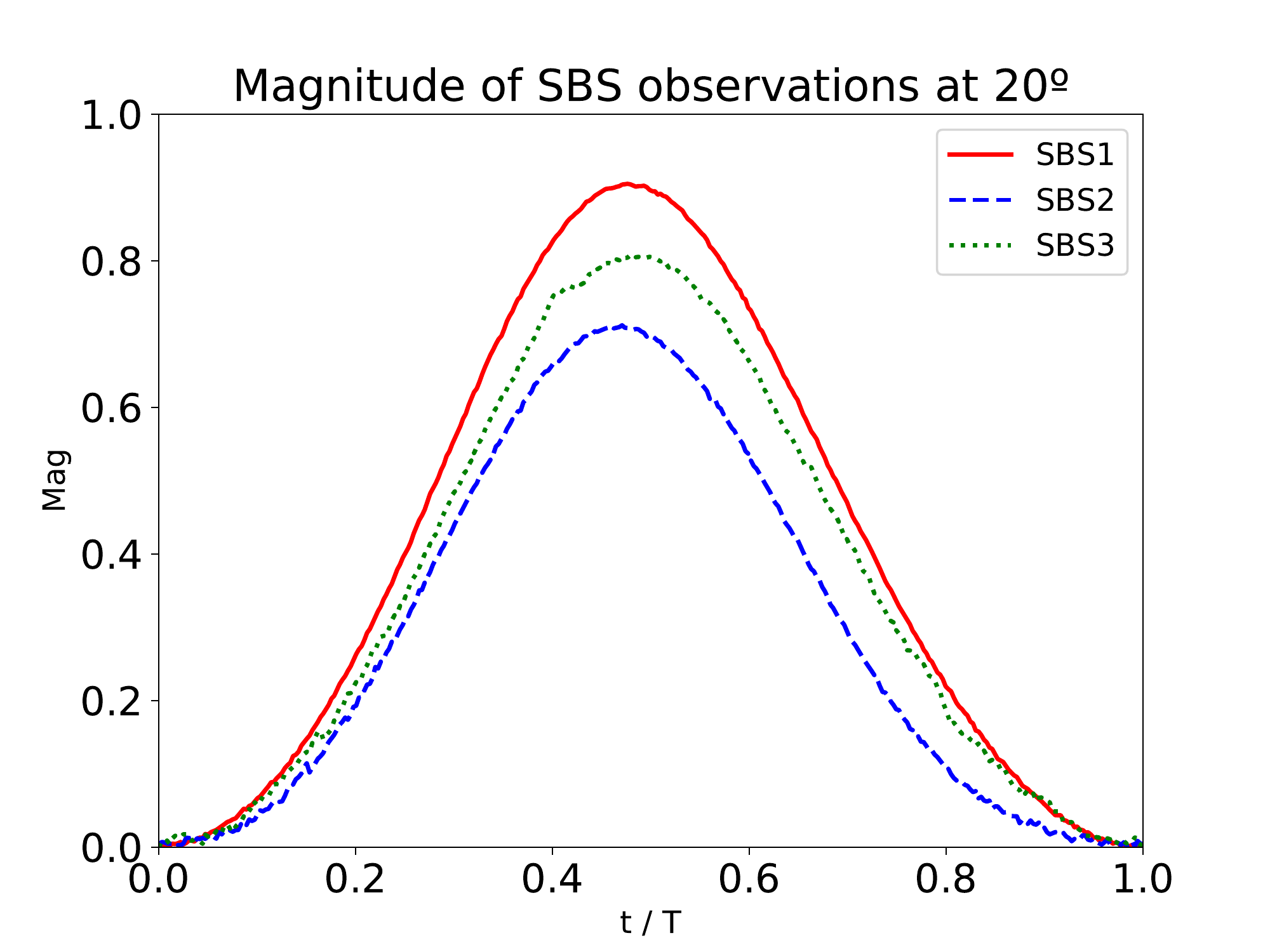}
\includegraphics[scale=0.28]{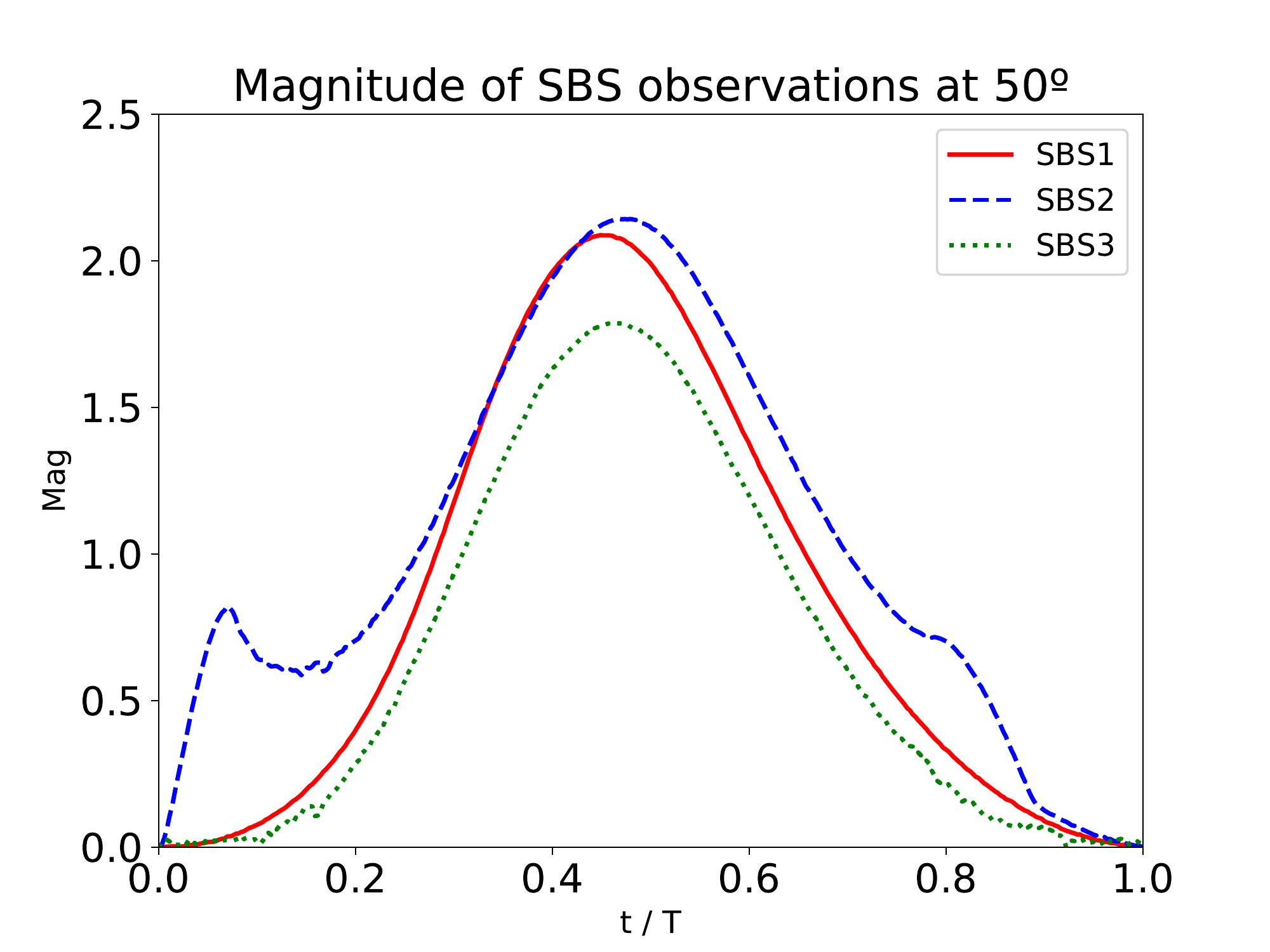}
\includegraphics[scale=0.28]{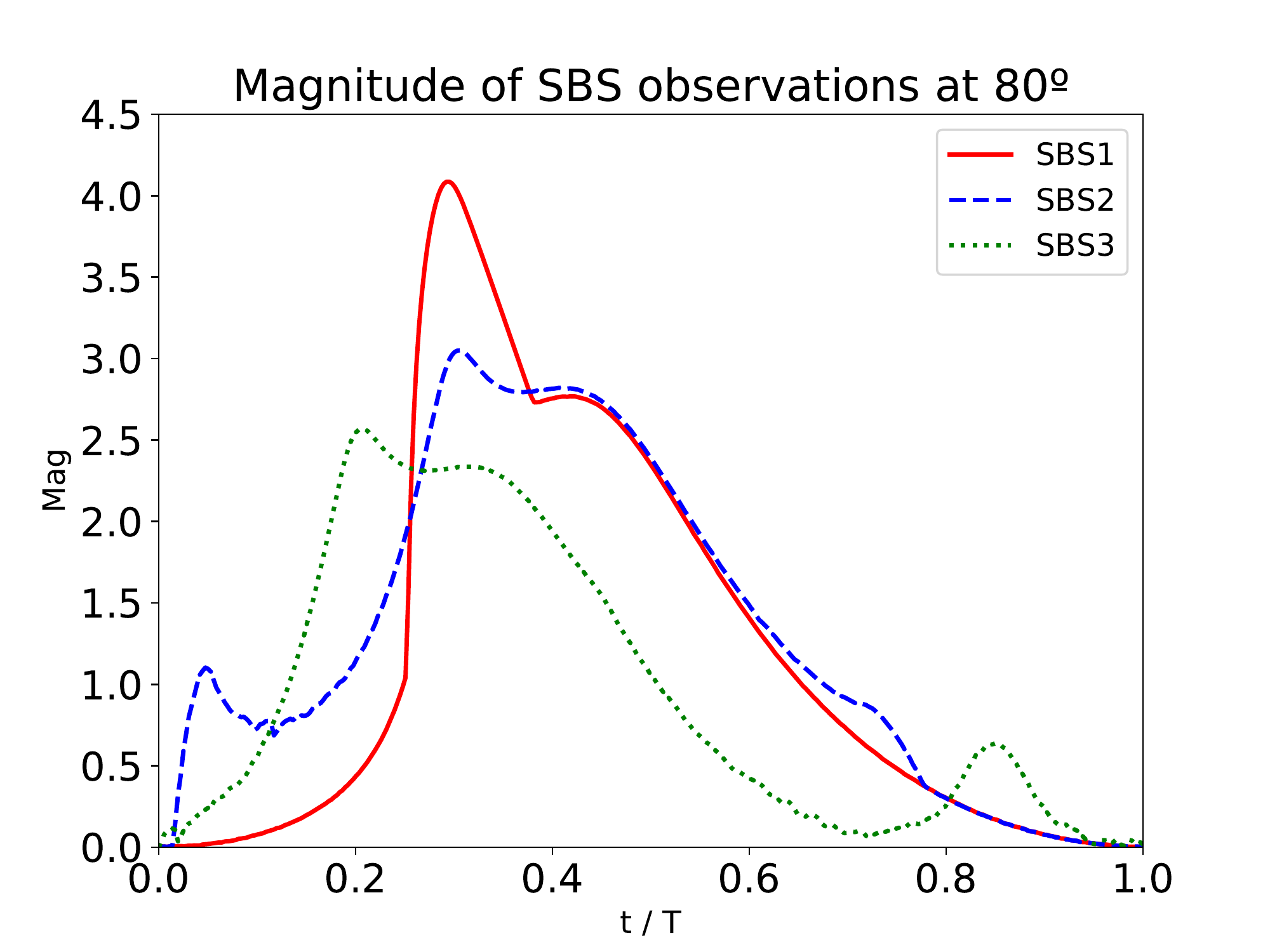}
\caption{Normalized magnitude of the observations in arbitrary units as a function of $t/T$ for the $\Lambda$BS models (top row) and the SBS models (bottom row) for an observation inclination of $\theta=20^\circ$ (left column), $\theta=50^\circ$ (middle column), and $\theta=80^\circ$ (right column).}
\label{fig:magnitudes}
\end{figure*}

\begin{figure*}[t!]
\includegraphics[scale=0.28]{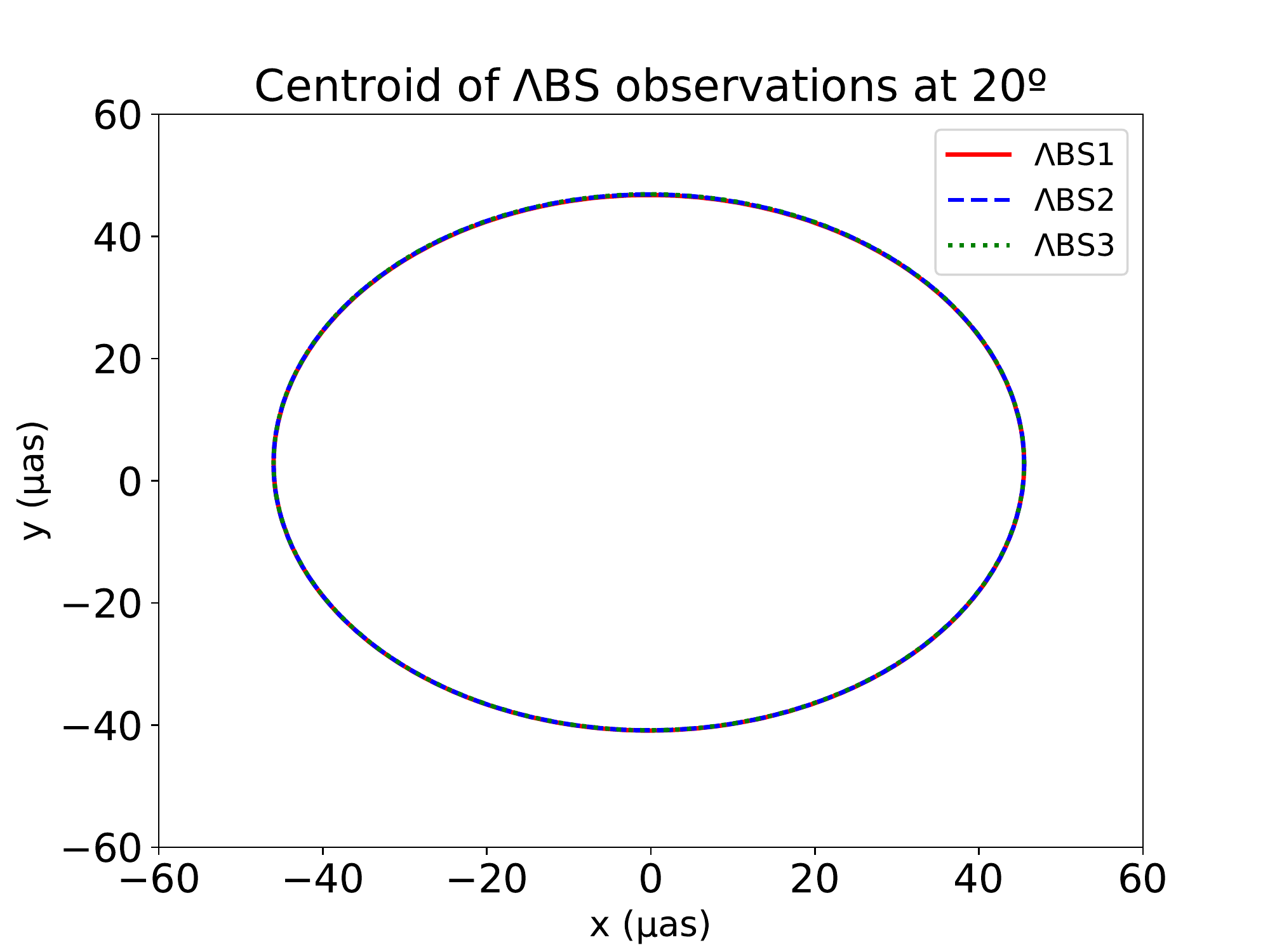}
\includegraphics[scale=0.28]{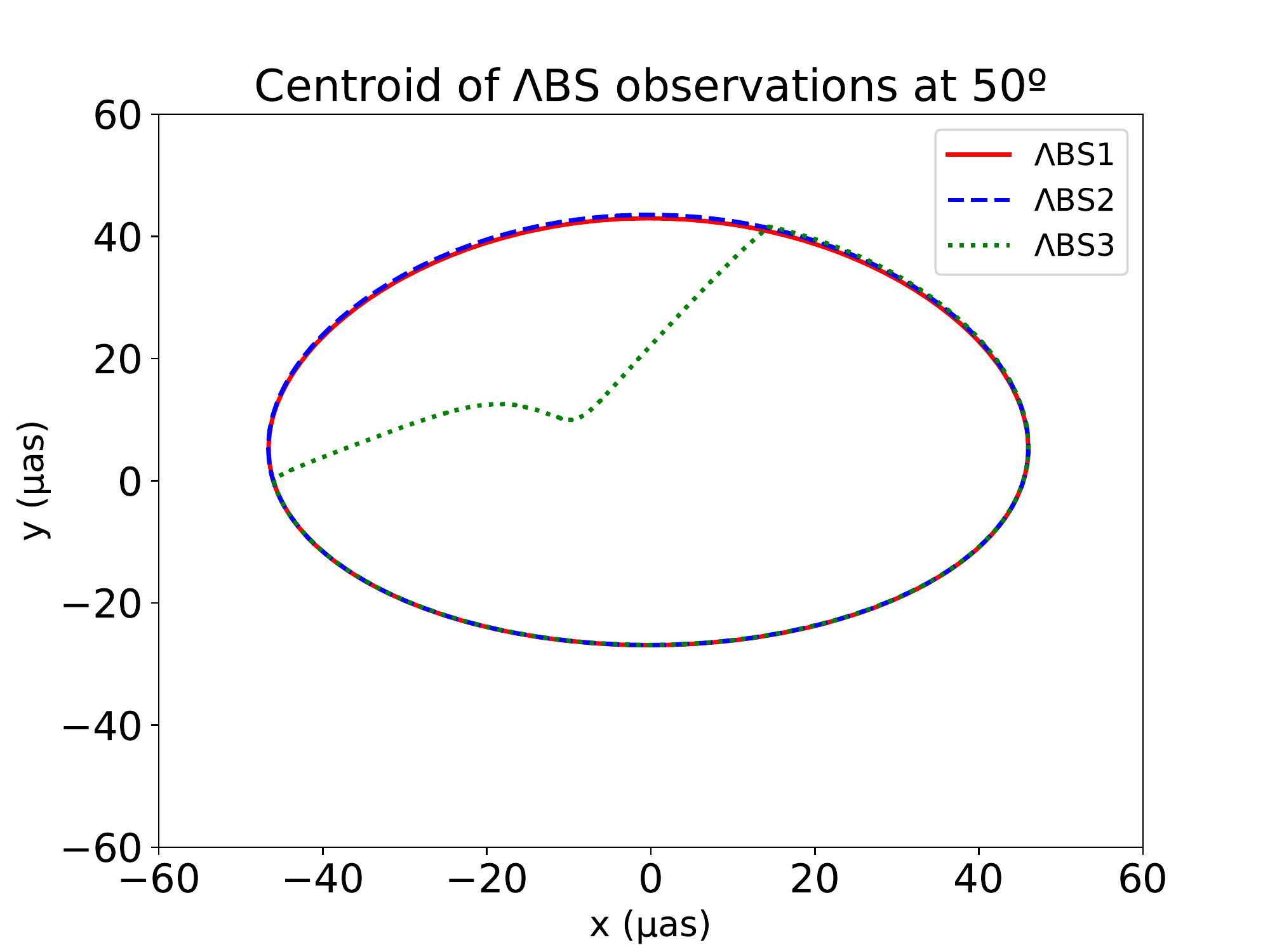}
\includegraphics[scale=0.28]{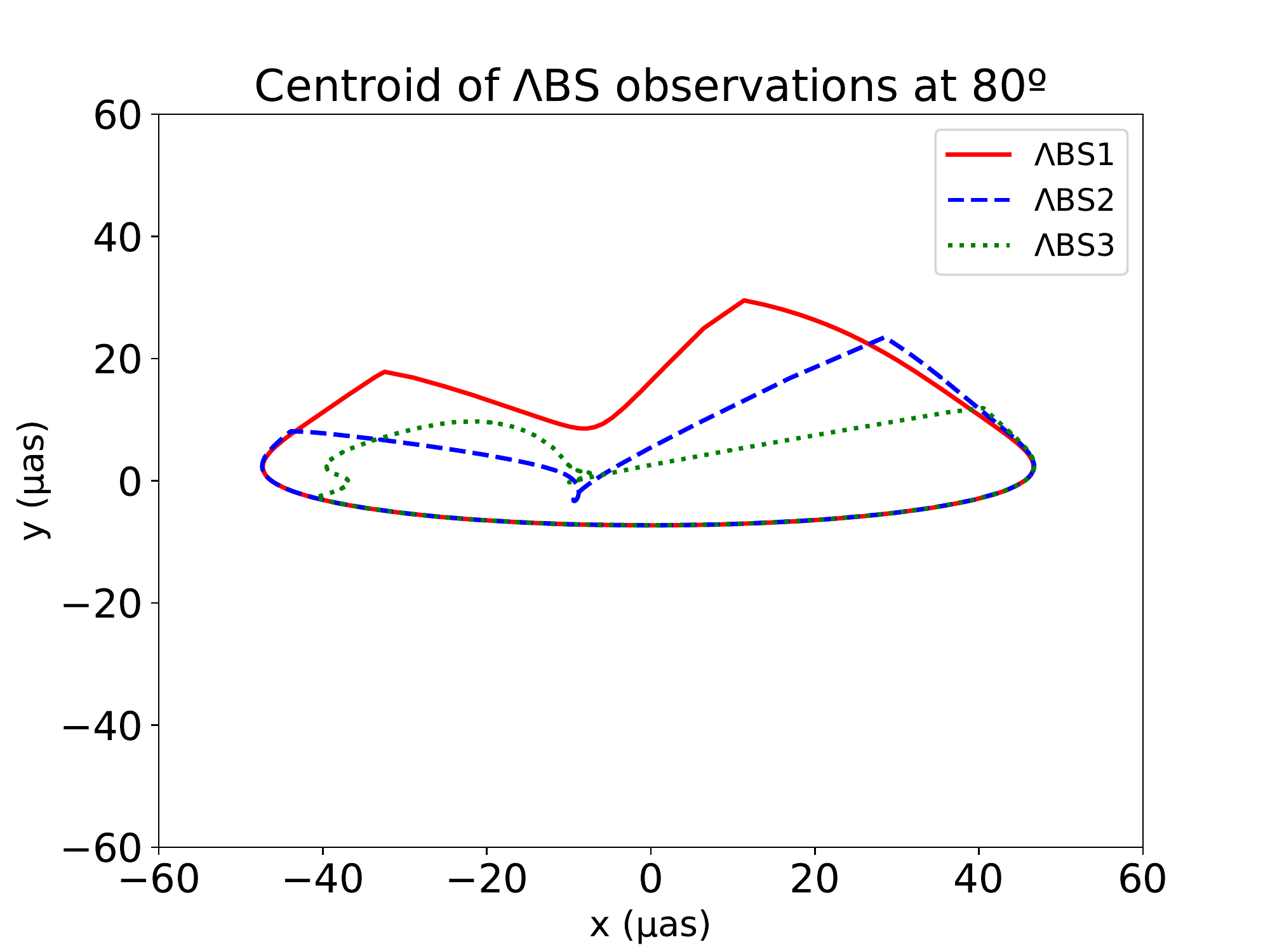}\\
\includegraphics[scale=0.28]{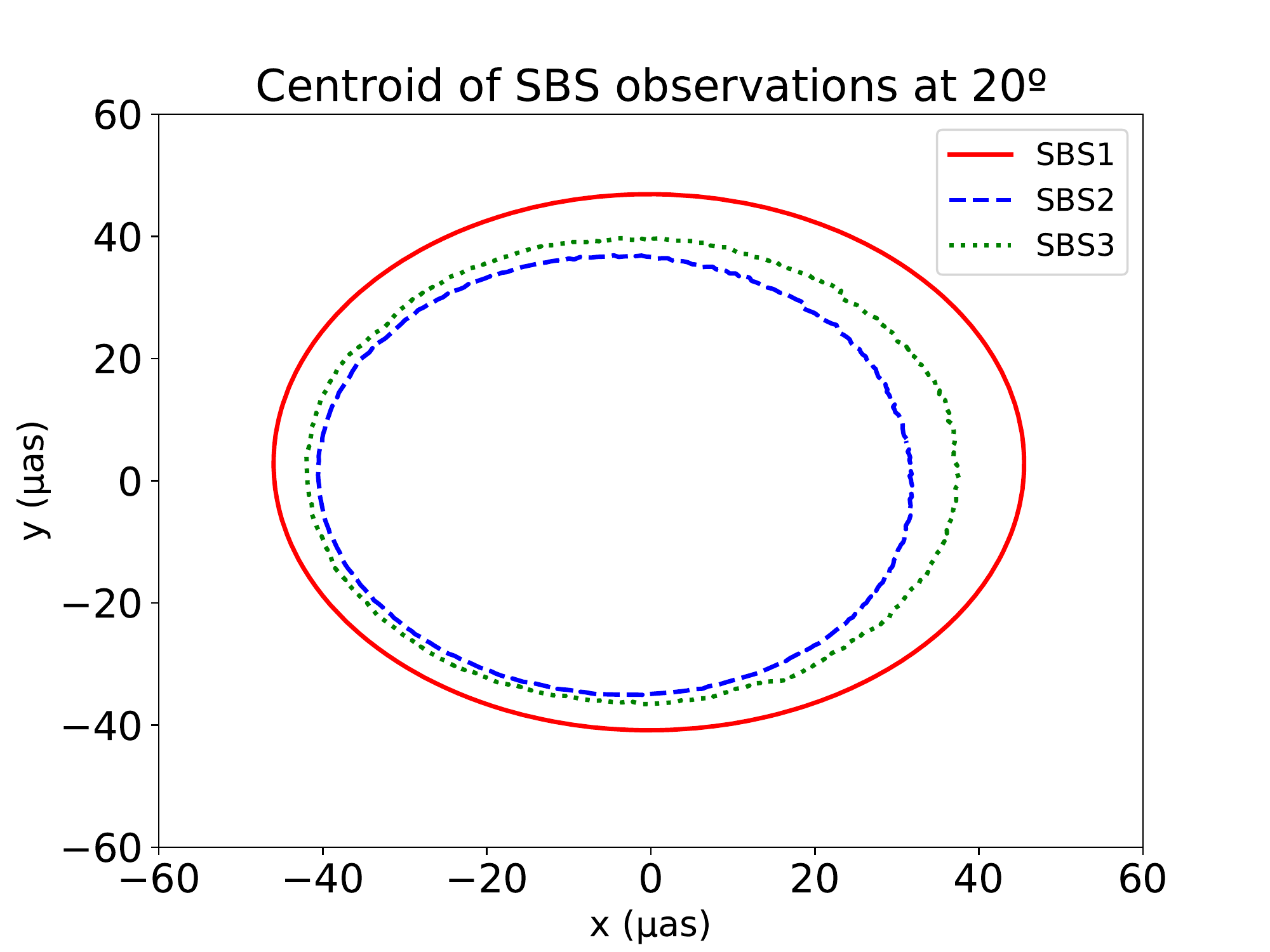}
\includegraphics[scale=0.28]{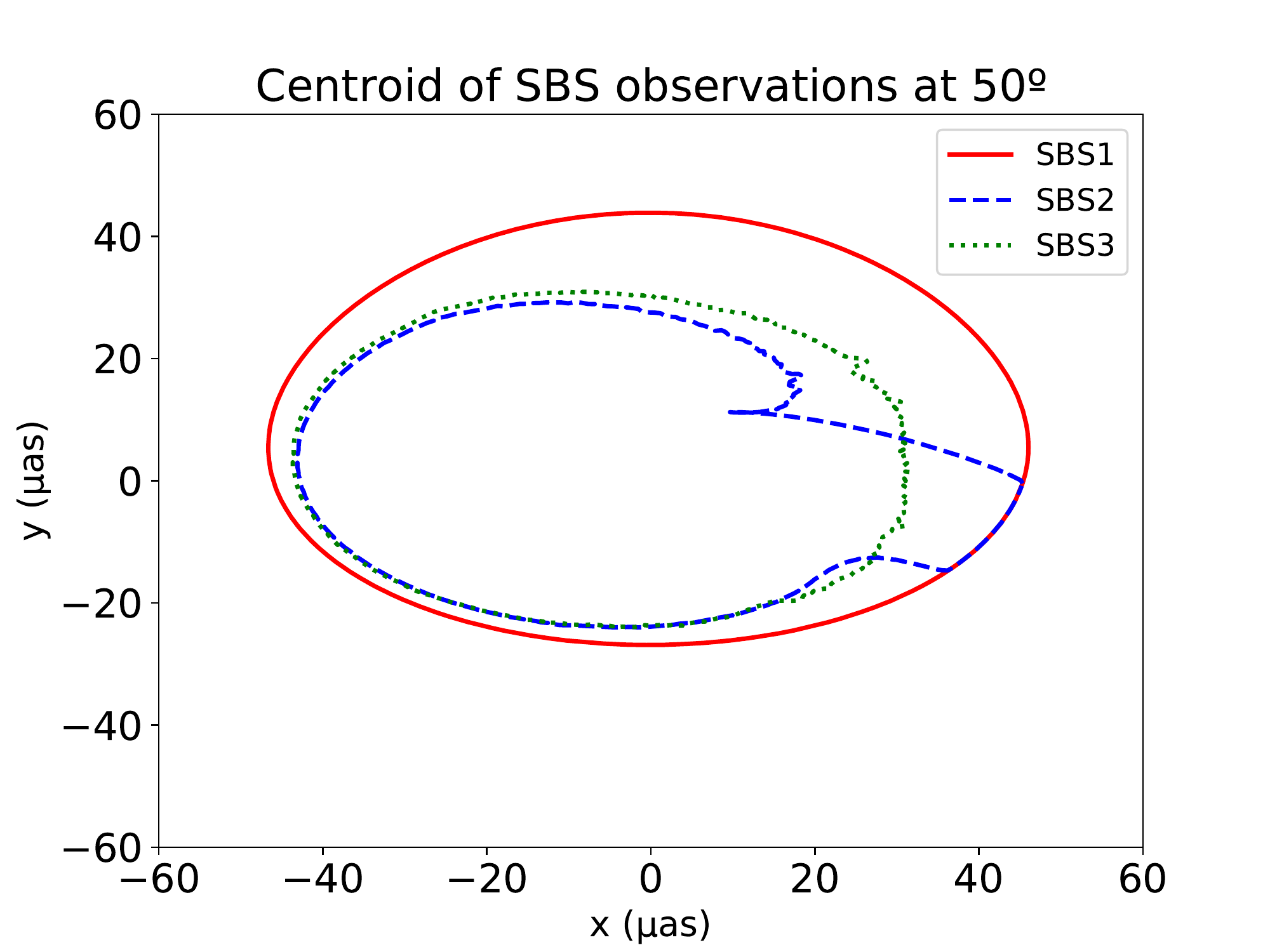}
\includegraphics[scale=0.28]{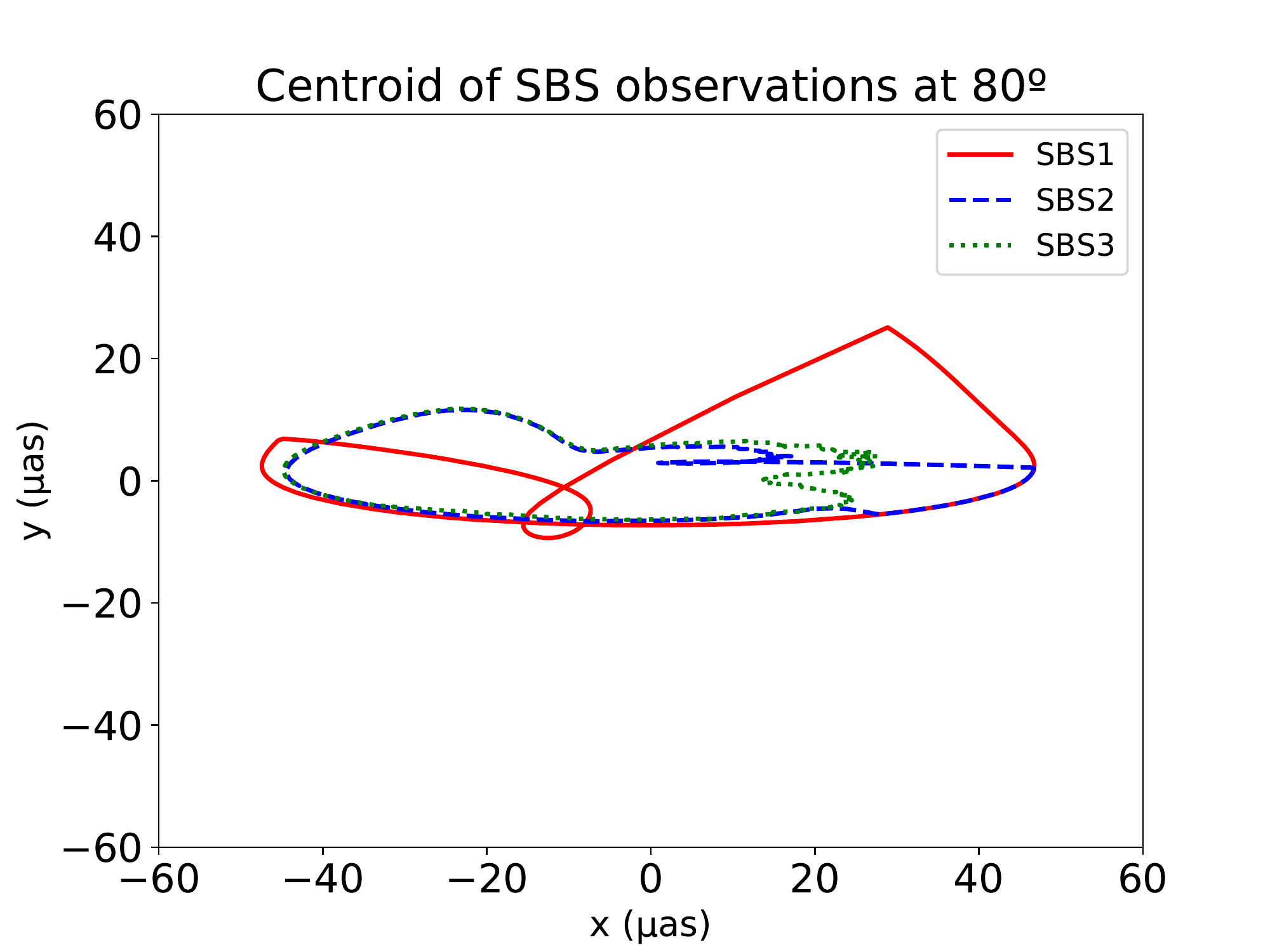}
\caption{Centroid of the observations in $\mu as$ for the $\Lambda$BS models (top row) and the SBS models (bottom row) for an observation inclination of $\theta=20^\circ$ (left column), $\theta=50^\circ$ (middle column), and $\theta=80^\circ$ (right column).}
\label{fig:centroids}
\end{figure*}

The qualitative behavior of both the magnitude $m_k$ (depicted in Fig. \ref{fig:magnitudes} for both the $\Lambda$BS and SBS configurations) and the centroid $\vec{c}_k$ (depicted in Fig. \ref{fig:centroids}) are strongly dependent on the sub-image structure of the observation. If a single track, i.e., the primary track, is observable, the magnitude of the observation features a single peak caused by the Doppler shifting due to the orbital motion, whereas the centroid of the observation follows the position of the primary image, as it happens for all of the $\Lambda$BS models and for an observation angle of $20^\circ$. The slight difference in the height of these peaks is caused by the differences in the angular velocity of the hot-spot, which is slightly larger for the more compact configurations.

If at some point of the orbit a secondary image appears, one observes an increase in the magnitude of the observation caused by the extra photons arising at the observer from the secondary image, and the centroid of the observation is shifted towards the secondary image. This effect can be clearly observed for the $\Lambda$BS3 model for an observation inclination of $\theta=50^\circ$, as well as the $\Lambda$BS1, $\Lambda$BS2, and SBS1 for an observation inclination of $\theta=80^\circ$. Note that if the effects of light deflection are strong enough to break the two components of the secondary track into two separated images, the secondary and the plunge-though, then the additional peak in the magnitude breaks into several sub-peaks, corresponding to the instants in which the secondary image appears and splits into two components, then both achieve a maximum of luminosity, and finally they merge and disappear. Depending on the relative intensity of these two components, the behavior of the centroid might follow a more complicated trajectory, as it happens for the $\Lambda$BS3 model at an observation angle of $\theta=80^\circ$ and for the SBS2 model at both $\theta=50^\circ$ and $\theta=80^\circ$.

When the light deflection is strong enough to induce the appearance of additional secondary tracks, the complexity of the behavior of the magnitude and the centroid increases. For low-inclination observations for which the additional tracks are present and do not merge, one observes that the centroid still follows an approximately elliptical curve, but this curve is smaller than in the case in which a single primary image is present, as the secondary contributions shift the centroid towards the center of the observation. Furthermore, for the magnitude, although it still features a single peak, the latter is smaller than in the case of a single primary image, as the photons corresponding to the secondary image arise to the observer from a trajectory crossing the central object through the opposite of the primary image, and thus contribute negatively to the Doppler shift. These effects are observed for the SBS2 model at $\theta=20^\circ$ and the SBS3 model for both $\theta=20^\circ$ and $\theta=50^\circ$. 

Finally, it is interesting to note that even though all observable tracks in the SBS3 model are visible independently of the observation angle, the contribution of the secondary tracks to the total flux increases with the observation angle, being particularly relevant in the region of the observers' screen opposite to the primary track. As a consequence, and even though the secondary tracks are always present, one can still observe the appearance of a additional peaks in the magnitude and consequent shifting of the centroid for the SBS3 model at an observation inclination of $\theta=80^\circ$. The main difference between this situation and the one described previously for which the secondary image appears at some point in the orbit, splits into two components, merges, and disappears again, is that in that situation one observes three additional peaks in the magnitude, whereas in this case only two additional peaks are present. Note that for all of the SBS3 model observations, both the magnitude and the centroid present a slight noise caused by the light-ring contribution.

\section{Accretion disks} \label{sec:IV}

\subsection{Intensity profiles}

Let us now turn to the observational properties of optically-thin accretion disks around the bosonic stars considered previously. For this purpose, we recur to a Mathematica-based ray-tracing code previously used in several other publications \cite{Guerrero:2021ues,Olmo:2023lil}, where the (infinitesimally-thin) accretion disk at the equatorial plane is modelled by a monochromatic intensity profile. To model these intensity profiles, we recur to the recently introduced Gralla-Lupsasca-Marrone (GLM) model \cite{Gralla:2020srx}, whose main interest is the fact that its predictions are in a close agreement with those of general relativistic magneto-hydrodynamics simulations of astrophysical accretion disks \cite{Vincent:2022fwj}. The intensity profile of the GLM model is given by
\begin{equation}\label{eq:GLMmodel}
I_e\left(r;\gamma,\beta,\sigma\right)=\frac{\exp\left\{-\frac{1}{2}\left[\gamma+\text{arcsinh}\left(\frac{r-\beta}{\sigma}\right)\right]^2\right\}}{\sqrt{\left(r-\beta\right)^2+\sigma^2}},
\end{equation}
where $\gamma$, $\beta$ and $\sigma$ are free parameters controlling the shape of the emission profile, namely the rate of increase, a radial translation, and the dilation of the profile, respectively. These parameters can be adjusted in order to select adequate intensity profiles for the models under study. For the purpose of this work, we select two different GLM models to model the intensity profile of the accretion disk, which we motivate in what follows.

\begin{itemize}

\item Given that all of the bosonic star configurations considered in this work feature stable orbital regimes close to the center of the star $r=0$, and under the assumption that the matter composing the accretion disk interacts only weakly with the fundamental fields composing the star, it is fair to assume that the intensity profile of the accretion disk increases monotonically from infinity downwards and peaks at the center. We denote this as the \textit{Central} accretion disk model, which is described by the parameters $\gamma=\beta=0$ and $\sigma=2M$ in the GLM model above. 

\item On the other hand, the SBS configurations feature marginally stable circular orbits at $r_{MS}\sim 6M$ (cf. Fig.~\ref{fig:ddpotential}). Given that circular orbits become unstable in a region of $r_o<r_{MS}$, and provided that the SBS models explored here have a maximum in $\Omega_o$, it is reasonable to consider an intensity profile of the accretion disk which increases monotonically down to $r=r_{MS}$, where it peaks, and then abruptly decreases for $r<r_{MS}$. This structure should be similar to the ones found in Ref.~\cite{Olivares:2018abq}, which shows through hydrodynamics simulations that some accretions disks in boson stars could have inner edges. We denote this as the \textit{ISCO} accretion disk model, given the similarity to the black hole case, being described by the parameters $\gamma=-2$, $\beta=6M$, and $\sigma=M/4$. Note that the $r_{ISCO}$ for the three SBS configurations is not exactly $6M$ and differs depending on the model. Nevertheless, to allow for a same-ground comparison of the results between these two models, we take $\beta=6M$ for every configuration. 

\end{itemize}

\begin{figure}
\includegraphics[scale=1]{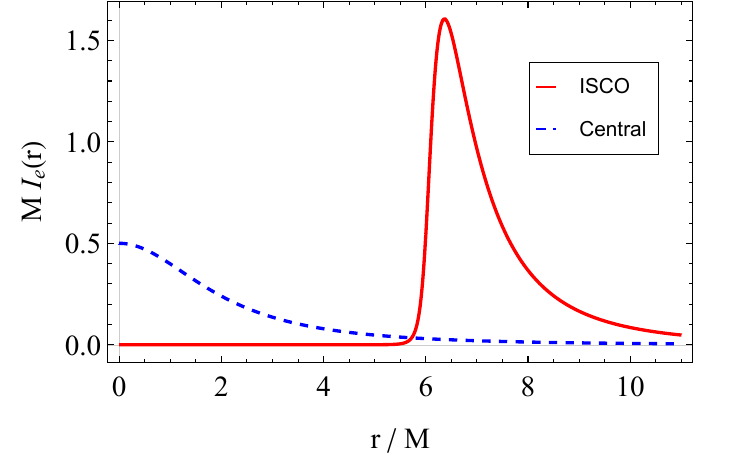}
\caption{Emitted intensity profiles for the GLM model given in Eq.\eqref{eq:GLMmodel} for two different combinations of parameters: the Central model with $\gamma=\beta=0$ and $\sigma=2M$; and the ISCO model with $\gamma=-2$, $\beta=6M$, and $\sigma=M/4$.}
\label{fig:disks}
\end{figure}

The emitted intensity profiles $I_e$ for the Central and ISCO disk models are plotted in Fig.~\ref{fig:disks}. In what follows, the Central disk model is used in the background of all bosonic star configurations, i.e., for both $\Lambda$BS and SBS, whereas the ISCO disk model is used only for those configurations which feature an ISCO, i.e., only the SBS ones.

\begin{figure*}[t!]
\includegraphics[scale=0.77]{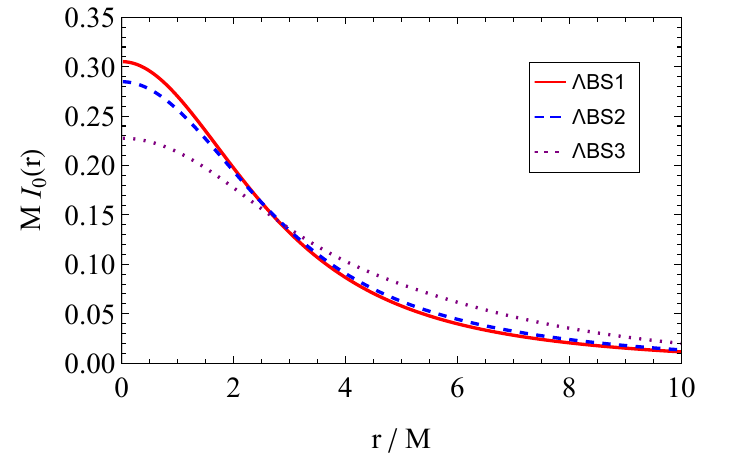}
\includegraphics[scale=0.77]{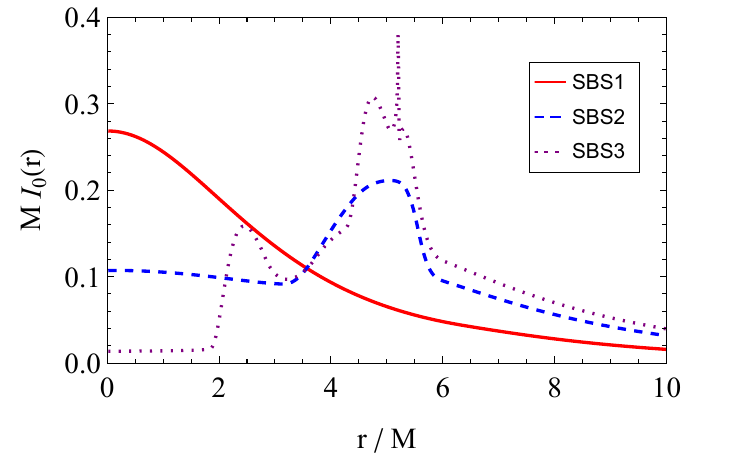}
\includegraphics[scale=0.77]{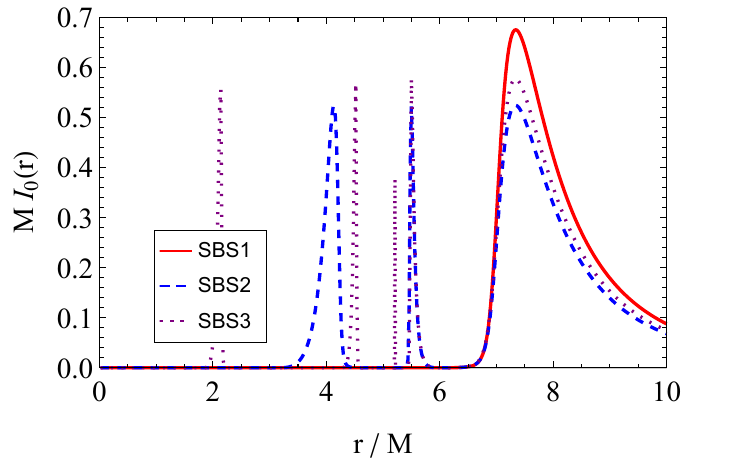}
\caption{Observed intensity profiles for the $\Lambda$BS configurations with the Central disk model (left panel), for the SBS configurations with the Central disk model (middle panel), and for the SBS boson stars with the ISCO disk model (right panel).}
\label{fig:intensity}
\end{figure*}

\subsection{Axial observations}

\begin{figure*}[t!]
\includegraphics[scale=0.5]{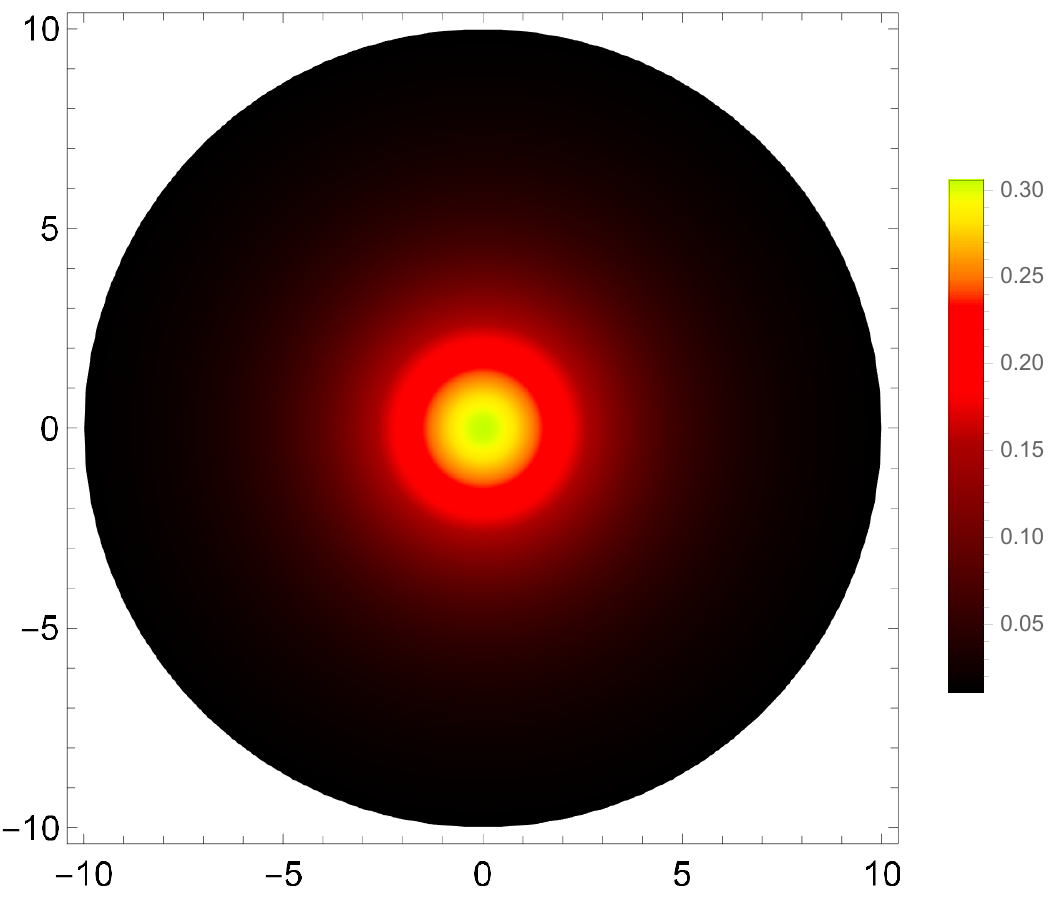}
\includegraphics[scale=0.5]{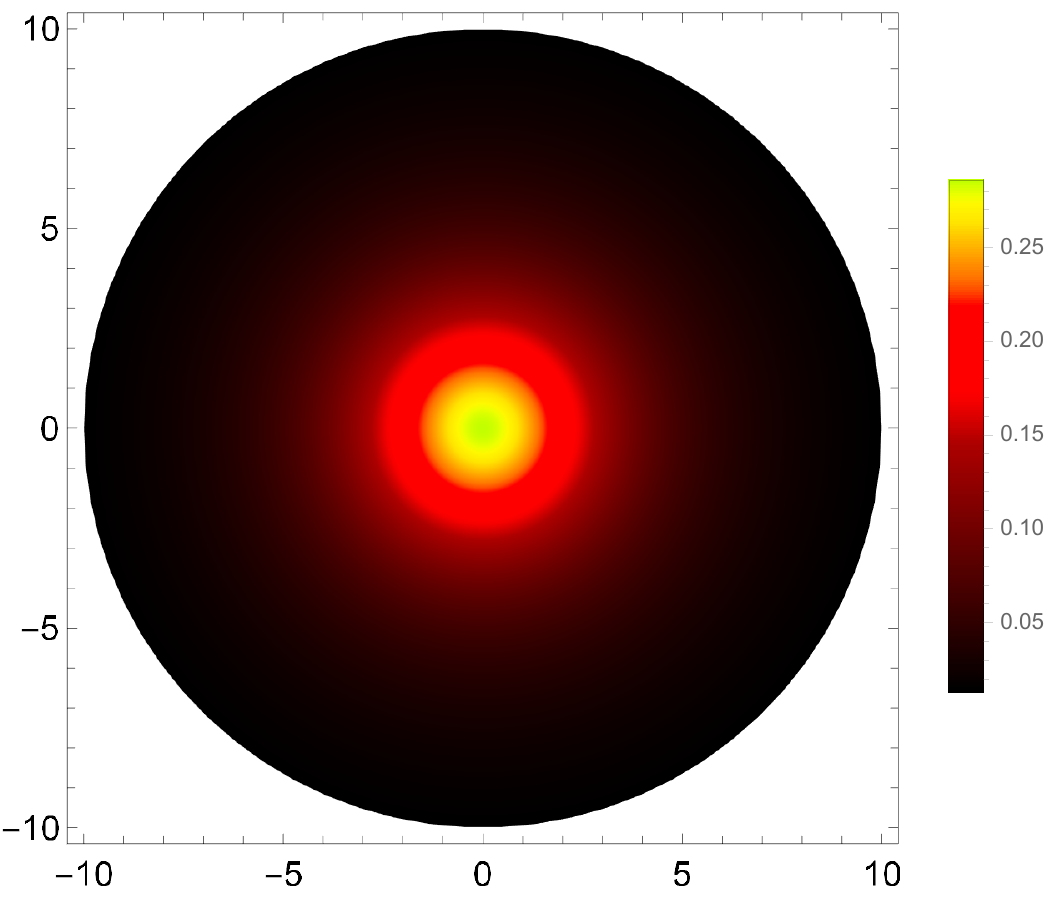}
\includegraphics[scale=0.5]{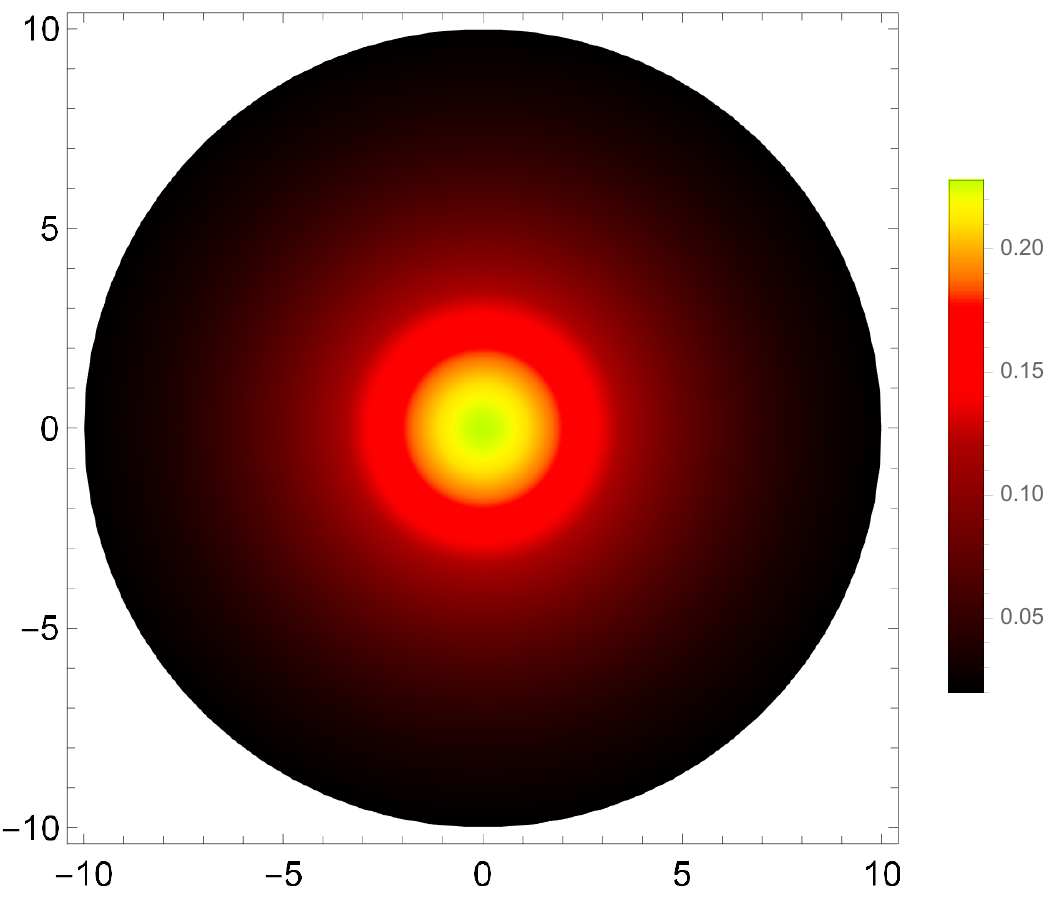}\\
\includegraphics[scale=0.5]{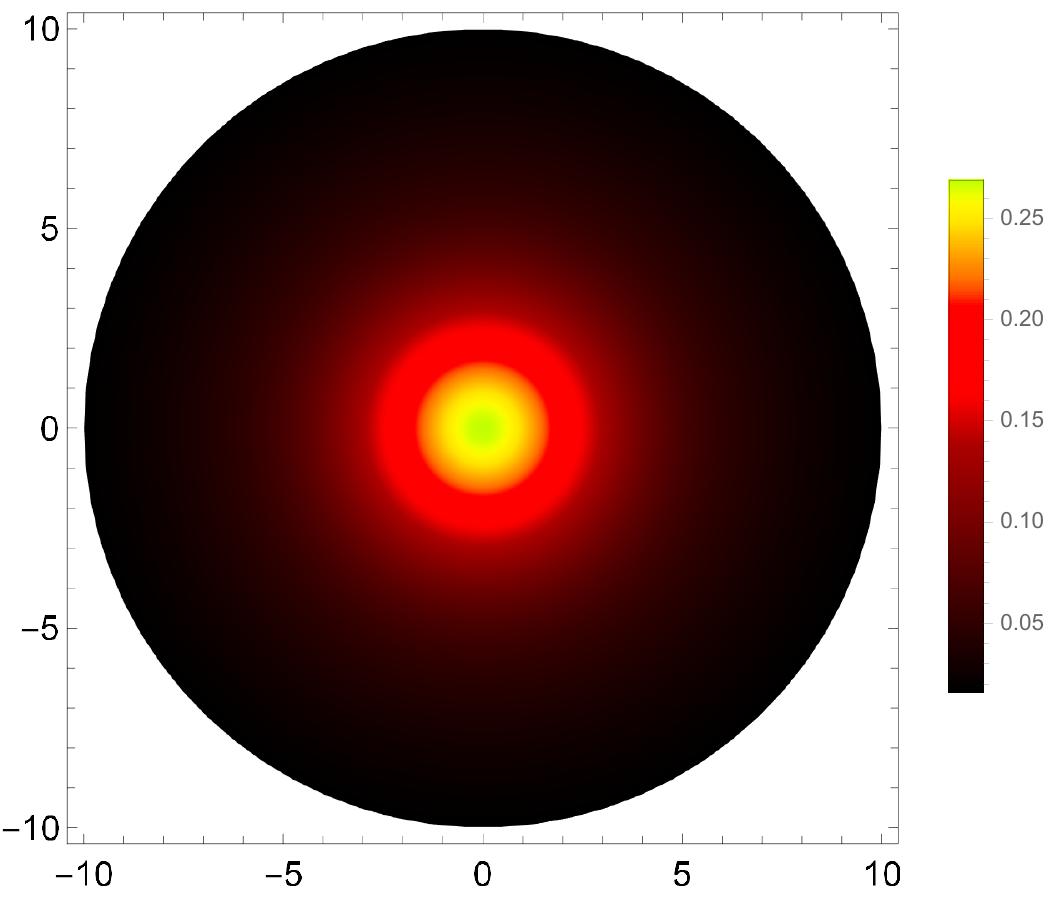}
\includegraphics[scale=0.5]{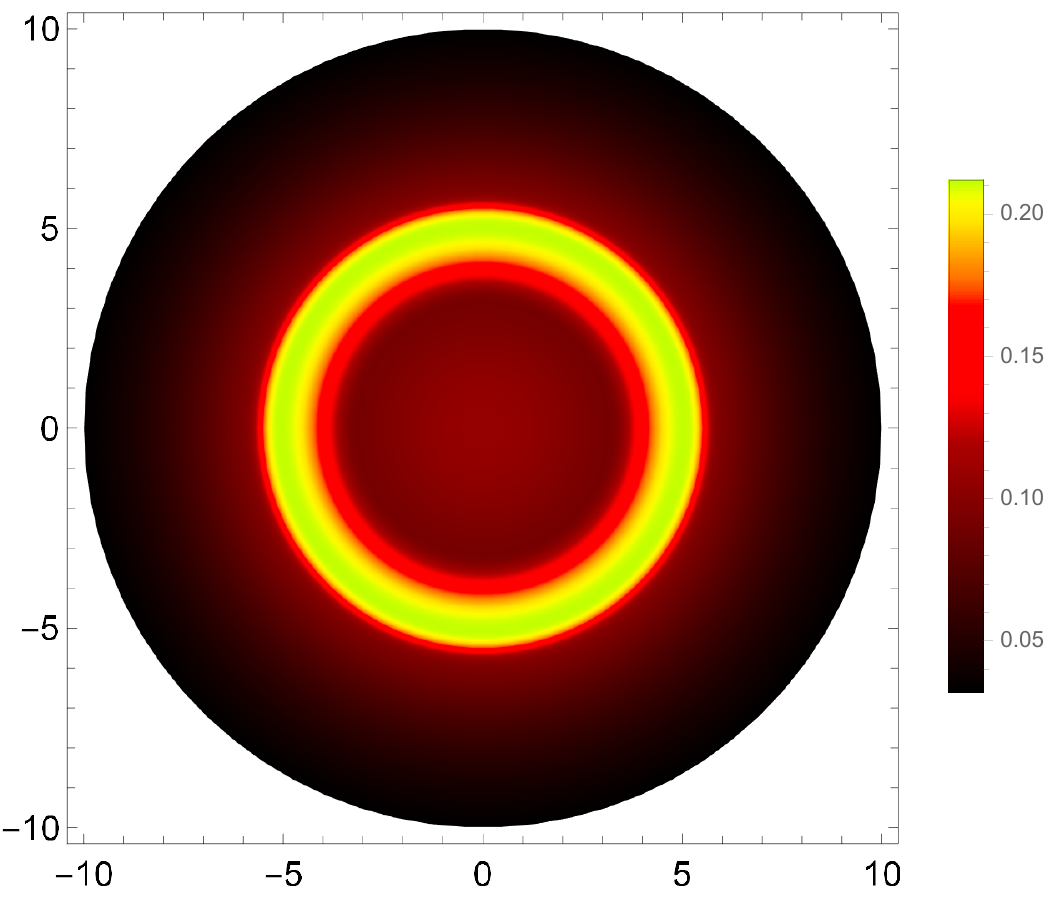}
\includegraphics[scale=0.5]{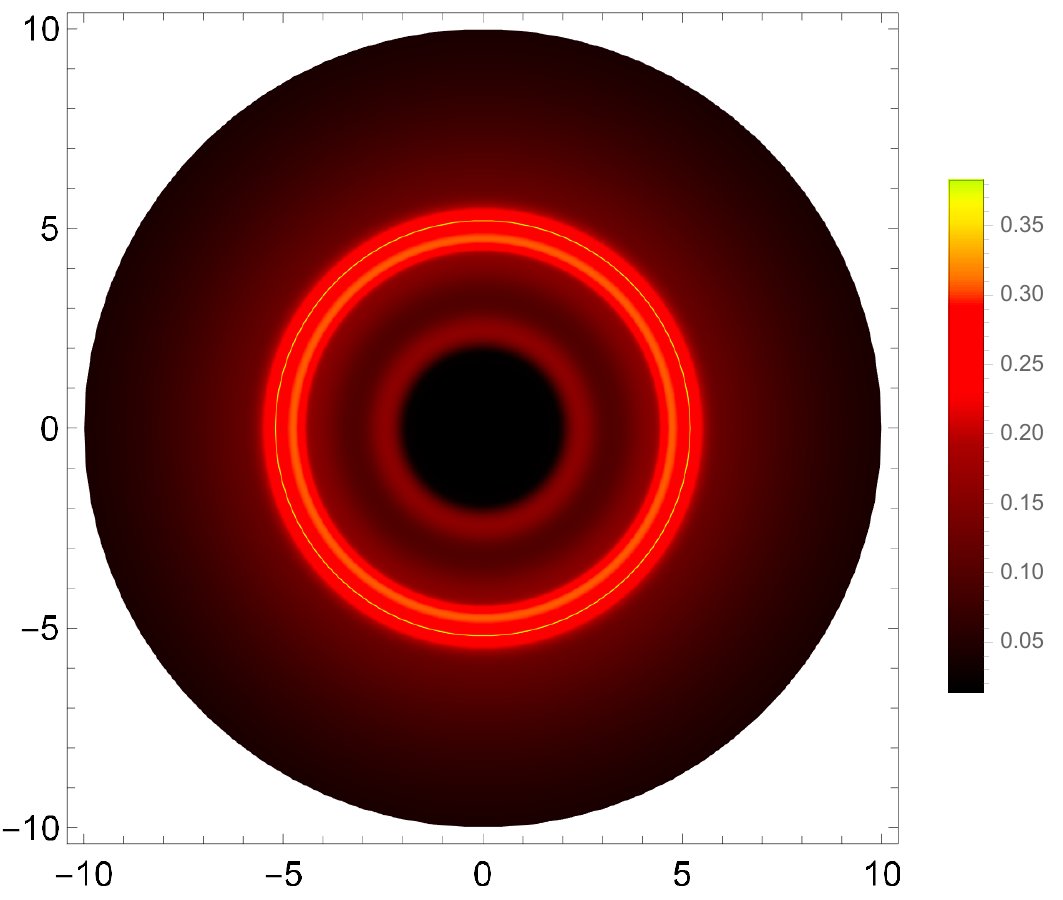}
\caption{Shadow images for the $\Lambda$BS (top row) and SBS (bottom row) configurations with the Central disk model, as seen from an inclination angle of $\theta=0^\circ$.}
\label{fig:shadows_axial_C}
\end{figure*}

\begin{figure*}[t!]
\includegraphics[scale=0.5]{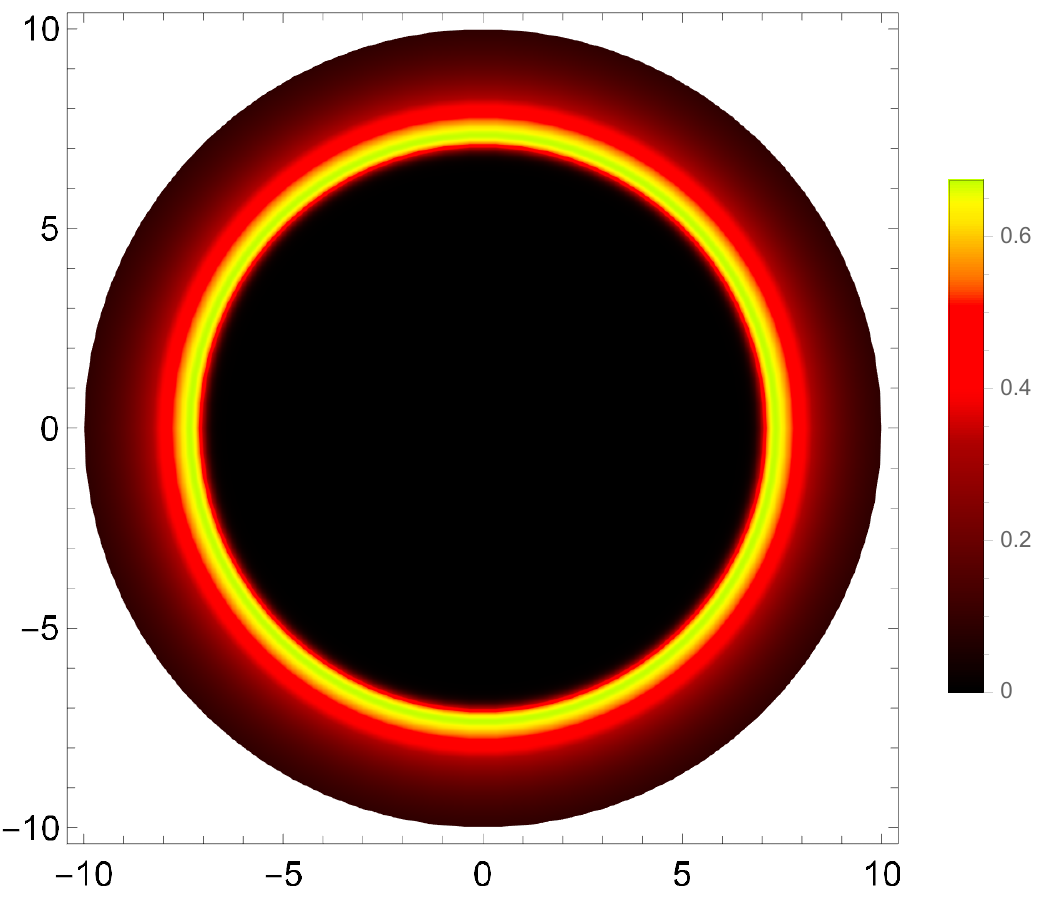}
\includegraphics[scale=0.5]{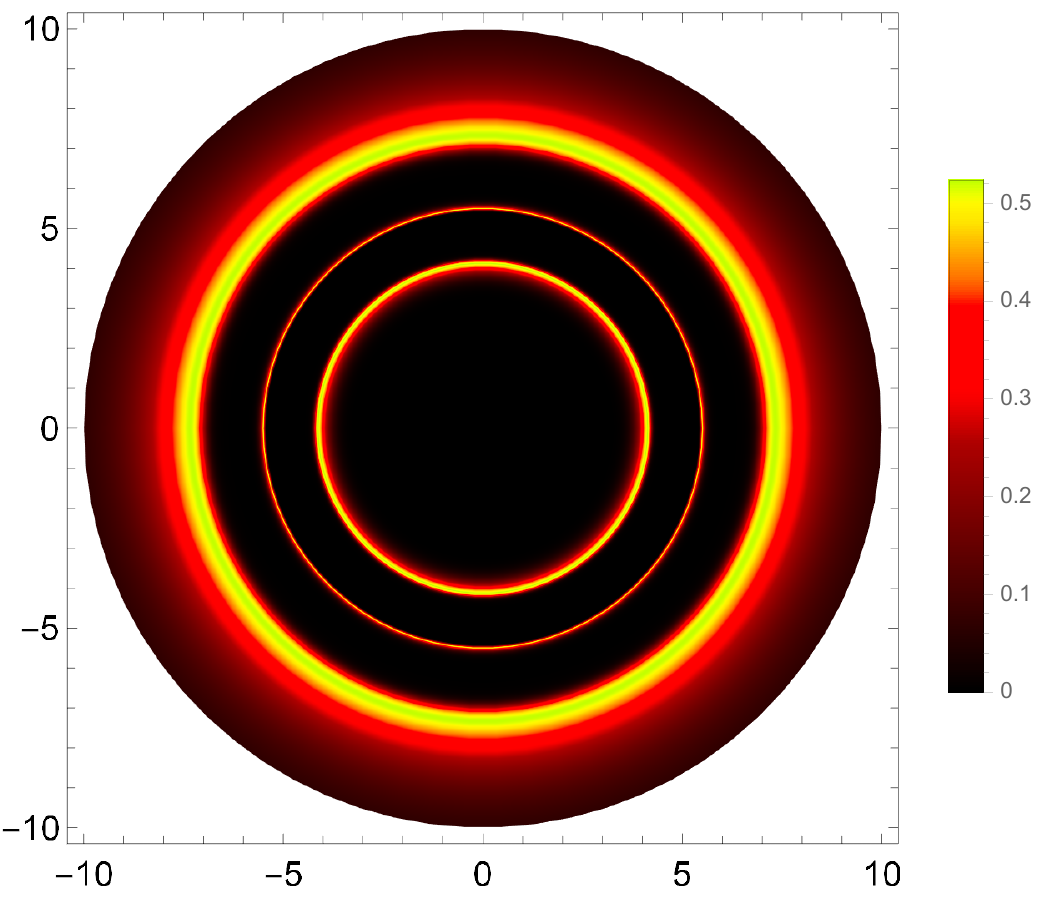}
\includegraphics[scale=0.5]{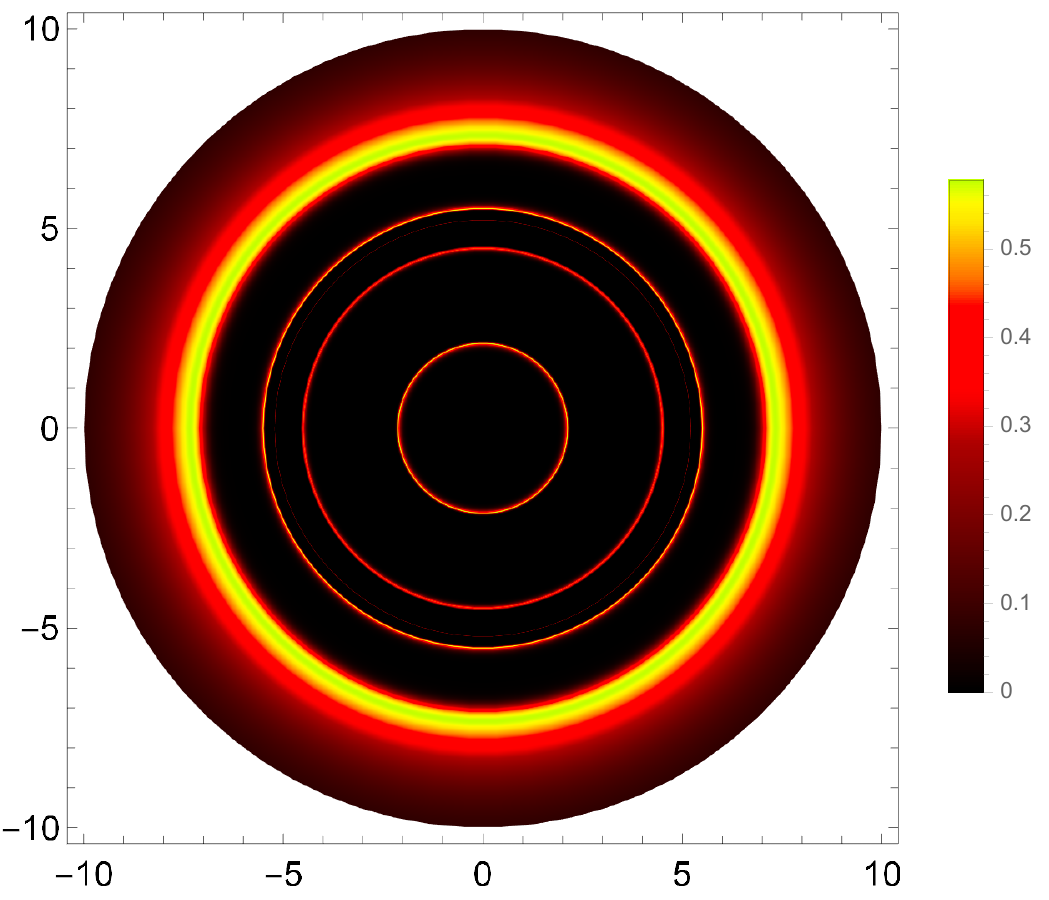}
\caption{Shadow images for the SBS configurations with the ISCO disk model, as seen from an inclination angle of $\theta=0^\circ$.}
\label{fig:shadows_axial_ISCO}
\end{figure*}

The intensity profiles given in Fig.\ref{fig:disks} correspond to the reference frame of the emitter $I_e$, i.e., the accretion disk, where the photons are emitted with a given frequency, say $\nu_e$. In the reference frame of the observer, the observed frequency $\nu_0$ is redshifted with respect to the emitted one, with $\nu_0=\sqrt{-g_{tt}}\nu_e$. Consequently, the intensity profile in the reference frame of the observer $I_0$ is affected by the shape of the background metric and takes the form
\begin{equation}
I_0\left(r\right)=A^2\left(r\right)I_e\left(r\right).
\end{equation}
The observed intensity profiles for the combinations of accretion disk models and bosonic star configurations outlined previously are given in Fig.\ref{fig:intensity}, whereas the corresponding observed axial images (i.e., as observed from the axis of symmetry of the accretion disk) for the Central and ISCO disk models are provided in Figs.\ref{fig:shadows_axial_C} and \ref{fig:shadows_axial_ISCO}, respectively. 


For the Central disk model, we verify that for every $\Lambda$BS configuration, as well as for the SBS1 configuration, the effects of the gravitational redshift are not strong enough to induce a decrease in the central intensity peak, and thus the observed images for these models present similar qualitative properties, more precisely a central blob of radiation. However, for the SBS2 and SBS3 configurations, one verifies that the effects of the gravitational redshift induce a strong dimming in the central peak, leading to a maximum of intensity away from the center. This dimming of intensity produces a shadow-like feature for these two models, inducing a brightness depression region in the center of the observed images. Furthermore, one also verifies that for these two models the light deflection is strong enough to produce additional contributions in the observed images caused by photons that have revolved around the central object more than a half orbit. These are known as the secondary images, similar to those already found in the hot-spots above, and produce the additional peaks of intensity visible for the SBS2 and SBS3 configurations. Furthermore, one can also observe a thin peak in the intensity profile of the SBS3 configuration, corresponding to the light-ring, which is also visible as a thin intense circle in the observed image. 

\begin{figure*}[t!]
\includegraphics[scale=0.4]{figs/SBS2_0deg_C.pdf}
\includegraphics[scale=0.4]{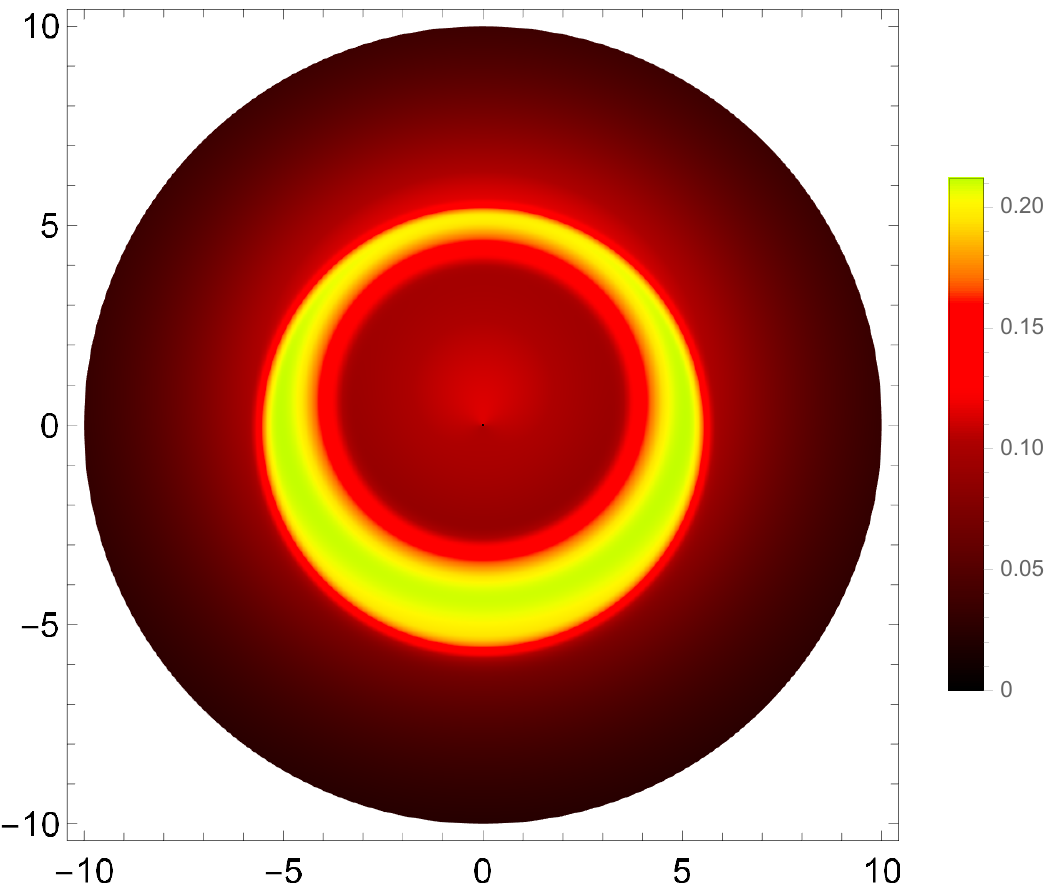}
\includegraphics[scale=0.4]{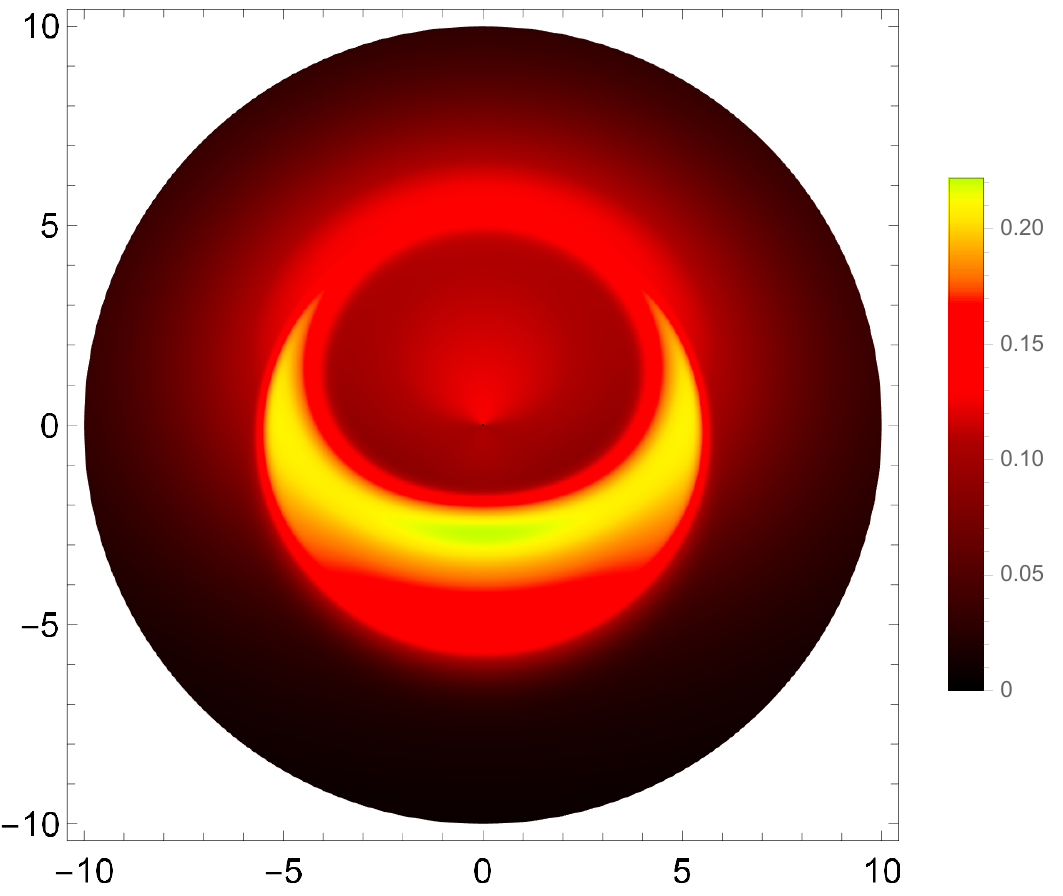}
\includegraphics[scale=0.4]{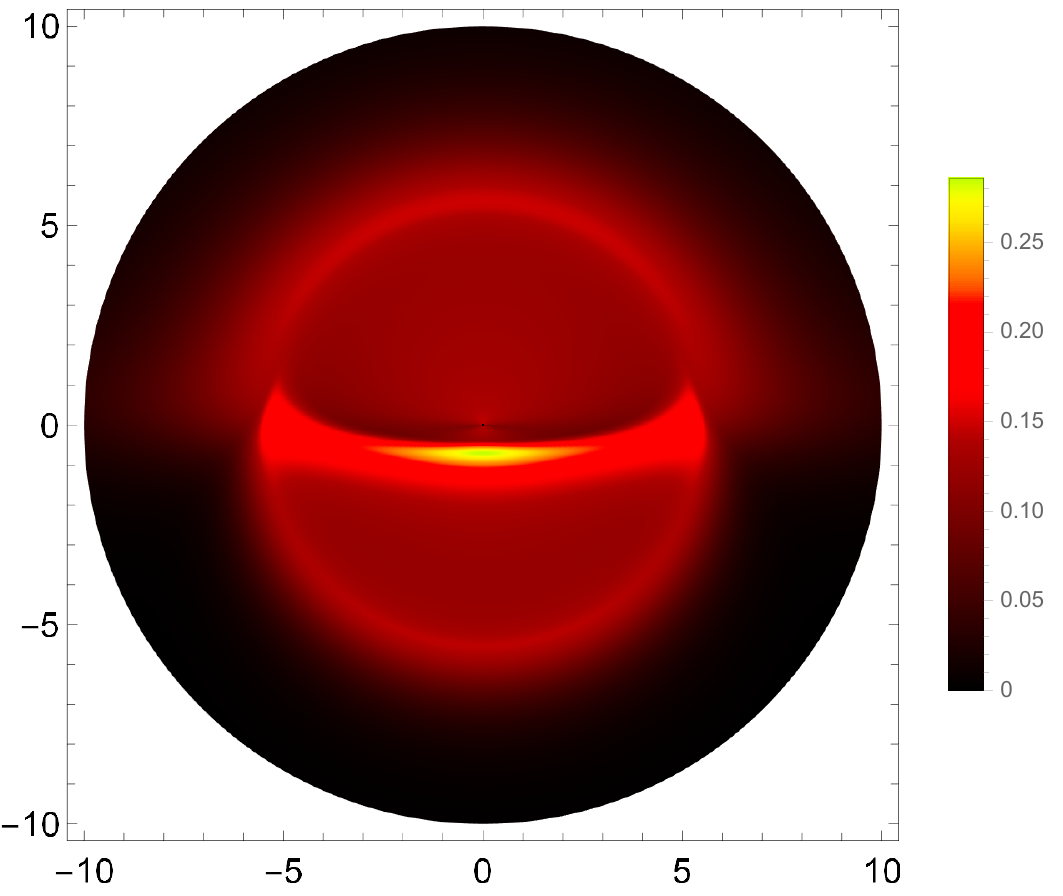}\\
\includegraphics[scale=0.4]{figs/SBS3_0deg_C.pdf}
\includegraphics[scale=0.4]{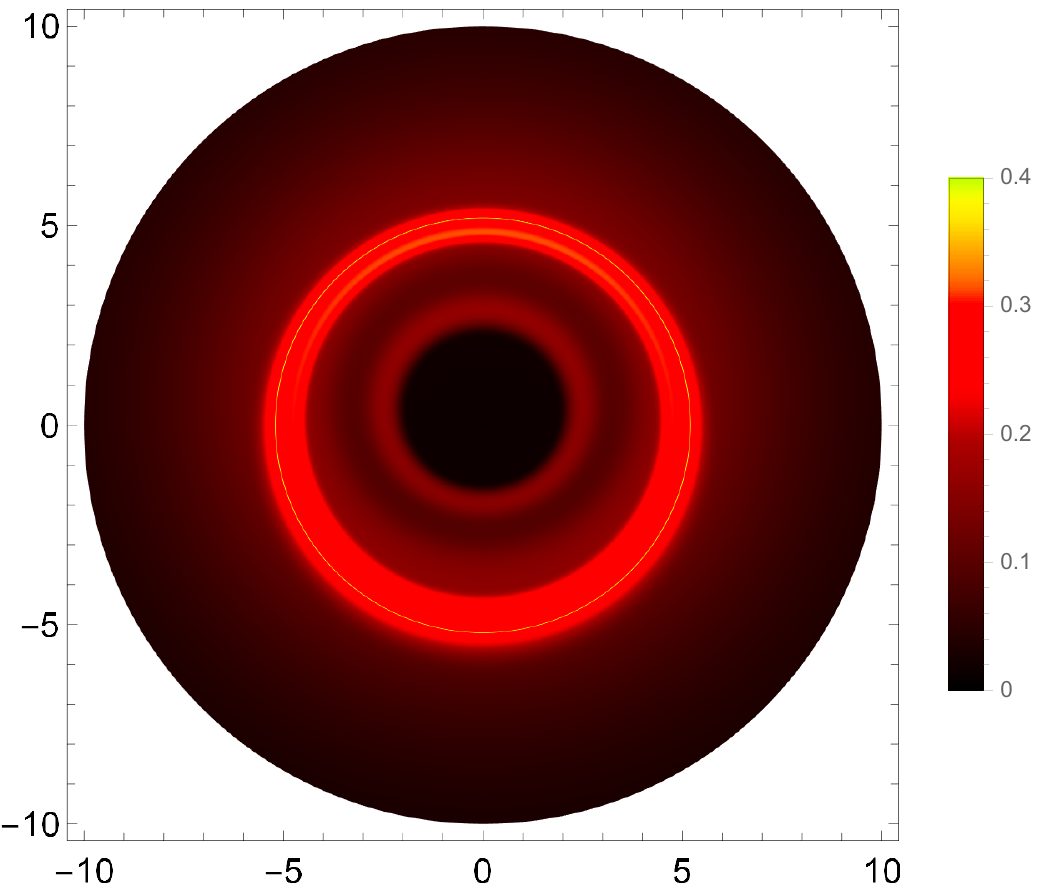}
\includegraphics[scale=0.4]{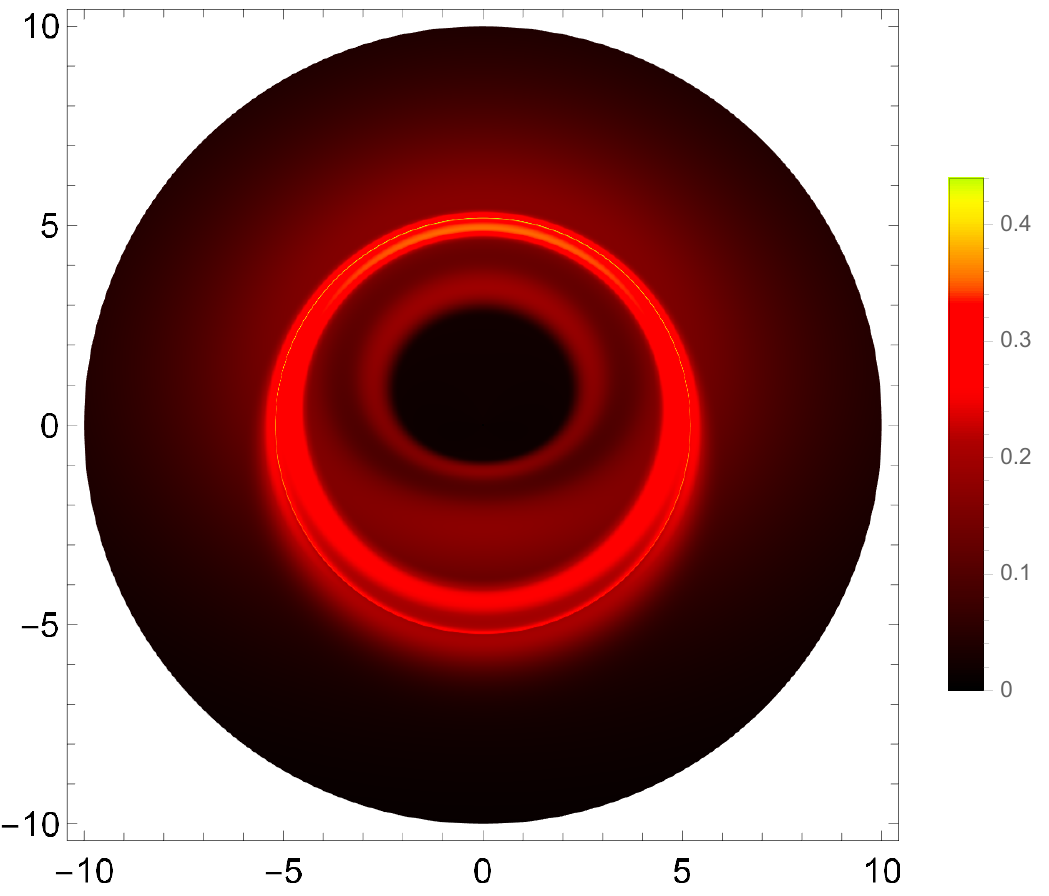}
\includegraphics[scale=0.4]{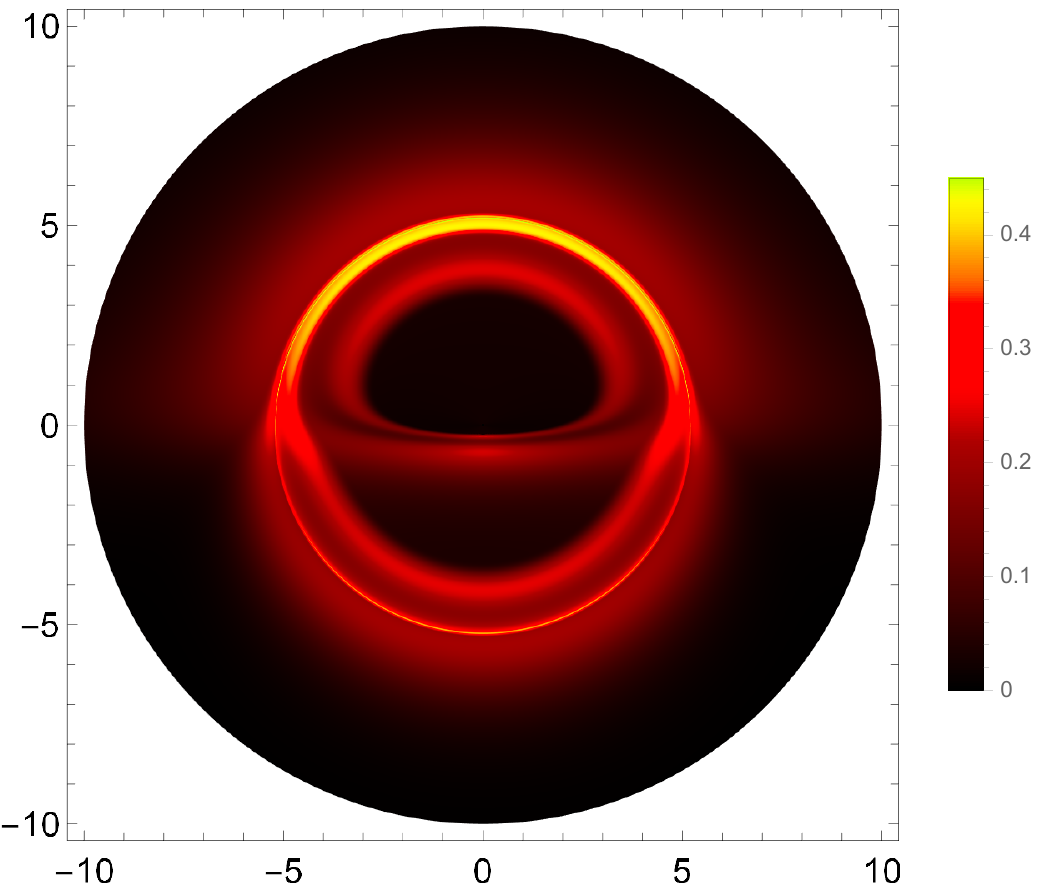}
\caption{Shadow images for the SBS2 (top row) and SBS3 (bottom row) bosonic stars with the Central disk model, as seen from inclination angles of $\theta=\{0^\circ, 20^\circ, 50^\circ, 80^\circ\}$, from left to right.}
\label{fig:shadows_inclined_C}
\end{figure*}

As for the ISCO disk model, given that the intensity profiles in the reference frame of the emitter are truncated at a finite radius, namely $r=6M$, all observed images produced with this model feature a central dark region independently of the bosonic star configuration considered as a background. Nevertheless, bosonic stars with different compactness and geodesic structures feature  qualitatively different behaviours. For the SBS1 configuration, the light deflection is not strong enough to produce secondary images, and thus the observed intensity profile features a single peak, which is translated into the observed image as a single ring and a dark shadow without any additional features. When the light deflection is strong enough to produce a secondary image, additional peaks of intensity start appearing in the observed intensity profiles, which contribute with extra circular contributions to the observed image inside the previous shadow. The number of secondary images depends on the metric chosen as a background, varying from a single secondary image \cite{Rosa:2023hfm} to several, as it happens for the SBS2 and SBS3 models. In particular, for the SBS3 configuration, one observes three additional secondary peaks in the observed intensity, as well as the light-ring contribution, which are translated as four additional circular contributions in the observed image, inside the shadow.

The results described above indicate that the $\Lambda$BS configurations, along with the SBS1 configuration with the Central disk model, are not compact enough to reproduce the expected observable properties of black hole space-times, more specifically, the shadow observed in the images of the supermassive objects in the center of M87 and Sgr A$^*$ galaxies, and thus they do not correspond to adequate models for black hole mimickers in this astrophysical context, provided that the universality of black hole metrics hold. On the other hand, the SBS2 and SBS3 configurations, along with the SBS1 configuration with the ISCO disk model, do produce shadow-like features in the observed images, and are thus potentially suitable candidates for black hole mimickers in this context and deserve a more careful analysis.

\subsection{Inclined observations}

\begin{figure*}[t!]
\includegraphics[scale=0.4]{figs/SBS1_0deg_ISCO.pdf}
\includegraphics[scale=0.4]{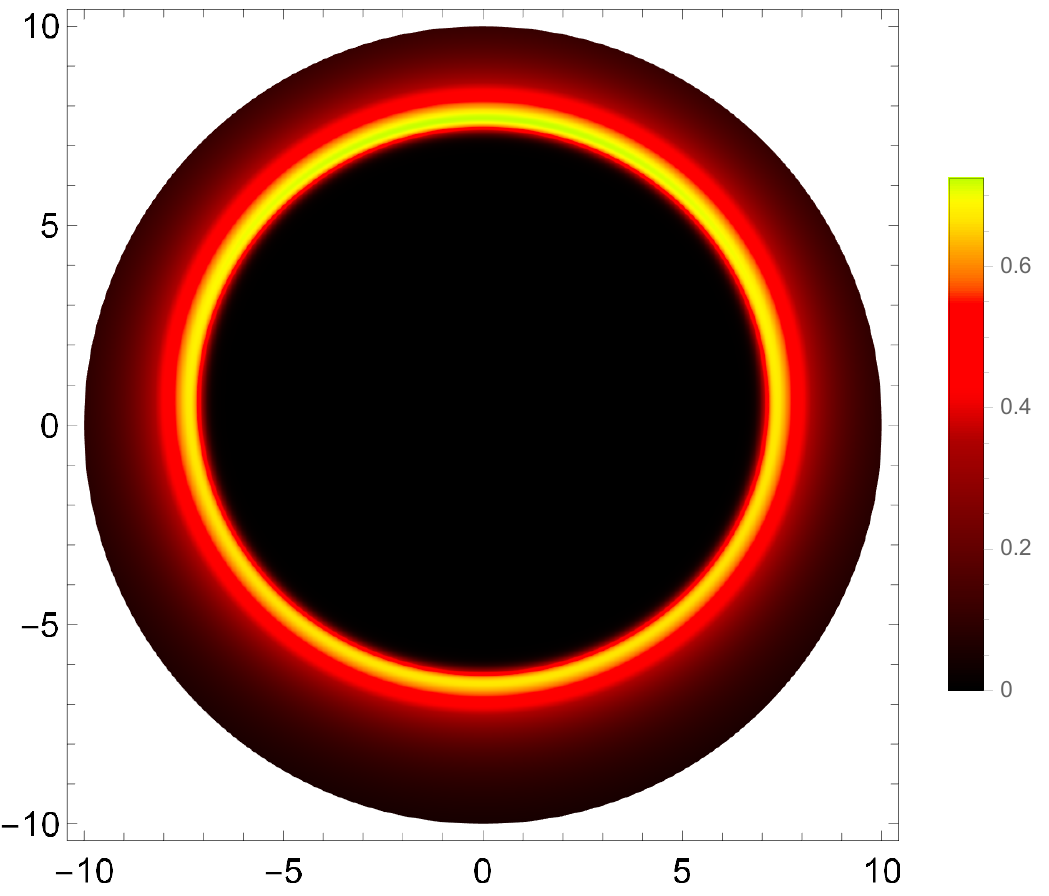}
\includegraphics[scale=0.4]{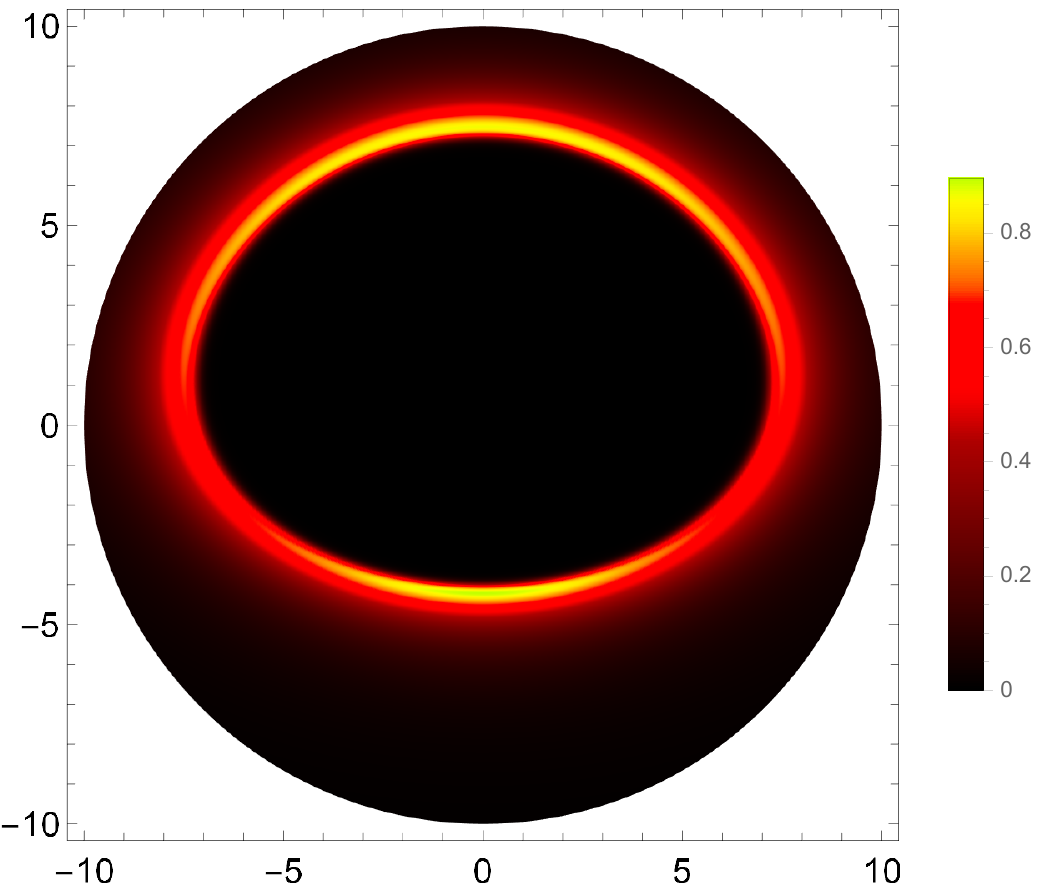}
\includegraphics[scale=0.4]{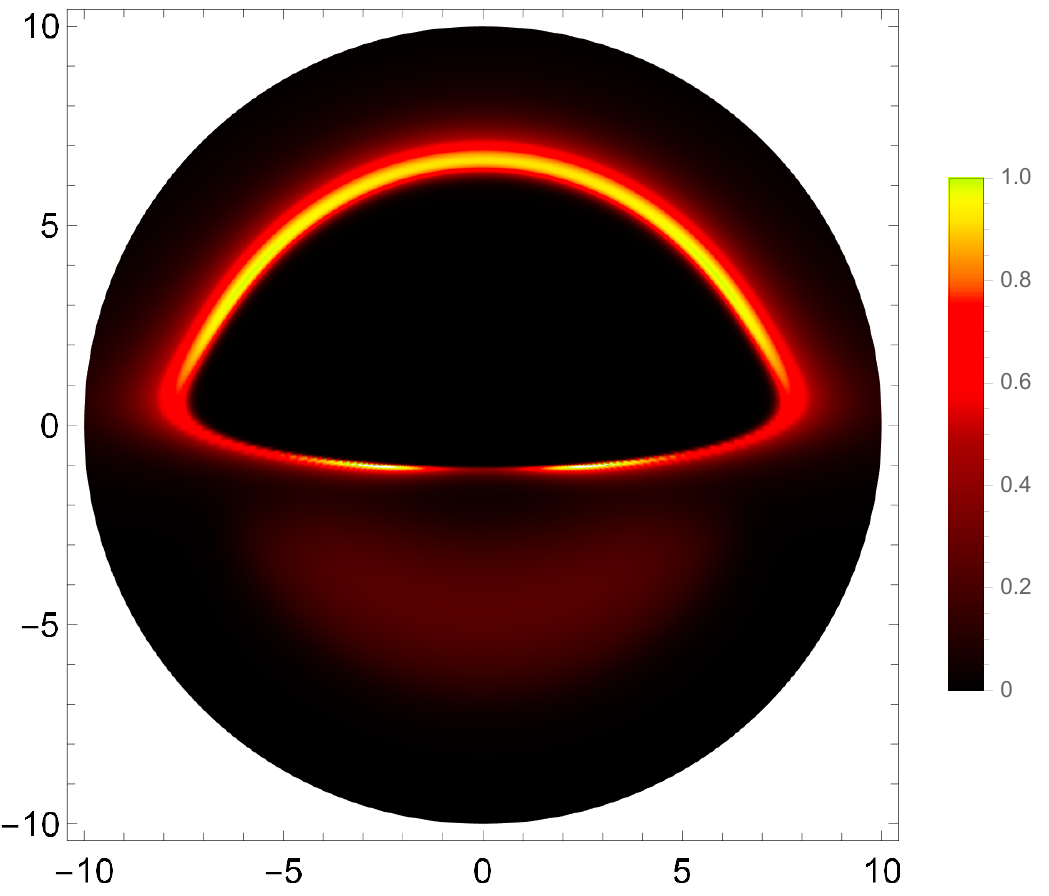}\\
\includegraphics[scale=0.4]{figs/SBS2_0deg_ISCO.pdf}
\includegraphics[scale=0.4]{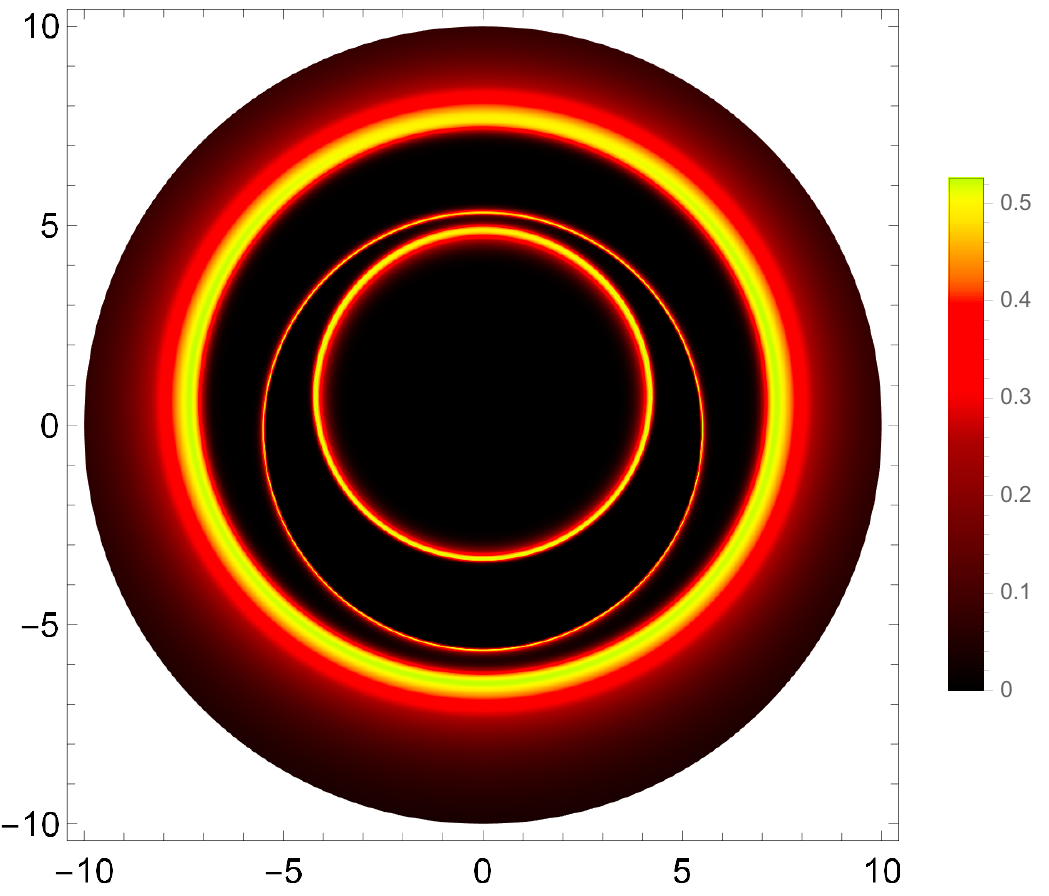}
\includegraphics[scale=0.4]{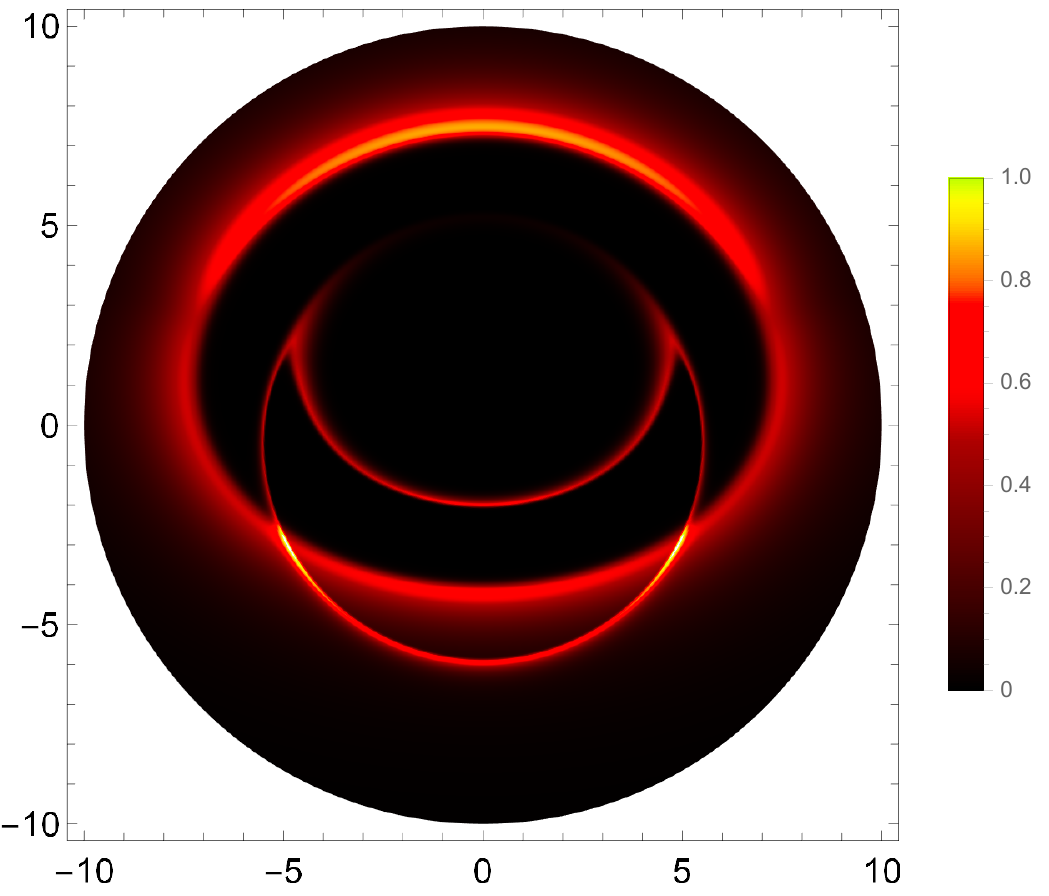}
\includegraphics[scale=0.4]{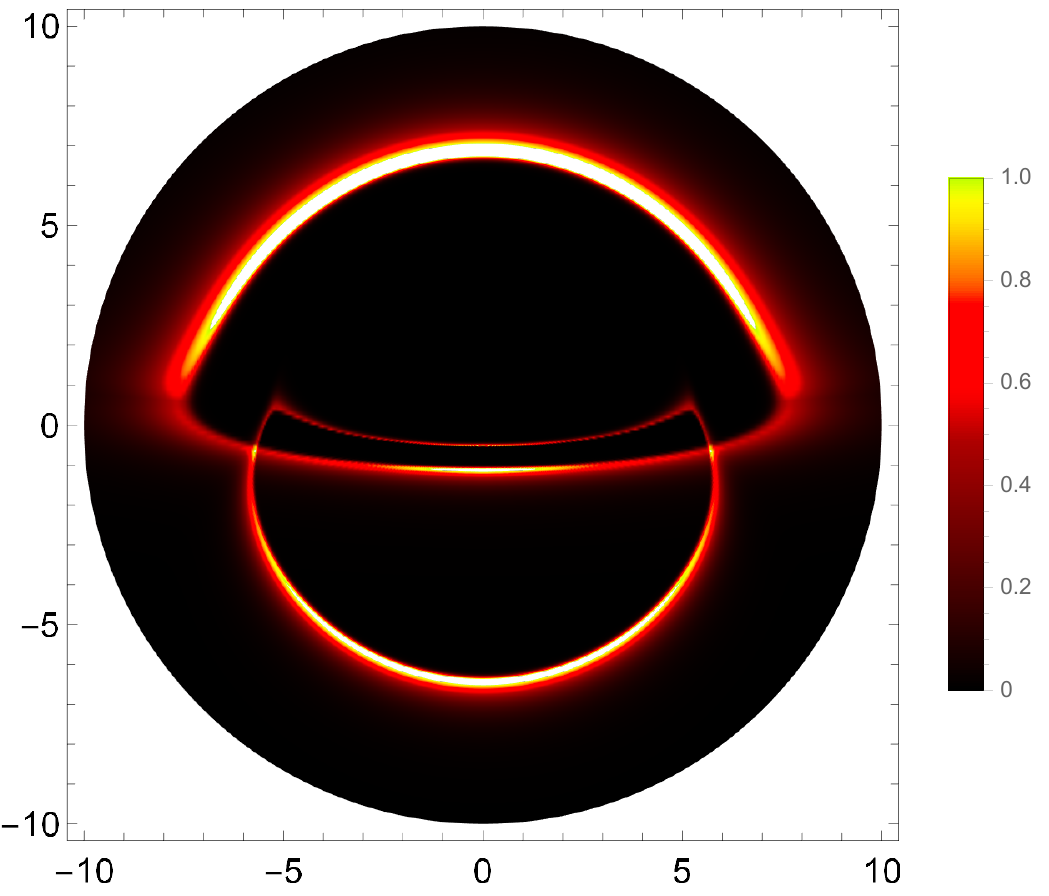}\\
\includegraphics[scale=0.4]{figs/SBS3_0deg_ISCO.pdf}
\includegraphics[scale=0.4]{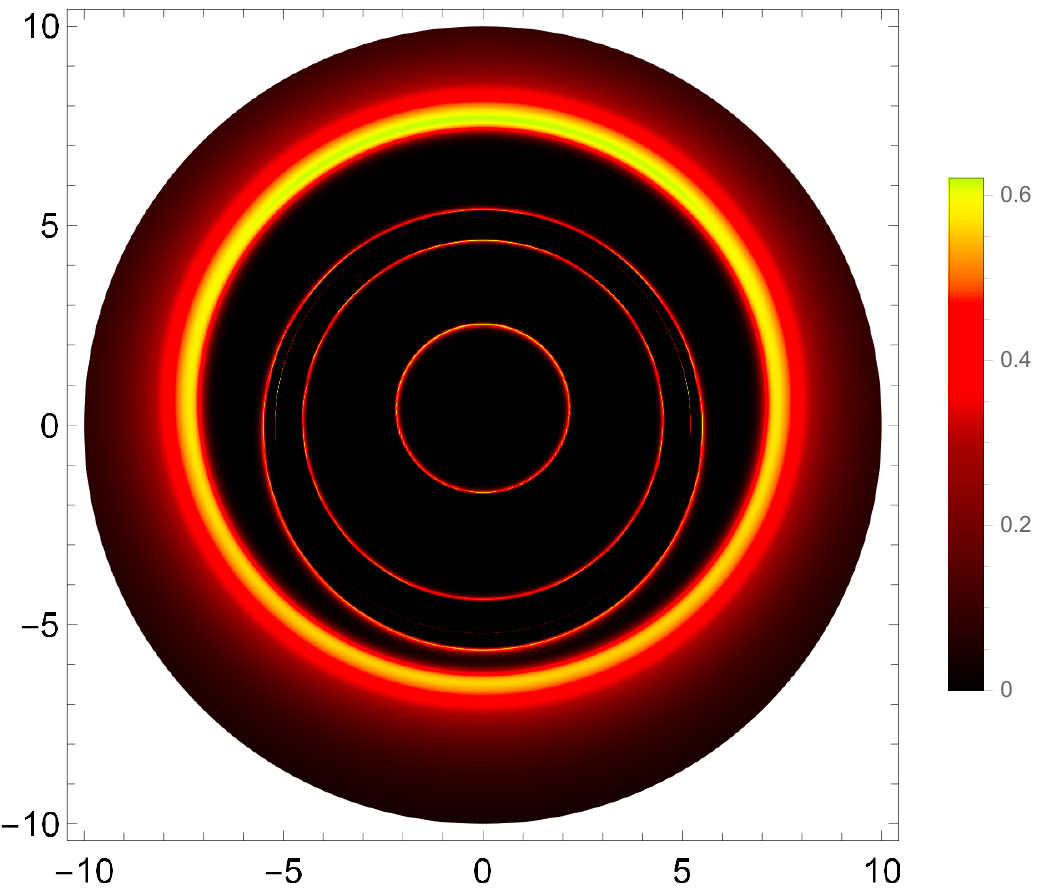}
\includegraphics[scale=0.4]{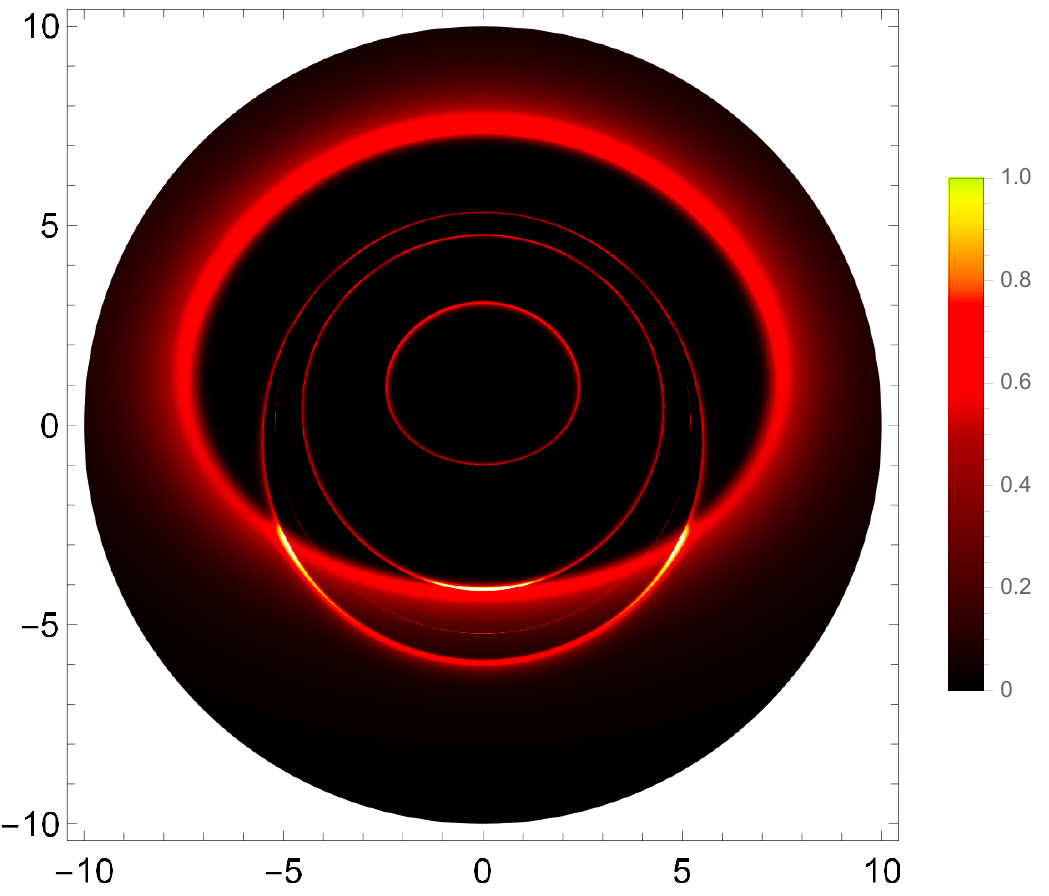}
\includegraphics[scale=0.4]{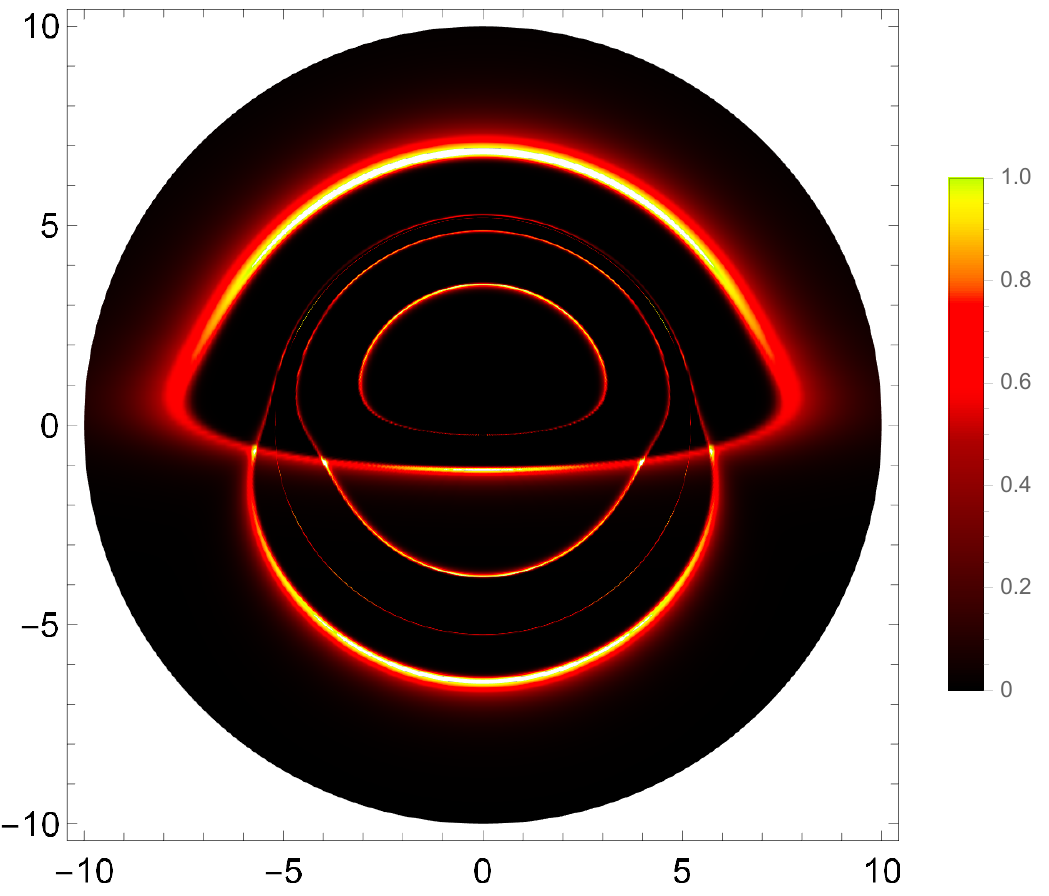}
\caption{Shadow images for the SBS1 (top row), SBS2 (middle row), and SBS3 (bottom row) configurations with the ISCO disk model, as seen from inclination angles of $\theta=\{0^\circ, 20^\circ, 50^\circ, 80^\circ\}$, from left to right.}
\label{fig:shadows_inclined_ISCO}
\end{figure*}

For the combinations of accretion disk models and bosonic star configurations deemed more astrophysically relevant as black hole mimickers in the previous section, we have produced additional images considering observers standing at inclination angles of $\theta=\{20^\circ, 50^\circ, 80^\circ\}$. As a comparison to our analysis here, observed images for the same inclination angles in the background of a Schwarzschild black hole can be found e.g. in Ref. \cite{Rosa:2023hfm}. The observed images for SBS2 and SBS3 with the Central disk model are given in Fig.\ref{fig:shadows_inclined_C}, whereas the observed images for all SBS configurations with the ISCO disk model are given in Fig.\ref{fig:shadows_inclined_ISCO}.

These results indicate that, even though the SBS2 configuration with a Central disk model presents a central dimming of radiation when observed axially, the contrast between the central dimming and the intensity of the secondary peak smoothens out as one increases the observation inclination, resulting in an observed image at high inclinations that differs drastically from the black hole scenario, see Ref. \cite{Rosa:2023hfm}. The same does not apply to the SBS3 model, for which a dark shadow-like region with a strong contrast with respect to the surrounding region near the light-ring is present independently of the inclination observation. It is worth to mention, however, that the size of the shadow of the SBS3 configuration is significantly smaller than its black hole scenario counterpart, mainly due to the secondary contributions that appear inside the light-ring, which in turn may trouble the compatibility of such models with calibrated observations of shadows' radius \cite{Vagnozzi:2022moj}. For the ISCO disk model, again one verifies that the SBS1, even though it produces a shadow similar to that of a black hole from an axial inclination perspective since light deflection is not strong enough to produce a secondary image, the resulting observation at high inclinations differs drastically from that of a black hole. The SBS2 and SBS3, on the other hand, do produce observed images featuring secondary images, and thus are more closely related to their black hole counterparts. Nevertheless, one can still enumerate several qualitative differences between these models and the black hole scenario, namely the plunge-through image in the absence of a light in the SBS2, a feature similar to what was previously found for boson stars without self interactions \cite{Rosa:2022tfv}, and several additional secondary tracks inside the light for SBS3, features that can effectively act as observational discriminators between boson stars of this kind and black hole space-times.

\section{Conclusion}  \label{sec:V}

In this work we have analyzed the observational properties of bosonic stars with self-interactions being orbited by isotropically emitting sources and optically-thin accretion disks. In particular, we studied bosonic stars with quartic interaction terms ($\Lambda$BS models), as well as solitonic boson stars with sixth-order interaction terms (SBS models). The latter models were proven to be the most interesting in an astrophysical context, as they more closely reproduce the observable predictions of black hole space-times and thus provide adequate models for black hole mimickers.

Indeed, we have shown that the light deflection effects in the $\Lambda$BS models are not strong enough to produce any qualitative differences with respect to the observations from boson and Proca stars without self-interactions, i.e., the same astrometric effects for orbital motion e.g. the shifting of the centroid and additional peaks of magnitude when the secondary tracks are present, as well as a weak central intensity dimming in accretion disk models that extend all the way down to the center of these configurations. We thus conclude that these models can hardly be taken as strong candidates to represent current observations from the EHT and GRAVITY collaborations, and are thus not adequate to describe supermassive compact objects found in galactic centres.

As for the SBS models, we verified that these are potentially relevant in this astrophysical context, provided that they are compact enough. For the least compact of these configurations, namely SBS1, the observational properties are similar to the ones of $\Lambda$BS configurations and bosonic stars without self-interactions, and thus inadequate to describe the images of the objects at the galactic centres. However, this is not true for the SBS2 and SBS3 models. Indeed, for the latter models one observes a dimming of intensity in the central region of the accretion disk caused by the gravitational redshift, resulting in a shadow-like feature similar to that of a black hole. For the SBS3 model, this dark region is more pronounced and remains visible for any inclination angle, although being slightly smaller than its black hole counterpart. Furthermore, both the SBS2 and SBS3 configurations present additional secondary images for both the orbital motion of a hot-spot and the optically-thin accretion disk, and the SBS3 configuration features also light-ring contributions. 

The qualitative differences between the SBS2 and SBS3 models with respect to the black hole scenario indicate that, even though these models are virtually indistinguishable from a black hole given the lack of enough resolution in current EHT observations to resolve secondary images in the main ring of radiation, an eventual upgrade in these observatories (via e.g. the ngEHT) and an increase in the quality and resolution of the observed images may allow the detection of these additional contributions to the image in order to conclusively infer the nature of these supermassive compact objects. 

Furthermore, it is worth to notice the similarities between the observed images for the ISCO model given in Fig.\ref{fig:shadows_inclined_ISCO} and the integrated fluxes of hot-spot orbits given in Fig.\ref{fig:SBSflux}, which emphasize the correctness of the results. Furthermore, the GYOTO software used to produce the results for the orbital motion of a hot-spot had been proven to produce high-accuracy results in several contexts \cite{Vincent:2015xta}, which emphasizes the validity of our Mathematica-based ray-tracing code as a high-precision tool for the study of light-deflection in the strong-field regime of gravity.

\section*{Acknowledgements}

J. L. R acknowledges the European Regional Development Fund and the programme Mobilitas Pluss for
financial support through Project No.~MOBJD647, and project No.~2021/43/P/ST2/02141 co-funded by the Polish National Science Centre and the European Union Framework Programme for Research and Innovation
Horizon 2020 under the Marie Sklodowska-Curie grant agreement No. 94533.; C.F.B.M. thanks Fundação Amazônia de Amparo a Estudos e Pesquisas (FAPESPA), Conselho Nacional de Desenvolvimento Científico e Tecnológico (CNPq)
and Coordenação de Aperfeiçoamento de Pessoal de Nível
Superior (CAPES) – Finance Code 001, from Brazil, for partial financial support; DRG is supported by Grant PID2019-108485GB-I00, funded by MCIN/AEI/10.13039/501100011033. This article is based upon work from COST Actions CA18108 and CA21136.

\bibliographystyle{unsrt}
\bibliography{main}

\end{document}